\documentclass[12pt]{article}
\usepackage{enumerate}
\usepackage{amsfonts}
\usepackage{amsmath}
\usepackage{amssymb}
\usepackage{amsthm}
\usepackage{latexsym}
\usepackage{color}

\def\H{\mathcal{H}}

\def\FF{\mathcal{F}}
\def\S{\mathfrak{S}}
\def\C{\mathfrak{C}}
\def\T{\mathfrak{T}}

\def\B{\mathfrak{B}}

\def\N{\mathbb{N}}

\newcommand{\supp}{\mathrm{supp}}
\newcommand{\rank}{\mathrm{rank}}
\newcommand{\id}{\mathrm{Id}}
\newcommand{\Tr}{\mathrm{Tr}}

\newcommand{\shs}{\hspace{1pt}}
\newcounter{defin}  \newcounter{lemma}  \newcounter{theorem}
\newcounter{property} \newcounter{corol}  \newcounter{remark} \newcounter{example}
\newenvironment{lemma}{\par\refstepcounter{lemma}     \textbf{Lemma \thelemma.} }{\rm\par}
\newenvironment{theorem}{\par\refstepcounter{theorem}     \textbf{Theorem \thetheorem.}\ }{\rm\par}
\newenvironment{property}{\par\refstepcounter{property}     \textbf{Proposition \theproperty.}\ }{\rm\par}
\newenvironment{corollary}{\par\refstepcounter{corol}     \textbf{Corollary \thecorol.} }{\rm\par}
\newenvironment{definition}{\par\refstepcounter{defin}     \textbf{Definition \thedefin.}\ }{\rm\par}
\newenvironment{remark}{\par\refstepcounter{remark}     \textbf{Remark \theremark.}}{\rm\par}
\newenvironment{example}{\par\refstepcounter{example}     \textbf{Example \theexample.}}{\rm\par}

\begin{document}

\title{Approximation of multipartite quantum states: revised version with new applications}
\author{M.E.~Shirokov\footnote{Steklov Mathematical Institute, Moscow, Russia, email:msh@mi.ras.ru}}
\date{}
\maketitle
\begin{abstract}
An universal approximation technique for analysis of different characteristics of states of composite
infinite-dimensional  quantum systems is proposed and used to prove  general results concerning the properties
of correlation and entanglement measures in such systems. Then these results are applied to the study of three important characteristics:
the relative entropy of $\pi$-entanglement, the  Rains bound (the unregularized and regularized versions of both characteristics are considered)
and the conditional entanglement of mutual information.

In particular, we analyse continuity and convexity properties of the above entanglement measures, prove several results simplifying their definitions
and establish a finite-dimensional approximation property for these characteristics
that allows us to generalize to the infinite-dimensional case the results proved in the finite-dimensional settings.
\end{abstract}

\tableofcontents

\section{Introduction}

A specific feature of infinite-dimensional quantum systems consists in the singular properties (discontinuity, infinite values, etc.) of basic characteristics
of quantum states \cite{ESP,H-SCI,W}. So, for the strict mathematical analysis of infinite-dimensional quantum systems it is necessary to apply special techniques, in particular, approximation
techniques to overcome the problems arising from these singularities.

In this article we develop the approximation technique proposed in \cite{FAP} in the one-party case to analysis of characteristics of multipartite infinite-dimensional quantum systems.
A central notion of this technique is the finite-dimensional approximation property of a state (briefly called the FA-property). According to \cite{FAP}
a quantum state $\rho$ with the spectrum $\{\lambda^{\rho}_i\}$ has the FA-property if there exists a sequence $\{g_i\}$ of nonnegative numbers such that
\begin{equation*}
\sum_{i=1}^{+\infty}\lambda^{\rho}_i g_i<+\infty\quad \textrm{and} \quad \lim_{\beta\to 0^+}\left[\sum_{i=1}^{+\infty}e^{-\beta g_i}\right]^{\beta}=1.
\end{equation*}

The FA-property of a state $\rho$ implies the finiteness of its entropy \cite{FAP}, but the converse
implication is not true (this is shown in \cite{Becker} by constructing a subtle example of a state with finite entropy not possessing the FA-property). In Section 2.2 the following simple sufficient condition for the FA-property is proved: \emph{the FA-property holds for a state $\rho$ with the spectrum $\{\lambda^{\rho}_i\}$ provided that}\footnote{This condition strengthens the condition obtained in \cite{FAP} which states that
the FA-property holds for a state $\rho$ with the spectrum $\{\lambda^{\rho}_i\}$ provided that
$\sum_{i=1}^{+\infty}\lambda^{\rho}_i \ln^q i<+\infty$ for some $q>2$.}
$\sum_{i=1}^{+\infty}\lambda^{\rho}_i \ln^2 i<+\infty$.
This condition shows that the FA-property holds for all states whose eigenvalues tend to zero faster than $[i\ln^q i]^{-1}$ as $i\to+\infty$ for some $q>3$, in particular, it holds
for all Gaussian states playing essential role in quantum information theory.\smallskip

Many characteristics of a state of a $n$-partite quantum system $A_1...A_n$ have the form of a function
$f(\rho\,|\,p_1,...,p_l)$
on the set $\S(\H_{A_1...A_n})$ of states of this system depending on some parameters  $p_1,...,p_l$ (other states, quantum channels, quantum measurements, etc.).
If $\rho$ is a state of an infinite-dimensional $n$-partite quantum system $A_1...A_n$ then we may approximate it by the sequence of states
\begin{equation}\label{rho-r}
\rho_r=[\Tr Q_r\rho\shs]^{-1}Q_r\rho\shs Q_r,\quad  Q_r=P_r^1\otimes...\otimes P_r^n,
\end{equation}
where $P_r^s$ is the spectral projector of $\rho_{A_s}$ corresponding to its $r$ maximal eigenvalues  (taking the multiplicity into account).
Naturally, the questions arise under what conditions
\begin{equation}\label{p-conv}
 f(\rho_r\,|\,p_1,...,p_l)\quad \textrm{tends to}\quad f(\rho\,|\,p_1,...,p_l)\quad \textrm{as} \;\; r\to\infty
\end{equation}
for given $p_1,...,p_l$ and under what conditions this convergence is uniform on $p_1,...,p_l$.  Our main technical result asserts that for given $m\leq n$  the FA-property of the marginal
states $\rho_{A_1}$,..., $\rho_{A_m}$ of a  state
$\rho$ in $\S(\H_{A_1...A_n})$ implies (\ref{p-conv}) for a wide class of functions $f$ (depending on $m$). It also gives a sufficient condition under which
the convergence in  (\ref{p-conv}) is uniform on $p_1,...,p_l$. In fact, a more general assertion is true in which the states $\rho_r$ are defined by formula (\ref{rho-r}) via the operators $Q_r=P_r^{s_1}\otimes...\otimes P_r^{s_l}\otimes I_{R}$, where  $\{s_1,...,s_l\}$ is any subset of $\{1,...,n\}$ and $R=A_1...A_n\setminus A_{s_1}...A_{s_l}$. For a smaller class of functions on $\S(\H_{A_1...A_n})$ the above result is valid with the FA-property of the
states $\rho_{A_1}$,..., $\rho_{A_m}$ replaced by the weaker condition of finiteness of their entropy  (Theorem  \ref{main} in Section 3).

The above result and its modifications are used in this article to  solve
the following tasks for a characteristic $f$ of composite quantum systems:
\begin{itemize}
  \item to obtain conditions for local lower semicontinuity and local continuity (convergence) of $f$, i.e. conditions under which the limit relations
  $$
  \liminf_{k\to+\infty} f(\rho_k)\geq f(\rho_0)\quad \textrm{and} \quad  \lim_{k\to+\infty} f(\rho_k)=f(\rho_0)
  $$
  hold for a given sequence $\{\rho_k\}$ of states converging to a state $\rho_0$;
  \item to obtain conditions under which a convex characteristic $f$ is integrable\footnote{This question is not trivial, since it is unclear how to prove the measurability of many important characteristics of infinite-dimensional quantum systems.} w.r.t. a Borel probability
  measure $\mu$ on $\S(\H_{A_1...A_n})$ and satisfies the Jensen inequality
  \begin{equation}\label{J-in}
  f(\bar{\rho}(\mu))\leq\int_{\S(\H_{A_1...A_n})}f(\rho)\mu(d\rho),
  \end{equation}
  where
  \begin{equation}\label{bar}
  \bar{\rho}(\mu)\doteq \int_{\S(\H_{A_1...A_n})}\rho\mu(d\rho)\qquad \textup{(Bochner integral);}
  \end{equation}
  \item to establish a finite-dimensional approximation property for $f$
that allows us to generalize to the infinite-dimensional case the results (equalities, inequalities, etc.) proved in the finite-dimensional settings;
  \item to simplify definition of $f$ for states satisfying certain regularity conditions;
  \item to obtain conditions under which $f(\rho)=0$ for a separable (in different senses) state $\rho$ (when $f$ is an entanglement measure).
 \end{itemize}

The article is organized as follows. In Section 2.1, we give a brief description of basic notation and describe several entropic characteristics
used below. Section 2.2 is devoted to the simple sufficient condition for the FA-property mentioned before. In Section 2.3, we introduce the classes
$I_n^m$ and $II_n^m$ ($1\leq m\leq n$) of functions on the set $\S(\H_{A_1...A_n})$ characterized by the existence of faithful uniform continuity bounds with two types ($I_n^m$ and $II_n^m$) of  energy constraint. These classes allow us to simplify the presentation of our main technical result (Theorem \ref{main}), obtained in Section 3.1. In Section 3.2, we analyse several corollaries of  Theorem \ref{main} concerning general properties of correlation and entanglement measures in infinite-dimensional composite quantum systems.  In Section 3.3, we consider a sufficient condition for local continuity (convergence) of functions on $\S(\H_{A_1...A_n})$ w.r.t. the strong information topology (obtained by modifying the main step of the proof of Theorem \ref{main}).

In Section 4 we apply the general results from Section 3 to three important characteristics of infinite-dimensional composite quantum systems \emph{which are difficult to
analyse by the methods proposed before}. In Section 4.1, we explore  properties of  the relative entropy of $\pi$-entanglement  and its regularization
extending the results from \cite{L&Sh-1,L&Sh-2}. Section 4.2 is devoted to the  Rains bound (unregularized and regularized) -- a function  of the relative entropy distance type for which the technique used in \cite{L&Sh-1,L&Sh-2} cannot be applied. In Section 4.3, we obtain several results concerning properties of the conditional entanglement of mutual information in the infinite-dimensional settings.

In Section 5 we summarize the results and discuss other possible applications of the technique proposed in the article.

\section{Preliminaries}

\subsection{Basic notation}

Let $\mathcal{H}$ be a separable Hilbert space,
$\mathfrak{B}(\mathcal{H})$ the algebra of all bounded operators on $\mathcal{H}$ with the operator norm $\|\cdot\|$ and $\mathfrak{T}( \mathcal{H})$ the
Banach space of all trace-class
operators on $\mathcal{H}$  with the trace norm $\|\!\cdot\!\|_1$. Let
$\mathfrak{S}(\mathcal{H})$ be  the set of quantum states (positive operators
in $\mathfrak{T}(\mathcal{H})$ with unit trace) \cite{H-SCI,N&Ch,Wilde}.

Write $I_{\mathcal{H}}$ for the unit operator on a Hilbert space
$\mathcal{H}$ and  $\id_{\mathcal{\H}}$ for the identity
transformation of the Banach space $\mathfrak{T}(\mathcal{H})$.\smallskip

We will use Greek letters to denote operators in $\T(\H)$ and Latin letters for other operators, in particular, unbounded operators on a Hilbert
space $\H$. The \emph{support} $\mathrm{supp}\rho$ of an operator $\rho$ in  $\T_{+}(\H)$ is the closed subspace spanned by the eigenvectors of $\rho$ corresponding to its positive eigenvalues.  The dimension of $\mathrm{supp}\rho$ is called \emph{rank} of $\rho$ and denoted by $\rank\rho$.\smallskip

The \emph{von Neumann entropy} of a quantum state
$\rho \in \mathfrak{S}(\H)$ is  defined by the formula
$S(\rho)=\operatorname{Tr}\eta(\rho)$, where  $\eta(x)=-x\ln x$ for $x>0$
and $\eta(0)=0$. It is a concave lower semicontinuous function on the set~$\mathfrak{S}(\H)$ taking values in~$[0,+\infty]$ \cite{H-SCI,W,L-2}.
The von Neumann entropy satisfies the inequality
\begin{equation}\label{w-k-ineq}
S(p\rho+(1-p)\sigma)\leq pS(\rho)+(1-p)S(\sigma)+h_2(p)
\end{equation}
valid for any states  $\rho$ and $\sigma$ in $\S(\H)$ and $p\in(0,1)$ with possible value $+\infty$ in both sides, where $\,h_2(p)=\eta(p)+\eta(1-p)\,$ is the binary entropy \cite{N&Ch,Wilde}.\smallskip

Lindblad's extension of the \emph{quantum relative entropy}
for any  operators $\rho$ and
$\sigma$ in $\mathfrak{T}_+(\mathcal{H})$ is defined as
\begin{equation}\label{qre-L}
D(\rho\shs\|\shs\sigma)=\sum_i\langle
\varphi_i|\,\rho\ln\rho-\rho\ln\sigma+\sigma-\rho\,|\varphi_i\rangle,
\end{equation}
where $\{\varphi_i\}$ is the orthonormal basis of
eigenvectors of the state $\rho$ and it is assumed that $D(0\shs\|\shs\sigma)=\Tr\sigma$ and that
$D(\rho\shs\|\shs\sigma)=+\infty$ if $\,\mathrm{supp}\rho\shs$ is not
contained in $\shs\mathrm{supp}\shs\sigma$ \cite{L-2}.\footnote{The support $\mathrm{supp}\rho$ of a positive trace class operator $\rho$ is the closed subspace spanned by the eigenvectors of $\rho$ corresponding to its positive eigenvalues.}

\smallskip

The \emph{quantum conditional entropy}
\begin{equation*}
S(A|B)_{\rho}=S(\rho)-S(\rho_{\shs B})
\end{equation*}
of a  state $\rho$ of a
bipartite quantum system $AB$ with finite marginal entropies $S(\rho_A)$ and $S(\rho_B)$ is essentially used in analysis of quantum systems \cite{H-SCI,Wilde}. It
can be extended to the set of all states $\rho$ with finite $S(\rho_A)$ by the formula
\begin{equation}\label{ce-ext}
S(A|B)_{\rho}=S(\rho_{A})-D(\rho\shs\Vert\shs\rho_{A}\otimes
\rho_{B})
\end{equation}
proposed in \cite{Kuz}, where it is shown that this extension  possesses all basic properties of the quantum conditional entropy valid in finite dimensions.  \smallskip

The \emph{quantum mutual information} (QMI) of a state $\,\rho\,$ of a $n$-partite quantum system $A_1 \ldots
A_n$ is defined as (cf.\cite{L-mi,Herbut,NQD})
\begin{equation}\label{mi-mpd}
     I(A_1\!:\ldots:\!A_n)_{\rho}\doteq
    D(\rho\shs\|\shs\rho_{A_{1}}\otimes\cdots\otimes\rho_{A_{n}})=\sum_{s=1}^n S(\rho_{A_{s}})-S(\rho),
\end{equation}
where the second formula is valid if $S(\rho)<+\infty$. It is easy to show (see, f.i., \cite{CBM}) that
\begin{equation}\label{nMI-UB}
  I(A_1\!:\ldots:\!A_n)_{\rho}\leq 2\sum_{s=1}^{n-1}S(\rho_{A_s}).
\end{equation}
Similar upper bound holds with any other $n-1$ marginal entropies of the state $\rho$.

Note also that the QMI satisfies the inequalities
\begin{equation}\label{QMI-LAA-1}
I(A_1\!:...:\!A_n)_{p\rho+(1-p)\sigma}\geq pI(A_1\!:...:\!A_n)_{\rho}+(1-p)I(A_1\!:...:\!A_n)_{\sigma}-h_2(p)
\end{equation}
and
\begin{equation}\label{QMI-LAA-2}
I(A_1\!:...:\!A_n)_{p\rho+(1-p)\sigma}\leq pI(A_1\!:...:\!A_n)_{\rho}+(1-p)I(A_1\!:...:\!A_n)_{\sigma}+(n-1)h_2(p)
\end{equation}
for any states  $\rho$ and $\sigma$ in $\mathfrak{S}(\mathcal{H}_{A_1...A_n})$ and $p\in(0,1)$, where $h_2(p)$ is the binary entropy
and both sides can be equal to $\,+\infty$. Indeed, if $\rho$ and $\sigma$ are states with finite
marginal entropies then inequality (\ref{QMI-LAA-1}) and (\ref{QMI-LAA-2}) can be proved by using the second expression in (\ref{mi-mpd}), concavity of the  entropy and  inequality (\ref{w-k-ineq}).
The validity of (\ref{QMI-LAA-1}) and (\ref{QMI-LAA-2}) for arbitrary states $\rho$ and $\sigma$  can be shown by
approximation using Proposition 5B in \cite{CMI}.
\bigskip

For a positive (semi-definite)  operator{\footnote{We assume that a positive operator is a self-adjoint operator \cite{R&S}.}} $H$ on a Hilbert space $\H$ and any operator $\rho$ in $\T_+(\H)$ we assume that
\begin{equation}\label{Tr-def}
\Tr H\rho=
\left\{\begin{array}{l}
        \sup_n \Tr P_n H\rho\;\; \textrm{if}\;\;  \supp\rho\subseteq {\rm cl}(\mathcal{D}(H))\\
        +\infty\;\;\qquad\qquad\textrm{otherwise}
        \end{array}\right.,
\end{equation}
where $P_n$ is the spectral projector of $H$ corresponding to the interval $[0,n]$  and ${\rm cl}(\mathcal{D}(H))$ is the closure of the domain $\mathcal{D}(H)$ of $\H$. Then
\begin{equation}\label{C-set}
\mathfrak{C}_{H,E}=\left\{ \rho\in\mathfrak{S}(\H)\,|\,\mathrm{Tr} H\rho\leq E \right\}
\end{equation}
is a closed convex nonempty subset of $\mathfrak{S}(\H)$ for any $E$ greater than the infimum of the spectrum of $H$. If
$H$ is treated as  Hamiltonian of a quantum system associated with the space $\H$ then
$\mathfrak{C}_{H,E}$  is the set of states with the mean energy not exceeding $E$.\smallskip

We will pay a special attention to the class of  unbounded positive operators on $\H$ with a discrete spectrum.
In Dirac's notation any such operator $H$ can be represented as
\begin{equation}\label{H-rep}
H=\sum_{i=1}^{+\infty} h_i|\tau_i\rangle\langle \tau_i|
\end{equation}
on the domain $\mathcal{D}(H)=\{ \varphi\in\H_\mathcal{T}\,|\,\sum_{i=1}^{+\infty} h^2_i|\langle\tau_i|\varphi\rangle|^2<+\infty\}$, where
$\mathcal{T}=\left\{\tau_i\right\}_{i=1}^{+\infty}$ is the orthonormal system of eigenvectors of $H$
corresponding to the nondecreasing sequence $\left\{\smash{h_i}\right\}_{i=1}^{+\infty}$ of eigenvalues
tending to $+\infty$ and $\H_\mathcal{T}$ is the linear span of $\mathcal{T}$.\smallskip

It is well known that the von Neumann entropy is continuous on the set $\mathfrak{C}_{H,E}$ for any $E$ if (and only if) the operator  $H$ satisfies  the condition
\begin{equation}\label{H-cond}
  \mathrm{Tr}\, e^{-\beta H}<+\infty\quad\textrm{for all}\;\beta>0
\end{equation}
and that the maximal value of the entropy on this set is achieved at the \emph{Gibbs state} $\gamma_H(E)\doteq e^{-\beta(E) H}/\mathrm{Tr} e^{-\beta(E) H}$, where the parameter $\beta(E)$ is determined by the equality $\mathrm{Tr} H e^{-\beta(E) H}=E\mathrm{Tr} e^{-\beta(E) H}$ \cite{W}. Condition (\ref{H-cond}) implies that $H$ is an unbounded operator with a  discrete spectrum, i.e. it has the form (\ref{H-rep}). So, by the Lemma in \cite{H-c-w-c} the set $\mathfrak{C}_{H,E}$ defined in (\ref{C-set}) is compact for any $E$.\smallskip

We will use the function $F_{H}(E)\doteq S(\gamma_H(E))=\max_{\rho\in\mathfrak{C}_{H,E}}S(\rho)$. By Proposition 1 in \cite{EC} condition (\ref{H-cond}) holds if and only if
\begin{equation}\label{H-cond-a}
  F_{H}(E)=o\shs(E)\quad\textrm{as}\quad E\rightarrow+\infty.
\end{equation}

We will use the following\smallskip

\begin{property}\label{S-finite} \emph{Let $\rho$ be a state in $\S(\H)$. The following properties are equivalent:}
\begin{enumerate}[(i)]
  \item \emph{$S(\rho)<+\infty$;}
  \item \emph{there is a positive operator $H$ on $\H$ satisfying condition (\ref{H-cond}) s.t. $\Tr H\rho<+\infty$;}
  \item $\sum_{i=1}^{+\infty}\lambda^{\rho}_i \ln i<+\infty$, \emph{where $\{\lambda^{\rho}_i\}_{i=1}^{+\infty}$ is the sequence of eigenvalues of $\rho$
  arranged in the non-increasing order.}
\end{enumerate}

\emph{If equivalent properties $(i)\textrm{-}(iii)$ hold then the positive operator $H$ in $(ii)$ can be chosen
diagonizable in any basis of eigenvectors of the state $\rho$ and such that $H |\varphi_1\rangle=0$, where $\varphi_1$ is
any eigenvector of $\rho$ corresponding to its maximal eigenvalue.}
\end{property}\smallskip

\emph{Proof.} The implication $\rm (ii)\Rightarrow(i)$ follows from the results mentioned before the proposition. To prove
$\rm (i)\Rightarrow(iii)$ it suffices to note that $\lambda^{\rho}_i\leq 1/i$ for all $i$.\smallskip

Assume that $\sum_{i=1}^{+\infty}\lambda^{\rho}_i \ln i<+\infty$. Then it is easy to show the existence of an
increasing sequence $\{c_i\}_{i=1}^{+\infty}$ tending to $+\infty$ such that $\sum_{i=1}^{+\infty}\lambda^{\rho}_i c_i\ln i<+\infty$. Then $\rm (ii)$ holds with the
operator
\begin{equation*}
H=\sum_{i=1}^{+\infty} c_i\ln i |\varphi_i\rangle\langle \varphi_i|,
\end{equation*}
where $\{ \varphi_i \}_{i=1}^{+\infty}$ is any basis of eigenvectors of $\rho$ corresponding to the sequence $\{\lambda^{\rho}_i\}_{i=1}^{+\infty}$ (if $\rho$
is not a faithful state then the basis $\{ \varphi_i \}_{i=1}^{+\infty}$  is obtained by extending the basis  of eigenvectors of $\rho$ in $\H_{\rho}\doteq\supp\rho$ to a basis in $\H$). $\Box$
\smallskip

We will often consider positive operators $H$ satisfying the condition
\begin{equation}\label{H-cond+}
  \lim_{\beta\rightarrow0^+}\left[\mathrm{Tr}\, e^{-\beta H}\right]^{\beta}=1,
\end{equation}
which is stronger than condition (\ref{H-cond}). By Lemma 1 below the condition (\ref{H-cond+}) is valid  if $\;\liminf\limits_{i\rightarrow+\infty} h_i/\ln^2 i=+\infty$, where
$\{h_i\}$ is the sequence of eigenvalues of $H$. By Lemma 1 in \cite{AFM} condition (\ref{H-cond+}) holds if and only if
\begin{equation}\label{H-cond++}
  F_{H}(E)=o\shs(\sqrt{E})\quad\textrm{as}\quad E\rightarrow+\infty
\end{equation}
(this should be compared with (\ref{H-cond-a})). It is essential that condition (\ref{H-cond+}) is valid for the Hamiltonians of many real quantum systems \cite{AFM,Datta}.\smallskip

We will use the following\smallskip

\begin{lemma}\label{sl} \emph{If $H$ is a positive operator on $\H$ satisfying condition (\ref{H-cond}) then
$H^2$ is a positive operator on $\H$ satisfying condition (\ref{H-cond+}).}
\end{lemma}\smallskip

\emph{Proof.} Since (\ref{H-cond})$\shs\Leftrightarrow\shs$(\ref{H-cond-a}) and (\ref{H-cond+})$\shs\Leftrightarrow\shs$(\ref{H-cond++}), to prove the lemma it suffices
to note that
$$
F_{H^2}(E)\leq F_{H}(\sqrt{E})\quad \forall E\geq h^2_1,
$$
where $h_1$ is the minimal eigenvalue of $H$. This inequality follows from the definitions of $F_{H^2}(E)$ and $F_{H}(\sqrt{E})$ due to the inequality
$$
\Tr H\rho\leq \sqrt{\Tr H^2\rho}
$$
valid for any state $\rho$ in $\S(\H)$ (with possible value $+\infty$ in one or both sides). It is  clear that $\mathcal{D}(H^2)\subseteq\mathcal{D}(H)$. So, the last inequality is a direct corollary of the concavity of the square root, since for any state $\rho$ with finite $\Tr H^2\rho$ we have
$$
\Tr H\rho=\sum_i h_i p_i\quad \textrm{and} \quad \Tr H^2\rho=\sum_i h^2_i p_i, \quad p_i=\langle\tau_i|\rho|\tau_i\rangle,
$$
where it is assumed that the operator $H$ has the form (\ref{H-rep}). $\Box$\smallskip

An interesting open question is whether it is possible to reverse the claim of Lemma \ref{sl}.

\subsection{Simple sufficient condition for the FA-property}

Following \cite{FAP} we will say that
a quantum state $\rho$ with the spectrum $\{\lambda^{\rho}_i\}_{i=1}^{+\infty}$ has the FA-property if there exists a sequence $\{g_i\}_{i=1}^{+\infty}$ of nonnegative numbers such that
\begin{equation*}
\sum_{i=1}^{+\infty}\lambda^{\rho}_i g_i<+\infty\quad \textrm{and} \quad \lim_{\beta\to 0^+}\left[\sum_{i=1}^{+\infty}e^{-\beta g_i}\right]^{\beta}=1.
\end{equation*}
We will write $\S_{\rm \textsf{FA}\!}(\H)$ for the set of all states in $\S(\H)$ having the FA-property. It follows from Corollary 3 in \cite{FAP} that
$\S_{\rm \textsf{FA}\!}(\H)$ is a face\footnote{It means that the set $\S_{\rm \textsf{FA}\!}(\H)$ is convex and contains any segment from $\S(\H)$ provided that
it contains at least one internal point of this segment.} of the convex set $\S(\H)$ contained in the face of all states in $\S(\H)$ with finite entropy (these two faces do not coincide by the examples in \cite{Becker}).\smallskip

The following proposition gives a simple sufficient condition for the FA-property.\smallskip

\begin{property}\label{FA-SC}
\emph{A quantum state $\rho$ with the spectrum $\{\lambda^{\rho}_i\}_{i=1}^{+\infty}$  has the FA-property provided that}
\begin{equation}\label{FA-SC+}
\sum_{i=1}^{+\infty}\lambda^{\rho}_i \ln^2 i<+\infty.
\end{equation}
\end{property}

\emph{Proof.} W.l.o.g. we may assume that $\rho$ is an infinite rank state. \smallskip

Let $\rho=\sum_{i=1}^{+\infty} \lambda^{\rho}_i|\varphi_i\rangle\langle \varphi_i|$ be a spectral decomposition of $\rho$.
If (\ref{FA-SC+}) holds then there exists an increasing sequence
$\{c_i\}_{i=1}^{+\infty}$ of positive numbers tending to $+\infty$ such that
\begin{equation}\label{FA-SC++}
\sum_{i=1}^{+\infty}\lambda^{\rho}_i c_i\ln^2 i<+\infty.
\end{equation}

It is easy to see that the positive operator
\begin{equation*}
H=\sum_{i=1}^{+\infty} \sqrt{c_i}\ln i|\varphi_i\rangle\langle \varphi_i|
\end{equation*}
on $\mathcal{D}(H)=\{ \psi\in\supp\rho\,|\,\sum_{i=1}^{+\infty} c_i\ln^2 i\shs|\langle\varphi_i|\psi\rangle|^2<+\infty\}$ satisfies condition (\ref{H-cond}). By Lemma \ref{sl} the operator $H^2$
satisfies condition (\ref{H-cond+}). Since (\ref{FA-SC++}) implies that
$\Tr H^2\rho <+\infty$, the state $\rho$ has the FA-property by the Theorem  in \cite[Section 2]{FAP}. $\Box$\smallskip

\begin{remark}\label{FA-SC-r} The condition given by Proposition \ref{FA-SC} is \emph{strictly stronger} than the last claim of the Theorem in \cite{FAP} which states
that a quantum state $\rho$ with the spectrum $\{\lambda^{\rho}_i\}_{i=1}^{+\infty}$ has the FA-property provided that
\begin{equation}\label{FA-SC-P}
\sum_{i=1}^{+\infty}\lambda^{\rho}_i \ln^q i<+\infty \quad\textrm{for some}\quad q>2.
\end{equation}
Indeed, it is easy to see that any state with the spectrum $\{c [i\ln^{p}(\ln i)\ln^3i\shs]^{-1}\}_{i=10}^{+\infty}$, $p>1$, where $c$ is a normalizing constant,
satisfies condition (\ref{FA-SC+}) but does not satisfy condition (\ref{FA-SC-P}).
\end{remark}\smallskip

\begin{remark}\label{FA-SC-rr} If $\rho$ is a state in $\S(\H)$ with the spectrum $\{\lambda^{\rho}_i\}_{i=1}^{+\infty}$ arranged in the non-increasing order
then Proposition \ref{FA-SC} show that
\begin{equation}\label{3-imp}
\sum_{i=1}^{+\infty}\lambda^{\rho}_i \ln^2\lambda^{\rho}_i<+\infty\;\Rightarrow\; \sum_{i=1}^{+\infty}\lambda^{\rho}_i \ln^2 i<+\infty\;\Rightarrow\;\rho\in\S_{\rm \textsf{FA}\!}(\H) \;\Rightarrow\;S(\rho)<+\infty.
\end{equation}
The first implication in (\ref{3-imp}) follows
from the inequality $\lambda^{\rho}_i\leq 1/i,\;\forall i$, the third one is mentioned in the Theorem in \cite[Section 2]{FAP}. In \cite{Becker}
it is shown that the third implication in (\ref{3-imp}) cannot be replaced by $"\Leftrightarrow"$. S.Becker proved recently that the second  implication in (\ref{3-imp}) cannot be replaced by $"\Leftrightarrow"$ as well \cite{Becker-pc}. Therefore, the question of finding a simple criterion for the FA-property remains open.
\end{remark}

\subsection{On uniform continuity under the energy-type constraints}

To simplify description of the main technical result of this article (Theorem \ref{main} in Section 3.1) we have to introduce
the classes $I_n^m$ and $II_n^m$ of functions on the set of states of\break $n$-partite infinite-dimensional quantum system $A_1...A_n$. They are described below.
\smallskip

Assume that $H_{A_1}$,...,$H_{A_m}$ are arbitrary positive  operators on the spaces $\H_{A_1}$,...,$\H_{A_m}$, $m\leq n$.
For a given function $f$ on $\S(\H_{A_1...A_n})$ taking a finite value at any state $\rho$ with finite "energy" $\sum_{k=1}^{m}\Tr H_{A_k}\rho_{A_k}$,
given $\varepsilon\in(0,1]$ and $E>0$ introduce the quantity
$$
B_f(\varepsilon,E|H_{A_1},...,H_{A_m})\doteq \sup\left\{|f(\rho)-f(\sigma)| \,\left|\, \rho,\sigma\in\C(E|H_{A_1},...,H_{A_m}),\textstyle\frac{1}{2}\|\rho-\sigma\|_1\leq\varepsilon \right.\right\}\!,
$$
where\footnote{$\Tr H_{A_k}\varrho_{A_k}$ is defined according to the rule (\ref{Tr-def}).}
\begin{equation}\label{sc-def}
\C(E|H_{A_1},...,H_{A_m})\doteq\left\{\varrho\in \S(\H_{A_1...A_n})\,\left|\,\sum_{k=1}^{m}\Tr H_{A_k}\varrho_{A_k}\leq mE\right.\right\}.
\end{equation}

The  quantity $B_f(\varepsilon,E|H_{A_1},...,H_{A_m})$ is the (optimal) modulus of continuity of the function $f$ under the energy-type constraint induced by
the operators  $H_{A_1}$,...,$H_{A_m}$.\smallskip

\begin{definition}\label{I-def}
A function $f$ on $\S(\H_{A_1...A_n})$ belongs to the class $I_n^m$ if
$$
\lim_{\varepsilon\to0^+}B_f(\varepsilon,E|H_{A_1},....,H_{A_m})=0
$$
for any $E>0$ and arbitrary positive operators $H_{A_1}$,....,$H_{A_m}$
satisfying condition (\ref{H-cond}).
\end{definition}
\smallskip
\begin{definition}\label{II-def}
A function $f$ on $\S(\H_{A_1...A_n})$ belongs to the class $II_n^m$ if
$$
\lim_{\varepsilon\to0^+}B_f(\varepsilon,E|H_{A_1},...,H_{A_m})=0
$$
for any $E>0$ and arbitrary positive operators $H_{A_1}$,...,$H_{A_m}$
satisfying condition (\ref{H-cond+}).
\end{definition}
\smallskip

According to these definitions the class $I_n^m$ (resp. the class $II_n^m$) consists of all functions
uniformly continuous on the set $\C(E|H_{A_1},...,H_{A_m})$ (defined in (\ref{sc-def}))
for any $E>0$ and arbitrary positive operators $H_{A_1}$,...,$H_{A_m}$
satisfying condition (\ref{H-cond}) (resp. condition (\ref{H-cond+})). Since condition (\ref{H-cond+}) is stronger than condition (\ref{H-cond}), we have $I_n^m\subseteq II_n^m$.\smallskip

Let $f$ be a given function from the class $I_n^m$ (resp. $II_n^m$). We will say that $\widehat{B}_f(\varepsilon,E|H_{A_1},...,H_{A_m})$ is a \emph{faithful $I_n^m$-type (resp. $II_n^m$-type) continuity bound}
for the function $f$ provided that
$$
B_f(\varepsilon,E|H_{A_1},...,H_{A_m})\leq \widehat{B}_f(\varepsilon,E|H_{A_1},...,H_{A_m})\quad \forall E>E_0,\, \varepsilon\in(0,1],
$$
and
$$
\lim_{\varepsilon\to0^+}\widehat{B}_f(\varepsilon,E|H_{A_1},...,H_{A_m})=0\quad \forall E>E_0/m
$$
for arbitrary positive operators  $H_{A_1}$,...,$H_{A_m}$
satisfying condition (\ref{H-cond}) (resp. condition (\ref{H-cond+})), where $E_0$ is the sum of the minimal eigenvalues of the operators $H_{A_1}$,...,$H_{A_m}$.

Among characteristics of a composite infinite-dimensional quantum system $A_1...A_n$
there are many that belong to the classes $I_n^m$ and $II_n^m$ for some $m\leq n$.\footnote{It follows from the definitions that if $f\in I_n^m$ (resp. $f\in II_n^m$) for some $m<n$ then $f\in I_n^{\tilde{m}}$ (resp. $f\in II_n^{\tilde{m}}$) for any $\tilde{m}\in(m,n]\cap\mathbb{N}$.} For example, the von Neumann
entropy of a state of $A_1$ and the quantum conditional entropy $S(A_1|A_2)$ belong, respectively,
to the classes $I_1^1$ and $I^1_2$:  $I_1^1$-type  and $I^1_2$-type   faithful continuity bounds for these characteristics are
obtained by Winter in \cite{W-CB}. The\break $I^1_2$-type  faithful continuity bounds for the quantum mutual information
$I(A_1\!:\!A_2)$, for the entanglement of formation $E_F$ and for the squashed entanglement $E_{\rm sq}$ in a bipartite system $A_1A_2$
obtained in \cite{SE} by using Winter's technique show that these characteristics
belong to the class $I^1_2$. Moreover, the last two entanglement measures have a common faithful $I^1_2$-type continuity bound.
The regularized entanglement of formation $E^{\infty}_F$ (coinciding with the entanglement cost $E_C$ at any state $\rho$ such that $\min\{S(\rho_{A_1}),S(\rho_{A_2})\}<+\infty$ \cite{Lami-new}) also belongs to the class  $I^1_2$: the faithful $I^1_2$-type continuity bound for $E^{\infty}_F$ obtained by
using Winter's technique is presented
in Proposition 18 in \cite{QC}.\footnote{We do not distinguish between the discrete and continuous versions of $E_F$ and $E^{\infty}_F$, since these versions coincide at any state $\omega\in\S(\H_{A_1A_2})$ such that at least one of the states $\omega_{A_1}$ and $\omega_{A_2}$ has finite energy w.r.t. a Hamiltonian satisfying condition (\ref{H-cond}) \cite{EM}.}

The universal version of the Alicki-Fannes-Winter technique presented in \cite{QC} allows us to show that
\begin{equation}\label{B-inc}
\bigcup_{C,D\geq0}\left[\widehat{L}^{m}_n(C,D)\cup N^{m}_{n,1}(C,D)\cup N^{m}_{n,2}(C,D)\cup N^{m}_{n,3}(C,D)\right]\subset II_n^m,
\end{equation}
where $\widehat{L}^{m}_n(C,D)$ and $N^{m}_{n,s}(C,D)$, $s=1,2,3$, are the classes of functions on
$\S(\H_{A_1..A_n})$ introduced in \cite{CBM,QC} which contain many
important characteristics of composite quantum systems. Moreover, Theorem  6 in \cite{QC}
implies that all functions in
$$
\widehat{L}^{m}_n(C,D)\cup N^{m}_{n,1}(C,D)\cup N^{m}_{n,2}(C,D)\cup N^{m}_{n,3}(C,D),\quad C,D\geq0,
$$
have the common faithful $II_n^m$-type continuity bound
\begin{equation}\label{fcb}
\!\widehat{B}_{C,D}(\varepsilon,E|H_{A_1},..,H_{A_m})
=C\sqrt{\varepsilon(2-\varepsilon)}F_{H_{A_1},..,H_{A_m}}\!\!\left[\frac{2mE}{\varepsilon(2-\varepsilon)}\right]+Dg(\sqrt{\varepsilon(2-\varepsilon)}),
\end{equation}
where
\begin{equation*}
F_{H_{A_1},...,H_{A_m}}(E)=\sup\left\{ S(\varrho)\,\left|\,\varrho\in\S(\H_{A_1..A_m}),\,\sum_{k=1}^{m}\Tr H_{A_k}\varrho_{A_k}\leq E \right.\right\}
\end{equation*}
and
\begin{equation*}
  g(x)\doteq(x+1)h_2\!\left(\frac{x}{x+1}\right)=(x+1)\ln(x+1)-x\ln x,\;\, x>0,\quad g(0)=0.
\end{equation*}
The faithfulness of the common continuity bound (\ref{fcb}) follows from the fact that the validity of condition (\ref{H-cond+}) for all the operators $H_{A_1},...,H_{A_m}$
implies that
$$
F_{H_{A_1},...,H_{A_m}}(E)=o(\sqrt{E})\quad\textrm{as} \quad E\to+\infty
$$
by Lemma 4 in \cite{CBM}.

We will see in the next Section 3 that for any given characteristic $f$ from the class $II_n^m$  one can realise its effective finite-dimensional approximation which allows us to prove nontrivial properties of $f$. If, moreover, this characteristic belongs to the
class $I_n^m$ then we may relax some requirements in such approximation (briefly speaking, to replace the FA-property of marginal states by the finiteness of the von Neumann entropy of these states).\smallskip

Practically, to show that a given characteristic $f$ belongs to the class $II_n^m$ we have either to use
general relation (\ref{B-inc}) or to directly prove the existence of a continuity bound $\widehat{B}_f(\varepsilon,E|H_{A_1},....,H_{A_m})$ for $f$ which
tends to zero as $\varepsilon \to0$  for arbitrary positive operators  $H_{A_1}$,....,$H_{A_m}$
satisfying condition (\ref{H-cond+}). More efforts are required to show that a given characteristic $f$ belongs to the class $I_n^m$:
it is necessary to prove the existence of a continuity bound $\widehat{B}_f(\varepsilon,E|H_{A_1},....,H_{A_m})$ for $f$ which
tends to zero as $\varepsilon \to0$  for arbitrary positive operators  $H_{A_1}$,....,$H_{A_m}$
satisfying condition (\ref{H-cond}). Sometimes one can use the following simple\smallskip

\begin{lemma}\label{tl} \emph{If $f$ and $g$ are functions from the class $I_n^m$ (resp. $II_n^m$) then
any their linear combitation is a function from the class $I_n^m$ (resp. $II_n^m$).}
\end{lemma}
\smallskip

\begin{example}\label{cl-exam} The one-way classical correlation $C_B$ and the quantum discord $D_B$
in a bipartite infinite-dimensional system $AB$ with measured system $B$ (cf.\cite{H&V,O&Z,Str}) belong to the class $II_2^1$ in both settings
$A_1=A,A_2=B$ and $A_1=B,A_2=A$. This follows from  (\ref{B-inc}), since
$C_B\in\widehat{L}_2^1(1,2)$ in both settings $A_1=A,A_2=B$ and $A_1=B,A_2=A$ and $D_B\in\widehat{L}_2^1(C,2)$,
where $C=2$ in the setting $A_1=A,A_2=B$ and $C=1$ in the setting $A_1=B,A_2=A$ \cite[Section 4.3]{QC}.
The $II_2^1$-type continuity bounds for the regularized  one-way classical correlation $C^{\infty}_B$ presented in \cite[Proposition 16]{QC}
implies that $C^{\infty}_B\in II_2^1$ in both settings $A_1=A,A_2=B$ and $A_1=B,A_2=A$.

A more detailed analysis shows that the functions $C_B$, $C^{\infty}_B$ and $D_B$ belong to the class $I_2^1$ in the  setting $A_1=A,A_2=B$.
Indeed, by the Koashi-Winter relations (cf.\cite{K&W}) we have
\begin{equation}\label{KWS-gen}
C_B(\omega_{AB})=S(\omega_{A})-E_F(\omega_{AC})\quad \textrm{and}\quad C_B^{\infty}(\omega_{AB})=S(\omega_{A})-E^{\infty}_F(\omega_{AC})
\end{equation}
for any pure state $\omega$ in $\S(\H_{ABC})$ such that $S(\omega_{A})<+\infty$, where $E_F$ and $E_F^{\infty}$ are the entanglement of formation and its regularization.

By applying the faithful $I_1^1$-type continuity bound for the entropy and the faithful $I_2^1$-type continuity bounds for the entanglement of formation and its regularization
mentioned before (\ref{B-inc}) it is easy to show that the functions $\,\omega\mapsto S(\omega_{A})$, $\omega\mapsto E_F(\omega_{AC})$ and $\,\omega\mapsto E^{\infty}_F(\omega_{AC})$
on $\S(\H_{ABC})$ have faithful $I_3^1$-type continuity bounds in the setting $A_1=A,A_2=B,A_3=C$ (because the trace norm does not increase under partial trace).

Since for any states $\rho$ and $\sigma$ in $\S(\H_{AB})$ such that $\frac{1}{2}\|\rho-\sigma\|_1<\varepsilon$ there exist pure states $\hat{\rho}$ and $\hat{\sigma}$ in $\S(\H_{ABC})$ such that $\frac{1}{2}\|\hat{\rho}-\hat{\sigma}\|_1\leq\sqrt{2\varepsilon}$ \cite{H-SCI,Wilde},
using the continuity bounds mentined before, the expressions in (\ref{KWS-gen}) and Lemma \ref{tl} one can prove that
$C_B$ and $C^{\infty}_B$ belong to the class $I_2^1$ in the  setting $A_1=A,A_2=B$.

To prove that $D_B$ belongs to the class $I_2^1$ in the  setting $A_1=A,A_2=B$, it suffices to note that
$D_B=I(A\!:\!B)-C_B$ and that $I(A\!:\!B)$ belongs to the class $I_2^1$ in the same setting (as mentioned before (\ref{B-inc})) and to use Lemma \ref{tl}.

The question of whether the functions  $C_B$, $C^{\infty}_B$ and $D_B$ belong to the class $I_2^1$ in the  setting $A_1=B,A_2=A$ remains open. $\Box$
\end{example}\smallskip

\begin{example}\label{cl-exam+} As noted before (\ref{B-inc}) the bipartite quantum mutual information\break
$I(A_1\!:\!A_2)$ belongs to the class $I_2^1$. By using this, the representation (cf.\cite{Y&C})
\begin{equation*}
I(A_1\!:\!...\!:\!A_n)_{\omega}=I(A_{n-1}\!:\!A_n)_{\omega}+ I(A_{n-2}\!:\!A_{n-1}A_n)_{\omega}+...+
     I(A_1\!:\!A_2...A_{n})_{\omega}
\end{equation*}
(valid for any state $\omega\in\S(\H_{A_1..A_n})$) and Lemma \ref{tl} it is easy to show that the $n$-partite quantum mutual information  $I(A_1\!:\!...\!:\!A_n)$ belongs to the class $I_n^{n-1}$.

By the same way one can show that the (extended) quantum conditional mutual information $I(A_1\!:\!...\!:\!A_n|A_{n+1})$ (defined by the rule described in \cite[Section 4.2]{QC}) belongs to the class $I_{n+1}^{n-1}$  (since $I(A_1\!:\!A_2|A_{3})$ lies in $I_{3}^{1}$ due to the $I_{3}^{1}$-type faithful continuity bound for this characteristic obtained in \cite{SE}).
$\Box$
\end{example}\smallskip

Besides the regularized entanglement of formation and the regularized one-way classical correlation
there are important characteristics in  $II_n^{m}$ not belonging to the classes  $\widehat{L}_n^{m}(C,D)$ and $ N_{n,s}^{m}(C,D)$.
One of them  is the regularized relative entropy
of\break $\pi$-entanglement  $E_{R,\pi}^{\infty}$ in a $n$-partite quantum system corresponding to any singleton set  $\pi$ of partitions of $\{1,\ldots, n\}$ described in Section 4.
A faithful $II_n^{n-1}$-type continuity bound for the function $E_{R,\pi}^{\infty}$ is presented in Proposition 39 in \cite{L&Sh-1}.\footnote{The unregularized  relative entropy
of $\pi$-entanglement  $E_{R,\pi}$ (with any  set of partitions $\pi$) belongs to the class $II_n^{n-1}$ due to (\ref{B-inc}), since it lies in $L_n^{n-1}(1,1)$ \cite[the proof of Proposition 39]{L&Sh-1}.}

\section{Approximation of multipartite quantum states}
\subsection{Basic theorem}

In this section we  obtain our main result about approximation of  characteristics of $n$-partite quantum systems belonging to  the classes $I_n^m$ and $II_n^m$ described in Section 2.3  (Definitions \ref{I-def} and \ref{II-def}).

Introduce the approximation map $\Lambda^{s_1,...,s_l}_r$ on the space
$\S(\H_{A_1...A_n})$ determined by a given set $\{s_1,...,s_l\}\subseteq \{1,...,n\}$ as follows
\begin{equation}\label{ap-map}
\Lambda^{s_1,...,s_l}_r(\rho)=[\Tr Q_r\rho\shs]^{-1}Q_r\rho\shs Q_r,\quad  Q_r=P_r^{s_1}\otimes...\otimes P_r^{s_l}\otimes I_{R},
\end{equation}
where $P_r^s$ is the spectral projector of $\rho_{A_s}$ corresponding to its $r$ maximal eigenvalues  (taking the multiplicity into account) and
$R=A_1...A_n\setminus A_{s_1}...A_{s_l}$.\footnote{Here and in what follows
speaking about the spectral projector $P_r$ of a state $\rho$ corresponding to its $r$ maximal eigenvalues we assume that $P_r$ is the projector onto the subspace $\supp\rho$ if $\,r>\rank\rho$.} \smallskip\pagebreak

\begin{theorem}\label{main} \emph{Let $A_1$,..,$A_n$ be arbitrary infinite-dimensional quantum systems, $n\geq1$}

\smallskip

A) \emph{If $f$ is a function on $\S(\H_{A_1..A_n})$ having  faithful $I^m_n$-type continuity bound $\widehat{B}_f$\footnote{See Section 2.3 (Definition \ref{I-def} and below).} then
for any state $\rho$ in $\S(\H_{A_1..A_n})$ such that $S(\rho_{A_s})<+\infty$ for $s=\overline{1,m}$
the following properties hold:}
\begin{itemize}
  \item \emph{$f(\rho)$ is finite;}
  \item \emph{for an arbitrary subset $\{s_1,...,s_l\}$ of $\,\{1,...,n\}$ there
is a sequence $\{Y_r\}_{r=1}^{+\infty}$ tending to zero such that}
\begin{equation}\label{main-r-A}
 |f(\Lambda^{s_1,...,s_l}_r(\rho))-f(\rho)|\leq Y_r\quad \forall r\in\N.
\end{equation}
\end{itemize}

B) \emph{If $f$ is a function on $\S(\H_{A_1..A_n})$ having  faithful $II^m_n$-type continuity bound $\widehat{B}_f$ then
the above properties hold for any state $\rho$ in $\S(\H_{A_1..A_n})$
such that $\rho_{A_s}\in\S_{\!\rm \textsf{FA}\!}(\H_{A_s})$ for $s=\overline{1,m}$.}\footnote{$\S_{\rm \textsf{FA}\!}(\H)$ is the set of all states in $\S(\H)$ having the FA-property (see Section 2.2).}\smallskip

\emph{The sequence $\{Y_r\}_{r=1}^{+\infty}$ in claims A and B is completely determined by the states $\rho_{A_s}$, $s\in \{s_1,...,s_l\}\cup\{1,...,m\}$, and by the continuity bound $\widehat{B}_f$.}
\end{theorem}\medskip

\begin{remark}\label{main-r++} Claim B of Theorem \ref{main} differs from claim A by the weaker assumption about the function $f$ (the class $II_n^m$ instead of $I_n^m$)
which is compensated by the stronger assumption about the marginal states $\rho_{A_s}$, $s=\overline{1,m}$ (the FA-property instead of the finiteness of the entropy).

Claim B is more applicable to practical tasks, since the list  of characteristics for which faithful $I_n^m$-type continuity bounds have been obtained (so far) is quite short, while relation (\ref{B-inc}) (along with the observation in Section V in \cite{CBM})  allows us to apply claim B to many characteristics of $n$-partite quantum systems (non-complete lists of characteristics belonging to the classes $I_n^m$ and $II_n^m$  are placed in the  Appendix).  Note that claim B can be applied
to all characteristics from the class $I_n^m$ as well, since  $I_n^m\subset II_n^m$.
\end{remark}\smallskip

\begin{remark}\label{main-r+++} The map $\Lambda^{s_1,...,s_l}_r$ defined in (\ref{ap-map}) is nonlinear. The arguments
from the below proof of  Theorem \ref{main} show that its claims remain valid if we replace the family of maps $\Lambda^{s_1,...,s_l}_r$
with the family of product-channels
\begin{equation}\label{ap-map+}
\widetilde{\Lambda}^{s_1,...,s_l}_r(\rho)=\Phi^{s_1}_r\otimes...\otimes\Phi^{s_l}_r\otimes\id_{R}(\rho),\quad \rho\in\S(\H_{A_1...A_n}),
\end{equation}
where $\Phi^s_r(\varrho)=P_r^s\varrho P_r^s+[\Tr(I_{A_s}-P_r^s)\varrho]\tau_s$ is a channel from $\T(\H_{A_s})$ to itself,
$P_r^s$ is the spectral projector of $\rho_{A_s}$ corresponding to its $r$ maximal eigenvalues  (taking the multiplicity into account),
$\tau_s$ is a pure state in $\S(\H_{A_s})$ corresponding to the maximal eigenvalue of $\rho_{A_s}$ and
$R=A_1...A_n\setminus A_{s_1}...A_{s_l}$.
\end{remark}\smallskip

\emph{Proof of Theorem \ref{main}.} A) For each natural $s$ in $[1,m]$ let $\mathcal{T}_s\doteq\{\varphi^s_i\}_{i=1}^{+\infty}$ be an orthonormal system of eigenvectors of the state $\rho_{A_s}$ corresponding to the non-increasing sequence $\{\lambda^s_i\}_{i=1}^{+\infty}$ of its eigenvalues (if $r\doteq\rank\rho_{A_s}<+\infty$
then $\mathcal{T}_s$ is an arbitrary extension of the basis $\{\varphi^s_i\}_{i=1}^{r}$ of eigenvectors of $\rho_{A_s}$ in $\,\supp\rho_{A_s}$ to a countable orthonormal system of vectors and $\lambda^s_i=0$ for all $i>r$). By Proposition \ref{S-finite} in Section 2.1 (and its proof) there is a positive operator
\begin{equation}\label{G-s-def}
G_{\!A_s}=\sum_{i=1}^{+\infty}g^s_i|\varphi^s_i\rangle\langle \varphi^s_i|
\end{equation}
on $\mathcal{D}(G_{\!A_s})=\{ \psi\in\supp\rho_{A_s}\,|\,\sum_{i=1}^{+\infty} g^2_i|\langle\varphi^s_i|\psi\rangle|^2<+\infty\}\subseteq\H_{A_s}$ satisfying condition (\ref{H-cond}) such that $g_1^s=0$ and
$$
E_s\doteq \Tr G_{\!A_s}\rho_{A_s}=\sum_{i=1}^{+\infty}\lambda^s_i g^s_i<+\infty.
$$

For each natural $s$ in $[1,n]$ let $P_r^s=\sum_{i=1}^{r}|\varphi^s_i\rangle\langle \varphi^s_i|$ be the spectral projector of $\rho_{A_s}$ corresponding to its $r$ maximal eigenvalues  (taking the multiplicity into account)
and $\bar{P}_r^s=I_{A_s}-P_r^s$. Then
\begin{equation}\label{p-ineq}
\Tr Q_r\rho\geq 1-\sum_{j=1}^l\Tr\bar{P}_r^{s_j}\rho_{A_{s_j}},
\end{equation}
where $Q_r$ is the projector defined in (\ref{ap-map}).  To prove this inequality  note that
$$
\left|\Tr Q_r^{j-1}\rho - \Tr Q_r^{j}\rho\shs\right|
\leq \|Q_r^{j-1}\|\Tr[\bar{P}_r^{s_j}\otimes I_{A_1...A_n\setminus A_{s_j}}]\rho
=\Tr\bar{P}_r^{s_j}\rho_{A_{s_j}},\quad j=\overline{1,l},
$$
where $Q_r^0=I_{A_1..A_n}$, $\,Q_r^j=P_r^{s_1}\otimes...\otimes P_r^{s_j}\otimes I_{R_j}$,  $R_j=A_1...A_n\setminus A_{s_1}...A_{s_j}$. Hence
$$
1-\Tr Q_r\rho\leq \sum_{j=1}^l \left|\Tr Q_r^{j-1}\rho - \Tr Q_r^{j}\rho\shs\right|\leq \sum_{j=1}^l\Tr\bar{P}_r^{s_j}\rho_{A_{s_j}}.
$$

By the construction of the state $\Lambda^{s_1,...,s_l}_r(\rho)$ we have
$$
c_r[\Lambda^{s_1,...,s_l}_r(\rho)]_{A_s}\leq \rho_{A_s},\quad c_r=\Tr Q_r\rho,
$$
for any $s$. Hence, it follows from (\ref{p-ineq}) that
\begin{equation}\label{e-cond}
  \sum_{s=1}^{m}\Tr G_{A_s}[\Lambda^{s_1,...,s_l}_r(\rho)]_{A_s}\leq c^{-1}_r\sum_{s=1}^{m}\Tr G_{A_s}\rho_{A_s}=c^{-1}_r E_S\leq \frac{E_S}{1-\sum_{j=1}^l\Tr\bar{P}_r^{s_j}\rho_{A_{s_j}}},
\end{equation}
where $E_{S}=E_1+...+E_m$.

By using Winter's gentle measurement lemma (cf.\cite{Wilde}) we obtain
\begin{equation}\label{n-est}
\|\rho-\Lambda^{s_1,...,s_l}_r(\rho)\|_1\leq2\sqrt{\Tr \bar{Q}_r\rho}\leq 2\sqrt{\sum_{j=1}^l\Tr\bar{P}_r^{s_j}\rho_{A_{s_j}}}=o(1)\quad\textrm{ as }\;r\to+\infty,
\end{equation}
where $\,\bar{Q}_r=I_{A_1...A_n}-Q_r\,$ and the last inequality follows from (\ref{p-ineq}).

Let $\widehat{B}_{f}(\varepsilon,E|H_{A_1},....,H_{A_m})$ be a faithful $I_n^m$-type continuity bound for the function $f$.
Since $\,\sum_{s=1}^{m}\Tr G_{A_s}\rho_{A_s}=E_S$ and all the operators $G_{A_1}$,...,$G_{A_m}$ satisfy condition  (\ref{H-cond}),
we conclude that $f(\rho)$ is finite and, by taking into account  (\ref{e-cond}) and (\ref{n-est}), we obtain
\begin{equation}\label{d-one}
  |f(\rho)-f(\Lambda^{s_1,...,s_l}_r(\rho))|\leq \widehat{B}_{f}(\varepsilon_r,2E_S|\shs G_{\!A_1},..,G_{\!A_m}), \quad  \varepsilon_r=\sqrt{\sum_{j=1}^l\Tr\bar{P}_r^{s_j}\rho_{A_{s_j}}},
\end{equation}
for all $r>r_0$, where $r_0$ is some natural number (not depending on $f$).

Since $\varepsilon_r\to0$ as $r\to+\infty$, the r.h.s. of (\ref{d-one}) tends to zero as $r\to+\infty$. By denoting the r.h.s. of (\ref{d-one}) by $Y_r$ for $r>r_0$ and setting $ Y_r=\widehat{B}_{f}(\varepsilon_r,E_r|\shs G_{\!A_1},..,G_{\!A_m})$ for $r\leq r_0$, where $E_r=E_S/(1-\varepsilon^2_r)$, we obtain  claim A of the theorem.\smallskip

B) By the definition of the FA-property for each natural $s$ in $[1,m]$ there exists a sequence  $\{g^s_i\}_{i=1}^{+\infty}$ of nonnegative numbers such that
\begin{equation*}
\sum_{i=1}^{+\infty}\lambda^s_i g^s_i<+\infty\quad \textrm{and} \quad \lim_{\beta\to 0^+}\left[\sum_{i=1}^{+\infty}e^{-\beta g^s_i}\right]^{\beta}=1,
\end{equation*}
where $\{\lambda^s_i\}_{i=1}^{+\infty}$ is the sequence of eigenvectors of the state $\rho_{A_s}$ arranged in the non-increasing order.
We may assume that $g_1^s=0$ for all $s$.

Then, by defining for each $s$ the positive operator
$G_{\!A_s}$ on $\H_{A_s}$ by the formula (\ref{G-s-def}) we may repeat all the arguments of the proof of claim A with the use
of the faithful $II_n^m$-type continuity bound $\widehat{B}_{f}(\varepsilon,E|H_{A_1},....,H_{A_m})$ for the function $f$ at the last step
(since now all the operators $G_{A_1}$,...,$G_{A_m}$ satisfy condition  (\ref{H-cond+})).

By noting that the r.h.s. of (\ref{d-one}) tends to zero as $r\to+\infty$ and by constructing the sequence $Y_r$ in the same way as in the proof of claim A  we obtain  claim B of the theorem.

The last claim of the theorem follows from the above proof.  $\square$ \smallskip

\begin{remark}\label{main-r+} If $\FF$ is a family of functions having a common faithful $I^m_n$-type (resp. $II^m_n$-type) continuity bound  then the last claim of Theorem \ref{main} implies that inequality (\ref{main-r-A})  holds for all functions $f\in\FF$ with the sequence $\{Y_r\}_{r=1}^{+\infty}$
 tending to zero and \emph{not depending on} $f$. So, in this case the approximation given by Theorem \ref{main} is uniform.
\end{remark}\smallskip

\begin{example}\label{m-ex} Let $f_{\Phi_1...\Phi_n}(\rho)=I(B_1\!:...:\!B_n)_{\Phi_1\otimes...\otimes\Phi_n(\rho)}$, where $\Phi_1:A_1\rightarrow B_1$,..., $\Phi_n:A_n\rightarrow B_n$
are arbitrary quantum channels. The arguments from Section V in \cite{CBM} show the function $f_{\Phi_1...\Phi_n}$ belongs to the class $L_{n}^{n-1}(2,n)$ for any channels $\Phi_1,...,\Phi_n$. It follows that the quantity $\widehat{B}_{2,n}(\varepsilon,E|H_{A_1},..,H_{A_{n-1}})$ defined in (\ref{fcb}) is a
common faithful $II_n^{n-1}$-type continuity bound for the family $\FF=\{f_{\Phi_1...\Phi_n}\}_{\Phi_1...\Phi_n}$ (where $\Phi_k$ runs over the set of all channels from $A_k$ to $B_k$, $k=\overline{1,n}$). Thus, if the marginal states $\rho_{A_1}$,...,$\rho_{A_{n-1}}$ have the FA-property then  Theorem
\ref{main}B and Remark \ref{main-r+} imply that for any subset $\{s_1,...,s_l\}$ of $\{1,...,n\}$ there exists a sequence $\{Y_r\}$  tending to zero such that
\begin{equation}\label{mi-uc}
|I(B_1\!:...:\!B_n)_{\Phi_1\otimes...\otimes\Phi_n(\rho_r)}-I(B_1\!:...:\!B_n)_{\Phi_1\otimes...\otimes\Phi_n(\rho)}|\leq Y_r  \qquad \forall r\in\mathbb{N},
\end{equation}
where $\rho_r=\Lambda^{s_1,...,s_l}_r(\rho)$, for any channels $\Phi_1,...,\Phi_n$. \smallskip

The last assertion of Theorem \ref{main} shows that inequality (\ref{mi-uc}) remains valid with
$\rho$ and $\rho_r=\Lambda^{s_1,...,s_l}_r(\rho)$ replaced by $\sigma$ and $\sigma_r=\Lambda^{s_1,...,s_l}_r(\sigma)$, where
$\sigma$ is any state in $\S(\H_{A_1..A_n})$ such that $\sigma_{A_k}=\rho_{A_k}$ for each $k\in\{s_1,...,s_l\}\cup\{1,...,n-1\}$.  \smallskip

\end{example}

\smallskip
\subsection{Some corollaries}

By using the results of the previous subsection one can
obtain the following observations that can  be applied to analysis of
entanglement and correlations measures in multipartite quantum systems (concrete applications are considered in Section 4).

\smallskip

\begin{corollary}\label{main-c} \emph{Let $E$ be a nonnegative function on the set $\,\S(\H_{A_1...A_n})$ such that
\begin{equation}\label{m-cond}
E(\Phi_1\otimes...\otimes\Phi_m\otimes \id_{A_{m+1}...A_n}(\rho))\leq E(\rho)
\end{equation}
for any $\rho\in\S(\H_{A_1...A_n})$ and any quantum channels $\Phi_s:A_s\to A_s$, $s=\overline{1,m}$, $m\leq n$. Let $\,\S_{0}(\H_{A_1...A_n})$ be the trace norm closure of the set $E^{-1}(0)\cap\S_{\rm f}(\H_{A_1...A_n})$, where}
$$
\S_{\rm f}(\H_{A_1...A_n})\doteq\left\{\rho\in\S(\H_{A_1...A_n})\,\left|\,\rank\rho_{A_s}<+\infty,\; s=1,2,..,n\right\}\right..
$$

\emph{If the function $E$ belongs to the class $I_n^m$ then}

\begin{enumerate}[(i)]
  \item \emph{the function $E$ is finite and lower semicontinuous on the set
\begin{equation}\label{S-I}
\S^m_{I}(\H_{A_1...A_n})=\left\{\shs\rho\in\S(\H_{A_1...A_n})\,|\,S(\rho_{A_{s}})<+\infty,\shs s=1,2,..,m\shs\right\}.
\end{equation}
Moreover,
\begin{equation}\label{E-LS}
\liminf_{k\to+\infty} E(\rho_k)\geq E(\rho_0)
\end{equation}
for arbitrary sequence $\{\rho_k\}\subset\S(\H_{A_1...A_n})$ converging to a state $\rho_0\in\S^m_I(\H_{A_1...A_n})$;}\smallskip

\item \emph{$E(\rho)=0$ for any state $\rho\in\S_{0}(\H_{A_1...A_n})\cap\S^m_{I}(\H_{A_1...A_n})$;}

\item \emph{if, in addition, condition (\ref{m-cond}) holds with $\,m=n$ then $E(\rho)>0$ for any state $\rho\in\S(\H_{A_1...A_n})$ not belonging to the set $\,\S_{0}(\H_{A_1...A_n})$;}

\item \emph{$E(\rho)=\widetilde{E}(\rho)$ for all  $\rho\in\S^m_{I}(\H_{A_1...A_n})$, where $\widetilde{E}$ is any function on $\S(\H_{A_1...A_n})$  belonging to the class $I_n^m$ such that $E(\rho)=\widetilde{E}(\rho)$ for all $\rho\in\S_{\rm f}(\H_{A_1...A_n})$;}
\item \emph{if, in addition, $E$ is a convex  function then}
\begin{itemize}
  \item \emph{$E(\rho)=\overline{\rm conv}E(\rho)$ for any  $\rho\in\S^m_{I}(\H_{A_1...A_n})$, where $\overline{\rm conv}E(\rho)$ is the convex closure of $E$;}\footnote{It means that $\overline{\rm conv}E(\rho)$ is the maximal convex lower semicontinuous function on $\S(\H_{A_1...A_n})$ not exceeding the function $E$.}
  \item \emph{$E$ is  $\mu$-integrable\footnote{It means that $E$ is measurable w.r.t.
the smallest $\sigma$-algebra on $\S(\H_{A_1...A_n})$ that contains all Borel sets and all the subsets of the zero-$\mu$-measure Borel sets.
The integral in the r.h.s. of (\ref{mu-convex}) is over the corresponding completion of the measure $\mu$ \cite{Rudin}.} w.r.t.  any Borel probability measure $\mu$ on $\S(\H_{A_1...A_n})$ with the barycenter (defined in (\ref{bar}))
belonging to the set $\S^m_{I}(\H_{A_1...A_n})$ and}
\begin{equation}\label{mu-convex}
E(\bar{\rho}(\mu))\leq\int_{\S(\H_{A_1...A_n})}E(\rho)\mu(d\rho).
\end{equation}
\end{itemize}
\end{enumerate}

\emph{If the function $E$ belongs to the class $II_n^m$ then the above claims $(i)\textup{-}(v)$
hold with the set $\S^m_{I}(\H_{A_1...A_n})$ replaced by
\begin{equation}\label{S-II}
\S^m_{II}(\H_{A_1...A_n})=\left\{\shs\rho\in\S(\H_{A_1...A_n})\,|\,\rho_{A_{s}}\!\!\in\S_{\rm\textsf{FA}\!}(\H_{A_{s}}),\shs s=1,2,..,m\shs\right\},
\end{equation}
where $\S_{\rm\textsf{FA}\!}(\H_{A_{s}})$ is the subset of $\S(\H_{A_{s}})$ consisting of states with the $FA$-property.}
\end{corollary}\medskip

\emph{Proof.} Assume first that the function $E$ belongs to the class $I_n^m$.

The finiteness of $E$ on $\S^m_I(\H_{A_1...A_n})$ follows directly from Theorem \ref{main}A. So,
to prove $\rm (i)$ it suffices to show that the relation (\ref{E-LS}) holds for any
sequence $\{\rho_k\}$ in $\S(\H_{A_1...A_n})$ converging to a state $\rho_0\in\S^m_I(\H_{A_1...A_n})$.

Consider the function
\begin{equation}\label{E-hat}
\widehat{E}(\rho)=\sup_{\Phi_1,...,\Phi_{m}}E(\Phi_1\otimes...\otimes\Phi_{m}\otimes\id_{A_{m+1}...A_n}(\rho))
\end{equation}
on $\S(\H_{A_1...A_n})$, where the supremum is over all channels $\Phi_1:A_1\to A_1$,...,$\Phi_m:A_m\to A_m$ with finite-dimensional outputs.
By the monotonicity condition (\ref{m-cond}) we have
\begin{equation}\label{m-ineq}
\widehat{E}(\rho)\leq E(\rho)\quad \forall\rho\in\S(\H_{A_1...A_n}).
\end{equation}

Since $E\in I_n^m$,  it follows from Theorem \ref{main}A along with
Remark \ref{main-r+++}  that
\begin{equation}\label{r-lim+}
\lim_{r\to+\infty} E(\widetilde{\Lambda}^{1,...,m}_r(\rho_0))=\lim_{r\to+\infty} E(\Phi^1_r\otimes...\otimes\Phi^{m}_r\otimes\id_{A_{m+1}...A_n}(\rho_0))=E(\rho_0),
\end{equation}
where $\{\widetilde{\Lambda}^{1,...,m}_r\}_r$ is the family of maps  defined in
(\ref{ap-map+}) via the channels  $\Phi^1_r$,...,$\Phi^m_r$ defined after (\ref{ap-map+}). Since the channels $\Phi^1_r$,...,$\Phi^m_r$ have
finite-dimensional outputs for all $r$, it follows from (\ref{m-ineq}) and (\ref{r-lim+}) that $\widehat{E}(\rho_0)=E(\rho_0)$.

Since for any given channels $\Phi_1:A_1\to A_1$,...,$\Phi_m:A_m\to A_m$ with finite-dimensional outputs the function
$\rho\mapsto E(\Phi_1\otimes...\otimes\Phi_{m}\otimes\id_{A_{m+1}...A_n}(\rho))$ is continuous on $\S(\H_{A_1...A_n})$ by  Lemma \ref{main-l} below, the function
$\widehat{E}$ is lower semicontinuous on $\S(\H_{A_1...A_n})$. So, it follows from (\ref{m-ineq}) and the equality $\widehat{E}(\rho_0)=E(\rho_0)$ mentioned before that
$$
\liminf_{k\to+\infty} E(\rho_k)\geq\liminf_{k\to+\infty} \widehat{E}(\rho_k)\geq \widehat{E}(\rho_0)=E(\rho_0).
$$

To prove $\rm (ii)$
assume that $\rho$ is a state in $\S_{0}(\H_{A_1...A_n})\cap\S^m_{I}(\H_{A_1...A_n})$. By the definition of $\S_{0}(\H_{A_1...A_n})$
the state $\rho$  is a limit of some sequence $\{\rho_k\}$ of states in
$\S(\H_{A_1...A_n})$ such that $\rank[\rho_k]_{A_s}<+\infty$ for $s=\overline{1,n}$ and  $E(\rho_k)=0$ all $k$.
Since $\rho$ is a state in $\S^m_{I}(\H_{A_1...A_n})$ it follows from  $\rm (i)$ that
$$
E(\rho)\leq\liminf_{k\to+\infty} E(\rho_k)=0.
$$

To prove $\rm (iii)$ take sequences of channels  $\Phi_k^1:A_1\to A_1$,...,$\Phi_k^n:A_n\to A_n$ with finite-dimensional outputs
strongly converging, respectively, to the identity channels $\id_{A_1}$,...,$\id_{A_n}$.
By the monotonicity condition (\ref{m-cond}) with $m=n$ the assumption $E(\rho)=0$ implies that $E(\rho_k)=0$ for all $k$, where $\rho_k\doteq\Phi^k_1\otimes...\otimes\Phi^k_{n}(\rho)$. Thus, all the states $\rho_k$ lie in $E^{-1}(0)\cap\S_{\rm f}(\H_{A_1...A_n})$. So,
the state $\rho$ belongs to the set $\S_{0}(\H_{A_1...A_n})$ because $\rho_k$ tends to $\rho$ as $k\to+\infty$.\smallskip

To prove $\rm (iv)$ assume that $\rho$ is a state in $\S^m_I(\H_{A_1...A_n})$. For each natural $s$ in $[1,n]$ let $P_r^s$ be the spectral projector of $\rho_{A_s}$ corresponding to its $r$ maximal eigenvalues (taking the multiplicity into account). Theorem \ref{main}A implies that
\begin{equation*}
\lim_{r\to+\infty} E(\rho_r)=E(\rho)<+\infty\quad\textrm{ and }\quad\lim_{r\to+\infty} \widetilde{E}(\rho_r)=\widetilde{E}(\rho)<+\infty,
\end{equation*}
where $\rho_r=c_r^{-1}Q_r\rho\shs Q_r$,  $\,Q_r=P_r^1\otimes...\otimes P_r^{n}$, $c_r=\Tr Q_r\rho$. Since $E(\rho_r)=\widetilde{E}(\rho_r)$ for any $r$ by the condition,
the above limit relations show that  $E(\rho)=\widetilde{E}(\rho)$.\smallskip

To prove the first part of $\rm (v)$ it suffices to show that  $\overline{\rm conv}E(\rho)$ coincides with the
function $\widehat{E}$ defined in (\ref{E-hat}), because the coincidence of $\widehat{E}$
with the function $E$ on $\S^m_I(\H_{A_1...A_n})$ is established in the proof of $\rm (i)$.

The lower
semicontinuity of $\widehat{E}$ is shown  in the proof of $\rm (i)$. So, since the convexity of $E$ implies the the convexity of $\widehat{E}$, we have only
to prove that $\overline{\rm conv}E\leq \widehat{E}$.

For a given arbitrary state $\rho_0\in\S(\H_{A_1...A_n})$ consider the sequence of states $\rho_r=\widetilde{\Lambda}^{1,...,m}_r(\rho_0)$
defined in the proof of $\rm (i)$ that converges to the state $\rho_0$. By the definition
of the function $\widehat{E}$ we have $\widehat{E}(\rho_r)\leq E(\rho_r)\leq \widehat{E}(\rho_0)$. So, the lower
semicontinuity of $\widehat{E}$ shows that $E(\rho_r)$ tends to $\widehat{E}(\rho_0)$ as $r\to+\infty$.
Hence, since
$\overline{\rm conv}E\leq E$, by the lower
semicontinuity of $\overline{\rm conv}E$ we have
$$
\overline{\rm conv}E(\rho_0)\leq\lim_{r\to+\infty}\overline{\rm conv}E(\rho_r)\leq\lim_{r\to+\infty}E(\rho_r)=\widehat{E}(\rho_0).
$$
The first part of  $\rm (v)$ is proved.

Let $\mu$ be a Borel probability measure on $\S(\H_{A_1...A_n})$ with the barycenter $\rho=\bar{\rho}(\mu)$ in $\S^m_I(\H_{A_1...A_n})$.
Following the proof of Theorem \ref{main}
for each natural $s$ in $[1,m]$ we construct a positive operator
$G_{\!A_s}$ on $\H_{A_s}$ satisfying condition (\ref{H-cond}) such that
$\Tr G_{\!A_s}\rho_{A_s}<+\infty$.

Since $E\in I_n^m$, the function $E$ is continuous on the closed subset $\C_{E}$  of $\S(\H_{A_1,..,A_n})$ consisting of states  $\varrho$ such that $\sum_{s=1}^{m}\Tr G_{\!A_s}\varrho_{A_s}\leq E$ for any $E>0$. It follows that the function $f_E$ coinciding with $E$ on the set $\C_{E}$ and equal
to $\,+\infty\,$ on the set $\S(\H_{A_1,..,A_n})\setminus\C_{E}$ is a Borel function on $\S(\H_{A_1,..,A_n})$ for each $E>0$.

Since
$$
\int_{\S(\H_{A_1...A_n})}\left[\sum_{s=1}^{m}\Tr G_{\!A_s}\varrho_{A_s}\right] \mu(d\varrho)=\sum_{s=1}^{m}\Tr G_{\!A_s}\rho_{A_s}<+\infty,
$$
we have $\mu(\S(\H_{A_1...A_n})\setminus\C_{*})=0$, where $\C_{*}=\cup_{E>0}\C_E$.
Hence, the function $E$ is\break $\mu$-integrable as it coincides with the Borel function $f_*=\inf_{E>0} f_E$
on the set $\C_{*}$.

The function $\overline{\rm conv}E$  is  convex and lower semicontinuous
on $\S(\H_{A_1...A_n})$ and does not  exceed the function $E$. Hence, by using the validity of Jensen's inequality for $\overline{\rm conv}E$ (cf.~\cite[Proposition A-2]{EM}) we obtain
$$
E(\bar{\rho}(\mu))=\overline{\rm conv}E(\bar{\rho}(\mu))\leq\int_{\S(\H_{A_1...A_n})} \overline{\rm conv}E(\varrho)\mu(d\varrho)\leq\int_{\S(\H_{A_1...A_n})} E(\varrho)\mu(d\varrho),
$$
where the equality follows from the first part of $\rm(v)$.

If the function $E$ belongs to the class $II_n^m$ then the
validity of  claims $\rm (i)\textup{-}(v)$  with $\S^m_{I}(\H_{A_1...A_n})$ replaced by the set $\S^m_{II}(\H_{A_1...A_n})$ defined in (\ref{S-II})
is proved by repeating the same arguments with the use of part $B$ of Theorem \ref{main}. $\square$ \medskip

\begin{lemma}\label{main-l} \emph{Let $\H_{A_1},...,\H_{A_n}$ be separable Hilbert spaces and
$\H^0_{A_1},...,\H^0_{A_m}$ be finite-dimensional subspaces of the spaces $\H_{A_1},...,\H_{A_m}$, respectively $(m\leq n)$.}

\emph{If  a function $f$ on $\,\S(\H_{A_1...A_n})$ belongs to the class $II_n^m$ then it is
uniformly continuous on the set of states in $\,\S(\H_{A_1...A_n})$ supported by the subspace}
\begin{equation}\label{subspace}
\H^0_{A_1}\otimes...\otimes\H^0_{A_m}\otimes\H_{A_{m+1}}\otimes...\otimes\H_{A_n}.
\end{equation}
\end{lemma}

\emph{Proof.} For each natural $s$ in $[1,m]$ it is easy to  construct a positive operator
$H_{A_s}$ on $\H_{A_s}$ satisfying condition (\ref{H-cond+}) such that $\,\ker H_{\!A_s}=\H^0_{A_s}$.
Since the set of states supported by the subspace in (\ref{subspace}) is contained
within the set of all states $\rho\in\S(\H_{A_1...A_n})$ such that $\sum_{s=1}^{m}\Tr H_{A_s}\rho_{A_s}\leq E$ for any $E>0$,
the assertion of the lemma follows from the existence of a faithful $II_n^m$-type continuity bound for $f$ (Definition 2 in Section 2.3). $\Box$ \smallskip

\begin{remark}\label{main-r+c}  Claim $\rm (ii)$ of Corollary \ref{main-c} gives a condition
under which an entanglement measure $E$ is equal to zero on the set of separable (or $\pi$-separable, see Section 4.1) states of
an infinite-dimensional composite quantum system  provided that this equality is proved for this measure in the finite-dimensional settings.
Claim $\rm (iii)$ of Corollary \ref{main-c} shows that to prove the faithfulness of an entanglement measure $E$ (its positivity on non-separable states) it suffices to
prove this property in the finite-dimensional settings.
\end{remark}\smallskip

\begin{example}\label{Sq-exam} The squashed entanglement of a state $\rho$ of a finite-dimensional bipartite quantum system $AB$ is defined as
\begin{equation}\label{se-def}
  E_{\rm sq}(\rho)=\frac{1}{2}\inf_{\hat{\rho}\in\mathfrak{M}_1(\rho)}I(A\!:\!B|E)_{\hat{\rho}},
\end{equation}
where  $E$ is a  quantum system of arbitrary finite dimension, $\mathfrak{M}_1(\rho)$ is the set of all extensions of $\rho$ to a state in $\S(\H_{ABE})$ and
$I(A\!:\!B|E)_{\hat{\rho}}=S(\hat{\rho}_{AE})+S(\hat{\rho}_{BE})-S(\hat{\rho}_{ABE})-S(\hat{\rho}_{E})$ is the quantum conditional mutual information
of $\hat{\rho}$ \cite{C&W,Tucci}. This definition is generalized to an arbitrary state $\rho$ of an infinite-dimensional bipartite quantum system $AB$
by using the extension of $I(A\!:\!B|E)$ to all states infinite-dimensional tripartite quantum system $ABE$ proposed in \cite[Theorem 2]{CMI} (see details in \cite{SE}).

The continuity bound for the function $E_{\rm sq}$ obtained in \cite[Proposition 22]{SE} shows that this function belongs to the class $I_2^1$ in both settings
$A_1=A,A_2=B$ and $A_1=B,A_2=A$. So, since the function $E_{\rm sq}$ satisfies condition (\ref{m-cond}) with $m=n=2$ claim $\rm (ii)$  of Corollary \ref{main-c}
shows that $E_{\rm sq}(\rho)=0$ for any separable state $\rho$ in $\S(\H_{AB})$ such that $\min\{S(\rho_A),S(\rho_B)\}<\infty$ (because this property obviously holds for any  separable state $\rho$ of a finite-dimensional system $AB$ \cite{C&W}). This claim is nontrivial because of the existence of  countably non-decomposable separable states in infinite-dimensional bipartite
quantum systems (see Remark 10 in \cite{SE}).  It is still unclear whether the equality $E_{\rm sq}(\rho)=0$ is true for all separable states in $\S(\H_{AB})$.

Claim $\rm (iii)$  of Corollary \ref{main-c} allows us to derive the faithfulness of the squashed entanglement $E_{\rm sq}$ in infinite-dimensional bipartite
quantum systems from the same property proved in \cite{Brandao} in the finite-dimensional settings. Note that the proof the faithfulness of the squashed entanglement $E_{\rm sq}$
given in \cite{Brandao} is very nontrivial and its direct generalization to the infinite-dimensional case would require serious efforts.
\end{example}\smallskip

\begin{remark}\label{main-r}  The aim of claim $\rm (iv)$ of Corollary \ref{main-c} is to simplify definitions of entanglement measures in infinite-dimensional composite quantum systems. Its concrete applications are considered in Section 4. The usefulness (and non-triviality) of claim $\rm (v)$ can be illustrated
by the following
\end{remark}\smallskip

\begin{example}\label{main-e}  The entanglement of formation (EoF) of a state of a finite-dimensional quantum system $AB$
is defined as a convex roof extension to the set $\S(\H_{AB})$ of the function $\rho\mapsto S(\rho_A)$ on the subset of pure states in $\S(\H_{AB})$ \cite{Bennett}.
In the infinite-dimensional case there are two versions $E_{F,d}$ and $E_{F,c}$ of the EoF defined, respectively, by means of
discrete and continuous convex roof extensions (see details in  \cite{EM},\cite[Section 4.4.2]{QC}). It is still unclear whether the functions $E_{F,d}$ and $E_{F,c}$ coincide or not.

Claim $\rm (v)$ of Corollary \ref{main-c} directly implies that $E_{F,d}(\rho)=E_{F,c}(\rho)$ for any state $\rho$ in $\S(\H_{AB})$ such that
$\min\{S(\rho_A),S(\rho_B)\}<\infty$. Indeed, the function $E=E_{F,d}$ is convex, satisfies the local monotonicity condition (\ref{m-cond}) and has a $I_2^1$-type faithful continuity bound presented in Proposition 22 in \cite{SE}. So, by the symmetry of $E_{F,d}$ w.r.t. the subsystems $A$ and $B$, claim $\rm (v)$ of Corollary \ref{main-c} show that
\begin{equation}\label{mu-convex+}
E_{F,d}(\rho)\leq\int_{\S(\H_{AB})}E_{F,d}(\varrho)\mu(d\varrho)=\int_{\S(\H_{AB})}S(\varrho_A)\mu(d\varrho)
\end{equation}
for any state $\rho\in\S(\H_{AB})$ such that $\min\{S(\rho_A),S(\rho_B)\}<\infty$ and any Borel probability measure $\mu$ on $\S(\H_{AB})$ supported by pure states with the barycenter $\rho$ (the last equality follows from the coincidence of $E_{F,d}(\varrho)$ with  $S(\varrho_A)$ for any pure state $\varrho$).
This inequality and the explicit expression for $E_{F,c}(\rho)$ (formula (132) in \cite{QC}) imply that
$E_{F,d}(\rho)\leq E_{F,c}(\rho)$. Since the converse inequality is obvious, we have $E_{F,d}(\rho)=E_{F,c}(\rho)$.

Note that the original proof of the above condition for coincidence of $E_{F,d}(\rho)$ and $E_{F,c}(\rho)$ (given in \cite{EM}) requires
more technical efforts and additional results. Note also that the above observation shows that \emph{to prove the global
coincidence of $E_{F,d}$ and $E_{F,c}$ it suffices to prove the $\mu$-integrability of $E_{F,d}$ for
any Borel probability measure $\mu$ on $\S(\H_{AB})$ supported by pure states\footnote{The $\mu$-integrability of $E_{F,d}$
is not obvious, since it is unclear whether the function $E_{F,d}$ is Borel or not (in contrast to the function $E_{F,c}$ which is lower semicontinuous on $\S(\H_{AB})$).} and the validity of the  inequality in (\ref{mu-convex+}) for
this measure.}
\end{example}
\medskip

Almost all correlation and entanglement measures used in quantum information theory are hardly computable functions defined by mathematically complex expressions
(including optimizations, regularizations, etc.). So, in the study of analytical properties of such characteristics it is desirable to
obtain conditions of local continuity (convergence) expressed in terms of local continuity of some easily defined functions.

It was observed recently (cf.\cite[Section 4]{LCA}) that for several important  correlation and entanglement measures $E$ in a composite quantum system
$A_1...A_n$ the following claim is valid: if $\{\rho_k\}\subset\S(\H_{A_1...A_n})$ is a sequence converging to a state $\rho_0\in\S(\H_{A_1...A_n})$
such that
\begin{equation}\label{mi-cont-g}
\lim_{k\to+\infty} I(A_1\!:...:\!A_n)_{\rho_k}=I(A_1\!:...:\!A_n)_{\rho}<+\infty
\end{equation}
then
\begin{equation}\label{E-LC}
\lim_{k\to+\infty} E(\rho_k)=E(\rho_0)<+\infty.
\end{equation}
This continuity condition has a clear physical sense, since it means that local continuity of total correlation $I(A_1\!:...:\!A_n)$ implies the local continuity
of the correlation (entanglement) measure $E$. Verifiable conditions for the validity of (\ref{mi-cont-g}) are presented in \cite[Section 4.2]{LCA}.\smallskip

By combining the approximation methods proposed in Section 3 and in \cite[Section 3.2]{LCA} one can prove
the following auxiliary result which simplifies the proof of the implication (\ref{mi-cont-g}) $\Rightarrow$ (\ref{E-LC}) for a wide class of characteristics.
\smallskip

\begin{corollary}\label{main-c+} \emph{Let $E$ be a nonnegative function on the set $\S(\H_{A_1...A_n})$
satisfying the local monotonicity condition (\ref{m-cond}) such that}
\begin{itemize}
  \item \emph{$E(\rho)\leq cI(A_1\!:...:\!A_n)_{\rho}$ for all $\rho$ in $\,\S(\H_{A_1...A_n})$ and some $\,c>0$;}
  \item \emph{for any states $\rho$  and $\sigma$ in $\S(\H_{A_1..A_n})$ and any $p\in(0,1)$ the inequality
\begin{equation}\label{w-convex}
E(p\rho+(1-p)\sigma)\leq p E(\rho)+(1-p)E(\sigma)+b(p)
\end{equation}
holds with possible value $+\infty$ in one or both sides, where $b(p)$ is
a continuous nonnegative function on $[0,1]$ with $\,b(0)=0$.}
\end{itemize}

\emph{If the function $E$ belongs to the class $I_n^m$ then to prove the implication (\ref{mi-cont-g}) $\Rightarrow$ (\ref{E-LC}) for any
sequence $\{\rho_k\}\subset\S(\H_{A_1...A_n})$  converging to a state $\rho_0$ from the set $\,\S^m_I(\H_{A_1...A_n})$
defined in  (\ref{S-I}) it suffices to show that
\begin{equation}\label{E-US}
\limsup_{k\to+\infty} E(\rho_k)\leq E(\rho_0)
\end{equation}
for any
sequence $\{\rho_k\}\subset\S(\H_{A_1...A_n})$  converging to a state $\rho_0\in\S^m_I(\H_{A_1...A_n})$ such that}
\begin{equation}\label{S-cont}
\lim_{k\to+\infty} S([\rho_k]_{A_i})=S([\rho_0]_{A_i}),\;\; i=\overline{1,m}.
\end{equation}

\emph{If the function $E$ belongs to the class $II_n^m$ then the above claim holds  with the set $\S^m_{I}(\H_{A_1...A_n})$ replaced by
the set $\S^m_{II}(\H_{A_1...A_n})$ defined in  (\ref{S-II}).}
\end{corollary}\medskip

\emph{Proof.} It suffices to prove the corollary in the case $c=1$. Assume that the function $E$ belongs to the class $I_n^m$  and
$\{\rho_k\}\subset\S(\H_{A_1...A_n})$ is a sequence converging to a state $\rho_0\in\S_I^m(\H_{A_1...A_n})$ such that (\ref{mi-cont-g}) holds.

By claim $\rm (i)$ of the $I_n^m$-part of Corollary \ref{main-c} to prove (\ref{E-LC}) it suffices to show that (\ref{E-US})
holds for the  sequence $\{\rho_k\}$ and the state $\rho_0$. For this purpose we will use the approximation technique proposed in \cite[Section 3.2]{LCA} (under the term "quantum Dini's lemma").

Let $\Psi_v(\rho_k)=P_v^{\rho_k}\rho_k$ for any $v\in\mathbb{N}$ and $k\geq0$, where $P_v^{\rho_k}$ is the spectral projector of
$\rho_k$ corresponding to its $v$ maximal eigenvalues (taking the multiplicity into account). It is shown at the end of Section 3.2.1 in \cite{LCA}
that the family $\{\Psi_v\}$  of maps from $\,\mathfrak{A}=\{\rho_k\}\cup\{\rho_0\}\,$ into $\T_+(\H_{A_1...A_n})$ satisfies all the conditions stated at the begin of that section (we use $v$ instead of $m$, since $m$ is occupied). In this case the set $M_{\rho_0}$ (introduced at the begin of Section 3.2.1 in \cite{LCA})
consists of all natural numbers $v$ such that either $\lambda^{\rho_0}_{v+1}<\lambda^{\rho_0}_{v}$ or $\lambda^{\rho_0}_{v}=0$ ($\{\lambda^{\rho_0}_i\}_i$ is the sequence of eigenvalues of $\rho_0$ arranged in the non-increasing order).

Denote the state  proportional to  the operator $\Psi_v(\rho_k)$ and the positive number $\Tr\Psi_v(\rho_k)$ by $\rho_k^v$ and $\mu^v_k$ correspondingly (by Lemma 3.2 in \cite{LCA} we may assume that $v$ is so large  that $\Tr\Psi_v(\rho_k)>1/2$ for all $k\geq0$). By definition of the set $M_{\rho_0}$ we have
\begin{equation}\label{M-pr}
  \lim_{k\to+\infty }\rho^v_k=\rho^v_0\quad \forall v\in M_{\rho_0}.
\end{equation}

The function $f(\varrho)=I(A_1\!:...:\!A_n)_{\varrho}$ is nonnegative  and lower semicontinuous on $\S(\H_{A_1...A_n})$. It satisfies the
inequality (\ref{QMI-LAA-1}). By the assumption  we have
\begin{equation*}
  \lim_{k\to+\infty }f(\rho_k)=f(\rho_0)<+\infty.
\end{equation*}
The function $g(\varrho)=E(\varrho)$ is nonnegative and satisfies inequality (\ref{w-convex}) on $\S(\H_{A_1...A_n})$.
By the assumption
$g(\varrho)\leq f(\varrho)$ for all $\varrho\in\S(\H_{A_1...A_n})$. Thus, to prove the validity of (\ref{E-US}) for the  sequence $\{\rho_k\}$ and the state $\rho_0$ it suffices, by Corollary 3.1 in \cite[Section 3.2.1]{LCA}, to show that
\begin{equation}\label{d-l-r+}
 \limsup_{v\to+\infty}g(\rho^v_0)\leq g(\rho_0)\quad \textrm{and} \quad \limsup_{k\to+\infty }g(\rho^v_k)\leq g(\rho^v_0)\quad \forall v\in M_{\rho_0}.
\end{equation}

To prove the first limit relation
in (\ref{d-l-r+}) note that the condition $\rho_0\in\S^m_{I}(\H_{A_1...A_n})$ implies the finiteness of
$S([\rho_0]_{A_i})$ for all $i=\overline{1,m}$. So, Simon's dominated convergence theorem \cite[the Appendix]{Ruskai}
implies that $S([\rho^v_0]_{A_i})$ tends to
$S([\rho_0]_{A_i})$ as $v\to+\infty$ (because $\mu_0^v[\rho_0^v]_{A_i}\leq [\rho_0]_{A_i}$ for all $v$ and $\mu_0^v$ tends to $1$ as $v\to+\infty$).
Hence, the condition (\ref{S-cont}) is valid for the sequence  $\{\rho^v_0\}_v$ converging to the state $\rho_0\in\S^m_{I}(\H_{A_1...A_n})$.
Thus, the assumption of the corollary implies the first limit relation in (\ref{d-l-r+}).

To prove the second limit relation in (\ref{d-l-r+}) note that (\ref{M-pr}) and (\ref{mi-cont-g}) imply
\begin{equation}\label{mi-cont-g+}
\lim_{k\to+\infty} I(A_1\!:...:\!A_n)_{\rho^v_k}=I(A_1\!:...:\!A_n)_{\rho_0^v}<+\infty\quad \forall v\in M_{\rho_0}
\end{equation}
by Corollary 4.1 in \cite{LCA} (because $\,\mu^v_k\rho^v_k\leq\rho_k\,$ for all $v$ and $k$ and $\mu^v_k\geq 1/2$ for all $k$  by the above assumption).
Since $\rank\shs \rho^v_k\leq v$ for all $k\geq0$, it follows from (\ref{M-pr}) that
\begin{equation}\label{2-l-r+1}
\lim_{k\to+\infty }S(\rho^v_k)=S(\rho^v_0)<+\infty\quad \forall v\in M_{\rho_0}.
\end{equation}

Using the lower semicontinuity of the von Neumann entropy and the second formula in (\ref{mi-mpd}) it is easy to show that the  limit relations  (\ref{mi-cont-g+}) and (\ref{2-l-r+1}) guarantee that
\begin{equation*}
\lim_{k\to+\infty}S([\rho^v_k]_{A_i})=S([\rho^v_0]_{A_i})<+\infty,\quad  i=1,2,...,n,\;\; \forall v\in M_{\rho_0}.
\end{equation*}

Hence, the condition (\ref{S-cont}) is valid for the sequence  $\{\rho^v_k\}_k$ converging to the state $\rho^v_0$ for each $v\in M_{\rho_0}$.
Since $\mu^v_0\rho^v_0\leq\rho_0\in\S^m_{I}(\H_{A_1...A_n})$, we have $\rho^v_0\in\S^m_{I}(\H_{A_1...A_n})$ for all $v$ (this follows from concavity
of the von Neumann entropy). Thus, the assumption of the corollary implies the second limit relation in (\ref{d-l-r+}).\smallskip

If the function $E$ belongs to the class $II_n^m$ then by using claim $\rm (i)$ of the $II_n^m$-part of Corollary \ref{main-c} and repeating the
above arguments we obtain the last claim of the corollary. It is necessary only no note
that the inequality  $\mu^v_0\rho^v_0\leq\rho_0$ and Corollary 2 in \cite{FAP} shows that
$\rho^v_0\in\S^m_{II}(\H_{A_1...A_n})$  for all $v\in M_{\rho_0}$ provided that $\rho_0\in\S^m_{II}(\H_{A_1...A_n})$. $\Box$\smallskip

\begin{example}\label{Sq-exam+} As mentioned in Example \ref{Sq-exam} the squashed entanglement $E_{\rm sq}(\rho)$ of a state $\rho$  of a bipartite quantum system $AB$ (defined in (\ref{se-def}))
is a convex function on $\S(\H_{AB})$ belonging to the class $I_2^1$ in both settings
$A_1=A,A_2=B$ and $A_1=B,A_2=A$. So, by the $I_n^m$-part of Corollary \ref{main-c+} to prove that
\begin{equation}\label{AB-imp}
\lim_{k\to+\infty} I(A\!:\!B)_{\rho_k}=I(A\!:\!B)_{\rho_0}<+\infty\quad\Rightarrow\quad\lim_{k\to+\infty} E_{\rm sq}(\rho_k)=E_{\rm sq}(\rho_0)<+\infty
\end{equation}
for any
sequence $\{\rho_k\}\subset\S(\H_{AB})$  converging to a state $\rho_0$ such that
either $S([\rho_0]_A)$ or $S([\rho_0]_B)$ is finite it suffices, by symmetry, to prove
the implication
\begin{equation}\label{AB-imp+}
\lim_{k\to+\infty} S([\rho_k]_A)=S([\rho_0]_A)<+\infty\quad\Rightarrow\quad\limsup_{k\to+\infty} E_{\rm sq}(\rho_k)\leq E_{\rm sq}(\rho_0)<+\infty.
\end{equation}
The last implication is easily verified (see the proof of Proposition 8D in \cite{SE}), while
the direct proof of (\ref{AB-imp}) presented in Section V-A in \cite{SE} requires serious technical efforts.

The use of Corollary \ref{main-c+} is a simple way to establish the implication (\ref{AB-imp}) for the c-squashed entanglement  $E_{\rm c\textup{-}sq}$ -- another entanglement measure
defined by the expression (\ref{se-def}) with the set $\mathfrak{M}_1(\rho)$ replaced by the set of all extensions of $\rho$ to a quantum-classical state
in $\S(\H_{ABE})$ w.r.t. the partition $AB|E$  \cite{Tucci,N&R,Y&C}. To do this, one should simply modify the proofs of Propositions 8D and 22
in \cite{SE} to show, respectively, that the implication (\ref{AB-imp+}) holds with $E_{\rm sq}$ replaced by $E_{\rm c\textup{-}sq}$ and that $E_{\rm c\textup{-}sq}\in I_2^1$ in the setting
$A_1=A,A_2=B$.
\end{example}\smallskip

The main  aim  of Corollary \ref{main-c+} is to simplify the proof of the implication (\ref{mi-cont-g}) $\Rightarrow$ (\ref{E-LC}) for entanglement measures --
convex functions on the set of states of a composite quantum system. Nevertheless, it can be also applied to
some important non-convex characteristics.\smallskip

\begin{example}\label{C-B} One of the important non-convex characteristics satisfying the inequality (\ref{w-convex})
is the one-way classical correlation $C_B$ (with measured system $B$) \cite{H&V}: using Lemma 1 in \cite[Section 4.3]{QC} one can show that the inequality (\ref{w-convex}) holds for $E=C_B$ with $b(p)=h_2(p)$ (the binary entropy). Since it was mentioned in Example \ref{cl-exam} that $C_B$ belongs to class $I_2^1$ in the  setting $A_1=A,A_2=B$, Corollary \ref{main-c+} can be applied to prove  the implication (\ref{AB-imp}) with $E_{\rm sq}$ replaced by $C_B$ under the condition $S([\rho_0]_A)<+\infty$. Indeed, the  implication (\ref{AB-imp+}) with $E_{\rm sq}$ replaced by $C_B$
can be easily deduced from  the Koashi-Winter relation (\ref{KWS-gen}) between $C_B$ and $E_F$ by using Proposition 4.8B in \cite{LCA}.

The implication (\ref{AB-imp}) with $E_{\rm sq}$ replaced by $C_B$ without any restrictions on the state $\rho_0$ is proved in \cite[Proposition 4.7]{LCA} by completely different method.
\end{example}\smallskip

\begin{remark}\label{E-P} The condition $E\leq cI(A_1\!:...:\!A_n)$ in Corollary \ref{main-c+} can not be omitted. This can be shown by considering the entanglement of purification -- an important characteristic of a state of bipartite system $AB$ defined as
\begin{equation}\label{ep-def}
  E_{P}(\rho)=\displaystyle\inf_{\hat{\rho}\in\mathfrak{M}_1(\rho)}S(\hat{\rho}_{AE}),\quad \rho\in\S(\H_{AB}),
\end{equation}
where  $E$ is an  infinite-dimensional quantum system and $\mathfrak{M}_1(\rho)$ is the set of all extensions of $\rho$ to a state in $\S(\H_{ABE})$ \cite{EoP}.

It is not hard to show\footnote{A faithful $I_2^1$-type continuity bound for $E_P$ can be obtained by using Winter's results in \cite{W-CB}.} that the function $E_P$ belongs to the class $I_2^1$ and satisfies the inequality (\ref{w-convex})  with $b(p)=h_2(p)$, but \emph{these is no $c>0$ such that $E_P(\rho)\leq cI(A\!:\!B)_{\rho}$ for all $\rho$ in $\S(\H_{AB})$}.
To prove the last claim it suffices to mention the state $\rho_*\in\S(\H_{AB})$ constructed recently by Lami and Winter such that $E_{P}(\rho_*)=S([\rho_*]_A)=+\infty$, while $I(A\!:\!B)_{\rho_*}<+\infty$ \cite{L&W}. As a result, \emph{the claim
of the $I_n^m$-part of Corollary \ref{main-c+} is not valid for the function $E_P$.} Indeed, the implication (\ref{AB-imp+})
with $E_{\rm sq}$ replaced by $E_P$ can be easily proved by noting that $E_{P}(\rho)=\inf_{m\in\mathbb{N}} E^m_{P}(\rho)$ for any state $\rho$ with finite $S(\rho_A)$, where
$E^m_{P}$ is the function defined by formula (\ref{ep-def}) with $\mathfrak{M}_1(\rho)$ replaced by  the set of all extensions of $\rho$ to a state in $\S(\H_{ABE_m})$,
where $E_m$ is a $m$-dimensional quantum system. Nevertheless, the implication (\ref{AB-imp}) with $E_{\rm sq}$ replaced by $E_P$ is not valid
for a sequence $\{\rho_k\}\subset\S(\H_{AB})$  converging to a state $\rho_0\in \S^1_{I}(\H_{AB})$.

To show the last claim take any state $\sigma_0$ in $\S(\H_{AB})$ supported by the antisymmetric subspace in $\H_{AB}$ such that  $S([\sigma_0]_A)<+\infty$.
Consider the sequence of states $\sigma_k\doteq(1-1/k)\sigma_0+(1/k)\rho_*$, where $\rho_*$ is the state constructed by Lami and Winter (mentioned before). By  construction the state  $\rho_*$ is supported by the antisymmetric subspace in $\H_{AB}$. It follows that the state $\sigma_k$
has the same property for any $k\geq0$ and hence $E_P(\sigma_k)=S([\sigma_k]_A)$ for any $k\geq0$ by Proposition 7 in \cite{Cr&W}. Since
$E_P(\rho_*)=S([\rho_*]_A)=+\infty$ by the construction, we have $E_P(\sigma_k)=S([\sigma_k]_A)=+\infty$ for any $k>0$ by the concavity of the entropy. Thus, we have
$$
E_P(\sigma_k)=+\infty \quad\forall k>0,\quad  E_P(\sigma_0)<+\infty \quad \textrm{and} \quad \lim_{k\to+\infty} I(A\!:\!B)_{\sigma_k}=I(A\!:\!B)_{\sigma_0}<+\infty
$$
The last limit relation is due to the inequality   $I(A\!:\!B)_{\sigma_k}\leq (1-1/k) I(A\!:\!B)_{\sigma_0} +(1/k)I(A\!:\!B)_{\rho_*}+h_2(1/k)$ (which follows from (\ref{QMI-LAA-2})) and the lower semicontinuity of QMI (the finiteness of $I(A\!:\!B)_{\sigma_0}$ follows from the finiteness of $S([\sigma_0]_A)$ due to (\ref{nMI-UB})).
\end{remark}

\subsection{On convergence w.r.t. the strong information topology}

Since the relative entropy  for two quantum states can be considered as a
measure of divergence of these states, it may be used to define a special convergence of a sequence  $\{\rho_k\}$
of states to a state $\rho_0$ via the relation
\begin{equation}\label{SIC}
 \lim_{k\to+\infty} D(\rho_k\|\rho_0)=0.
\end{equation}
In the classical case this convergence and the corresponding sequential topology are
 studied in details in \cite{Harrem} where it is called  \emph{the strong information topology} and denoted by $I$.
We will follow this notation in the quantum case.

By Pinsker's inequality the $I$-convergence implies the trace norm convergence, i.e.

\begin{equation*}
  I\,\textup{-}\!\lim_{k\to+\infty}\rho_k=\rho_0\quad\Rightarrow \quad \|\!\cdot\!\|_1\textup{-}\lim_{k\to+\infty}\rho_k=\rho_0,
\end{equation*}
and it is easy to see that the converse implication is not true.  \smallskip

Proposition 2 in \cite{EC} (which generalizes Proposition 21 in \cite{Harrem} to the quantum case) claims  that
\begin{equation}\label{S-imp}
   I\,\textup{-}\!\lim_{k\to+\infty}\rho_k=\rho_0\quad\Rightarrow \quad \lim_{k\to+\infty} S(\rho_k)=S(\rho_0)<+\infty
\end{equation}
provided that $\Tr\rho^\lambda_0<+\infty$ for some $\lambda<1$. Following \cite{Harrem}, we will call states possessing the last property
\emph{power dominated}. One can show that the above implication is not valid if $\rho_0$ is not a power dominated state \cite[Proposition 2]{EC}.\smallskip

Applying the idea used in  the proof of Theorem \ref{main} one can establish the following \smallskip

\begin{property}\label{SIT} \emph{Let $E$ be a function on the set $\,\S(\H_{A_1...A_n})$ belonging to the class $I_n^m$, $1\leq m\leq n$,
and $\{\rho_n\}$ be  a sequence in $\,\S(\H_{A_1...A_n})$  converging to a state $\rho_0$ in the strong information topology (i.e. relation (\ref{SIC}) holds). If the states $[\rho_0]_{A_1}$,...,$[\rho_0]_{A_m}$ are power dominated (in the sense described before) then}
\begin{equation}\label{SIT+}
 \lim_{k\to+\infty} E(\rho_k)=E(\rho_0)<+\infty.
\end{equation}
\end{property}

\emph{Proof.} Note first that the monotonicity of the quantum relative entropy implies  that
\begin{equation}\label{D-k}
   I\,\textup{-}\,\lim_{k\to+\infty}[\rho_k]_{A_i}=[\rho_0]_{A_i}, \qquad i=1,2,...,m.
\end{equation}
Hence, implication (\ref{S-imp}) shows that
\begin{equation}\label{SIT++}
 \lim_{k\to+\infty} S([\rho_k]_{A_i})=S([\rho_0]_{A_i})<+\infty,\qquad i=1,2,...,m.
\end{equation}
It is easy to see that  (\ref{D-k}) and (\ref{SIT++}) imply that
$$
\lim_{k\to+\infty}\Tr[\rho_k]_{A_i}(-\ln [\rho_0]_{A_i})=\Tr[\rho_0]_{A_i}(-\ln [\rho_0]_{A_i}), \qquad i=1,2,...,m.
$$
Thus, it follows from the implication $\rm(iii)\Rightarrow\rm(i)$ in Proposition 4 in \cite{EC} that
for each $i=\overline{1,m}$ there exists a positive operator $G_i$ on $\H_{A_i}$ satisfying
condition (\ref{H-cond}) such that  $\sup_{k\geq0}\Tr G_i[\rho_k]_{A_i}<+\infty$. Using
this and repeating the steps of the proof of Theorem \ref{main}A based on applying any faithful $I_n^m$-type continuity bound $\widehat{B}_{E}(\varepsilon,E|H_{A_1},....,H_{A_m})$ for the function $E$
(with the operators $G_1$,...,$G_m$ in the role of $H_{A_1}$,...,$H_{A_m}$) it is easy to establish (\ref{SIT+}). $\Box$\smallskip

\begin{remark}  The proof of Proposition \ref{SIT} shows that:
\begin{itemize}
  \item the condition  (\ref{SIC}) can be replaced by the weaker condition  (\ref{D-k}) provided that
the sequence $\{\rho_k\}$ converges to the state $\rho_0$ in the trace norm;
  \item the assumption that the states $[\rho_0]_{A_1}$,...,$[\rho_0]_{A_m}$ are power dominated
can replaced by the condition (\ref{SIT++}).
\end{itemize}
\end{remark}

\begin{remark}\label{SIT-r} Since the von Neumann entropy  belongs to the class $I_1^1$,
Proposition \ref{SIT}  can be treated as a deep generalization
of the main claim of Proposition 2 in \cite{EC} (which states the implication (\ref{S-imp})).

The second claim of Proposition 2 in \cite{EC} shows that \emph{(\ref{SIC}) does not imply (\ref{SIT+})
in general} (when $E\in I_n^m$ and the  states $[\rho_0]_{A_1}$,...,$[\rho_0]_{A_m}$ are not power dominated).
\end{remark}

\section{Applications}

In this section we apply the results of Section 3 to analysis of important characteristics of composite infinite-dimensional quantum system:
the (unregularized and regularized) relative entropy of $\pi$-entanglement, the (unregularized and regularized) Rains bound
and the conditional entanglement of mutual information.

\subsection{The relative entropy of $\pi$-entanglement and its regularization}

\subsubsection{Definitions and basic properties}

Let $A_1...A_n$ be a $n$-partite infinite-dimensional quantum system. Denote the set of all partitions of $[n]\doteq\{1,\ldots, n\}$
by $P(n)$.\footnote{A partition is a finite collection of non-empty sets
$\pi(j)\subseteq[n]$ that do not intersect each other and together cover the whole $[n]$.}
For a given $\pi=\{\pi_k\}_{k\in K_{\pi}}\subseteq P(n)$
we define the set of $\pi$-separable states by (cf.\cite{Szalay2015})
$$
\S_{\rm sep}^{\pi}(\H_{A_1...A_n})\doteq \mathrm{cl}\left( \mathrm{conv}\left(\,\bigcup_{k\in K_{\pi}}\left\{ \bigotimes_{j=1}^{|\pi_k|} |\psi_j\rangle\langle\psi_j|,\; |\psi_j\rangle\in \H_{A_{\pi_k(j)}},\; \langle\psi_j | \psi_j\rangle = 1 \right\} \right) \right),
$$
where $A_{\pi_k(j)}$ is the system obtained by joining those $A_i$ such that $i\in \pi_k(j)$, $|\pi_k|$ -- the number of elements in $\pi_k$, $\mathrm{conv}(\cdot)$ and $\mathrm{cl}(\cdot)$ denote, respectively, the convex hull and the closure w.r.t. the trace norm.

The (unregularized) relative entropy of $\pi$-entanglement of a state $\rho$ of a $n$-partite system $A_1\ldots A_n$
corresponding to a (non-empty) set $\pi\subseteq P(n)$ is defined as
\begin{equation}\label{ER-def}
E_{R,\pi}(\rho) =\inf_{\sigma\in\S_{\rm sep}^{\pi}(\H_{A_1...A_n})}D(\rho\shs\|\shs\sigma).
\end{equation}
It is shown in~\cite{L&Sh-1} that this infimum is always attained and that the state at which it is attained is
unique provided that $\rho$ is a faithful state. \smallskip

If $\pi$ consists of the single  partition $\left\{\{1\},\ldots, \{n\}\right\}$ then $\S_{\rm sep}^{\pi}(\H_{A_1...A_n})$
coincides with the convex set $\S_{\rm sep}(\H_{A_1...A_n})$ of all (fully) separable states in $\S(\H_{A_1...A_n})$ and hence  $E_{R,\pi}$ is
the relative entropy of entanglement $E_R$ in this case. \smallskip

The relative entropy of $\pi$-entanglement possesses the basic properties of entanglement measures (convexity, selective LOCC-monotonicity, asymptotic continuity, etc.) and satisfies
the  inequality
\begin{equation}\label{RE-LAA}
p E_{R,\pi}(\rho)+(1-p)E_{R,\pi}(\sigma)\leq E_{R,\pi}(p\rho+(1-p)\sigma)+h_2(p)
\end{equation}
valid for any states $\rho$  and $\sigma$ in $\S(\H_{A_1...A_n})$ and any $p\in[0,1]$, where $h_2$ is the binary entropy \cite{W-CB,L&Sh-1}.
It is bounded above by the sum of any $n-1$ marginal entropies, i.e.
\begin{equation}\label{ER-UB}
E_{R,\pi}(\rho)\leq  \sum_{j=1}^{n-1}S(\rho_{A_{s_j}}),
\end{equation}
for any set $A_{s_1},..., A_{s_{n-1}}$ of $n-1$ subsystems \cite{ERUB},\cite[the Appendix]{L&Sh-1}.

The inequalities (\ref{RE-LAA}) and (\ref{ER-UB}) along with the convexity and nonnegativity of $E_{R,\pi}$  imply that
$E_{R,\pi}\in L_n^{n-1}(1,1)$ and hence $E_{R,\pi}$ has a faithful $II_n^{n-1}$-type continuity bound due to the relation (\ref{B-inc}) in Section 2.3.

The relative entropy of $\pi$-entanglement is nonadditive on tensor products. If $\pi$ consists of a single partition
then the relative entropy of $\pi$-entanglement is subadditive on tensor products, and hence one can define its regularization
\begin{equation}\label{E-F-r-def}
E^{\infty}_{R,\pi}(\rho)=\lim_{k\rightarrow+\infty}k^{-1}E_{R,\pi}(\rho^{\otimes k})=\inf_{k\in\mathbb{N}}k^{-1}E_{R,\pi}(\rho^{\otimes k}),
\end{equation}
where $\rho^{\otimes k}$ is treated as a state of the $n$-partite quantum system $A^{k}_1...A^{k}_n$, where $A^{k}_j$ denotes the system obtained
by joining $k$ copies of $A_j$. The above limit exists due to Fekete’s lemma \cite{Fekete}.

For the results below it is  essential that for any permutation  $\{s(i)\}_{i=1}^n$  of $[n]$ the following claims are valid:
\begin{itemize}
  \item the relative entropy
of $\pi$-entanglement  $E_{R,\pi}$ belongs to the class $II_n^{n-1}$ in the settings $A'_1=A_{s(1)}$,...,$A'_n=A_{s(n)}$ for any set of partitions $\pi$ (see Section 2.3);
  \item the regularized relative entropy
of $\pi$-entanglement  $E_{R,\pi}^{\infty}$ belongs to the class $II_n^{n-1}$ in the settings $A'_1=A_{s(1)}$,...,$A'_n=A_{s(n)}$ for any $\pi$ consisting of a single partition.
\end{itemize}
These claims follow from Proposition 39 in \cite{L&Sh-1} (where a common faithful $II_n^{n-1}$-type continuity bound for  $E_{R,\pi}$ and  $E_{R,\pi}^{\infty}$ is presented)
and  Remark \ref{sym} below.\smallskip

\begin{remark}\label{sym} In general, the functions $E_{R,\pi}$ and $E_{R,\pi}^{\infty}$ are not symmetrical w.r.t. the subsystems
$A_1$,...,$A_n$. But as long as we believe that $\pi$ is arbitrary, we can assume that the energy constraints are imposed on the first $n-1$ subsystems $A_1$,...,$A_{n-1}$.
Indeed, any permutation $A'_1=A_{s(1)}$,...,$A'_n=A_{s(n)}$ of the subsystems can be taken into account by the appropriate replacement $\pi\rightarrow\pi'$ of the set of partitions.
\end{remark}

\subsubsection{Simplified definition of $E_{R,\pi}$}

In this subsection and below we apply the results from Section 3 for exploring  properties of the functions $E_{R,\pi}$ and $E_{R,\pi}^{\infty}$
in the infinite-dimensional settings   (extending the results obtained in \cite{L&Sh-1,L&Sh-2}). \smallskip

The following proposition concerns the possibility to  take the infimum in the definition (\ref{ER-def}) of the relative entropy of $\pi$-entanglement
over a subset of $\S_{\rm sep}^{\pi}(\H_{A_1...A_n})$ consisting of  "simple" $\pi$-separable states.\smallskip

\begin{property}\label{one} \emph{Let $A_1,..,A_n$ be arbitrary quantum systems and  $\,\pi=\{\pi_k\}_{k\in K_{\pi}}$ a given set of partitions of $\{1,2,...,n\}$. Let
\begin{equation}\label{S-II-n-1}
\widetilde{\S}^{n-1}_{II}(\H_{A_1...A_n})=\left\{\shs\rho\in\S(\H_{A_1...A_n})\,|\,\rho_{A_{s}}\!\!\in\S_{\rm\textsf{FA}\!}(\H_{A_{s}}) \textrm{ for } \,n-1\, \textrm{ indexes } \shs s\shs\right\}
\end{equation}
be the set consisting of  states $\rho$ of the $n$-partite system $A_1...A_n$
such that the FA-property holds for at least $\shs n-1$
marginal states of $\rho$.}

\emph{If $\rho$ is a state in $\,\widetilde{\S}^{n-1}_{II}(\H_{A_1...A_n})$  then the infimum in the definition (\ref{ER-def}) of $E_{R,\pi}(\rho)$ can be taken only over all finitely-decomposable $\pi$-separable states in $\S(\H_{A_1...A_n})$, i.e. states having the form
\begin{equation}\label{f-d-s}
\sigma=\sum_{i=1}^m p_i\, \sigma^i_1\otimes\sigma^i_2\otimes...\otimes\sigma^i_{|\pi_{k(i)}|},\;\;m<+\infty,
\end{equation}
where $\{p_i\}_{i=1}^m$ is a probability distribution, $k(i)\in K_{\pi}$ and $\sigma^{\shs i}_j$ is a state in $\H_{A_{\pi_{k(i)}(j)}}$
for all $\,i=\overline{1,m}\,$ and $\,j=\overline{1,|\pi_{k(i)}|}$ ($A_{\pi_{k(i)}(j)}\,$ is the system obtained by joining those $A_l$ such that $\,l\in \pi_{k(i)}(j)$).}
\end{property}
\smallskip

Proposition \ref{one} implies, in particular, that we may define $E_{R,\pi}(\rho)$
for any state $\rho$ in $\,\widetilde{\S}^{n-1}_{II}(\H_{A_1...A_n})$ \emph{by ignoring the existence of countably-non-decomposable
$\pi$-separable states} in $\S(\H_{A_1...A_n})$, i.e. the states that can not represented as a countable convex mixture of
pure $\pi$-separable states. This claim  is not trivial, since in general there are no reasons to
assume that the infimum in definition (\ref{ER-def}) of $E_{R,\pi}(\rho)$ can be taken over the dense subset of $\S_{\rm sep}^{\pi}(\H_{A_1...A_n})$ consisting of countably decomposable
$\pi$-separable states (because the relative entropy is not a continuous function of its arguments).\smallskip

\emph{Proof.} By the arguments in Remark \ref{sym} we may assume that $\rho$ is a state
from the set $\S^{n-1}_{II}(\H_{A_1...A_n})$ defined in (\ref{S-II}).
Consider the function
\begin{equation*}
  \widetilde{E}_{R,\pi}(\rho)=\inf_{\sigma\in\widetilde{\S}^{\pi}_{\mathrm{sep}}(\H_{A_1...A_n})}D(\rho\shs\|\shs\sigma),
\end{equation*}
where $\widetilde{\S}^{\pi}_{\mathrm{sep}}(\H_{A_1...A_n})$ is the subset of $\S(\H_{A_1...A_n})$ consisting of all finitely-decomposable $\pi$-separable states. Since
the set $\widetilde{\S}^{\pi}_{\mathrm{sep}}(\H_{A_1...A_n})$ is convex, the joint convexity of the relative entropy and Proposition 5.24 in \cite{O&P} imply that the function $\widetilde{E}_{R,\pi}$ is convex and that the inequality (\ref{RE-LAA}) with $E_{R,\pi}$ replaced by $\widetilde{E}_{R,\pi}$ holds for any states $\rho$ and $\sigma$ in $\S(\H_{A_1...A_n})$ and any $p\in[0,1]$. Note also that
$$
 \widetilde{E}_{R,\pi}(\rho)\leq D(\rho\shs\|\shs\rho_{A_1}\otimes...\otimes\rho_{A_n})=I(A_1:...:A_n)_{\rho}\leq 2\sum_{s=1}^{n-1}S(\rho_{A_s}),
$$
where the second inequality follows from (\ref{nMI-UB}). Hence, the function $\widetilde{E}_{R,\pi}$ belongs to the class $L_n^{n-1}(2,1)\subset II_n^{n-1}$ (see Section 2.3).

Thus, the equality
\begin{equation}\label{E-eq}
E_{R,\pi}(\rho)=\widetilde{E}_{R,\pi}(\rho)
\end{equation}
for any state $\rho\in\S^{n-1}_{II}(\H_{A_1...A_n})$
can be derived from claim $\rm(iv)$ of the $II_n^m$-part of  Corollary \ref{main-c} in Section 3.2 applied to the function $E=E_{R,\pi}\in L_n^{n-1}(1,1)\subset II_n^{n-1}$
by proving that (\ref{E-eq}) holds for any state $\rho\in\S(\H_{A_1...A_n})$ such that $\rank \rho_{A_i}<+\infty$ for all $i=1,2,...,n$.
The validity of (\ref{E-eq}) for any such  state $\rho$ follows from Lemma \ref{loc-l} below, since it shows that in the definition of $E_{R,\pi}(\rho)$ we may assume that $\rho$ is a state of a finite-dimensional
$n$-partite quantum system. $\Box$

\smallskip

\begin{lemma}\label{loc-l}  \emph{For an arbitrary state $\rho$ in $\S(\H_{A_1...A_n})$ the infimum in (\ref{ER-def})
can be taken over the set of all $\pi$-separable states in $\,\S(\H_{A_1...A_n})$ supported by the subspace
$\H^1_{\rho}\otimes...\otimes \H^n_{\rho}$, where  $\H^s_{\rho}=\supp \rho_{A_s}$, $s=\overline{1,n}$.}
\end{lemma}\smallskip

\emph{Proof.} Consider the channel
$$
\Phi(\varrho)=Q\varrho\shs Q+[\Tr(I_{A_1...A_n}-Q)\varrho]\tau,\quad  Q=P_{1}\otimes...\otimes P_{n},
$$
where $P_s$ is the projector on the subspace $\H^s_{\rho}$, $s=\overline{1,n}$, and $\tau$ is any $\pi$-separable state in  $\S(\H_{A_1...A_n})$ supported by the subspace
$\H^1_{\rho}\otimes...\otimes \H^n_{\rho}$. By monotonicity of the relative entropy we have
$$
D(\rho\shs\|\Phi(\sigma))=D(\Phi(\rho)\|\Phi(\sigma))\leq D(\rho\shs\|\shs\sigma)
$$
for any $\pi$-separable state $\sigma$ in $\S(\H_{A_1...A_n})$. Since $\Phi(\sigma)$ is
a $\pi$-separable state supported by the subspace
$\H^1_{\rho}\otimes...\otimes \H^n_{\rho}$, the above inequality implies the assertion of the lemma.  $\square$

\begin{corollary}\label{one-cor} \emph{For an arbitrary $\pi$-separable state $\rho$  in $\,\widetilde{\S}^{n-1}_{II}(\H_{A_1...A_n})$
there is a sequence $\{\rho_k\}$ of $\pi$-separable states having the form
(\ref{f-d-s}) such that}
$$
\lim_{k\to+\infty}D(\rho\shs\|\rho_k)=0.
$$
\end{corollary}

\subsubsection{Continuity and convexity properties of $E_{R,\pi}^{\infty}$ in infinite-dimensions}

It is shown in~\cite{L&Sh-1} that the relative entropy of $\pi$-entanglement $E_{R,\pi}$ is a lower semicontinuous function on $\S(\H_{A_1...A_m})$ for any set of partitions $\pi$.
Then a criterion of local continuity of $E_{R,\pi}$ is obtained in~\cite{L&Sh-2}, which allows us to derive several verifiable conditions of local continuity of
$E_{R,\pi}$ for any set partitions $\pi$, in particular, to prove that continuity of $E_{R,\pi}$ on a subset of $\S(\H_{A_1...A_n})$ follows from
continuity of the quantum mutual information (QMI) on this subset (this is important as there are easy verifiable conditions for local continuity of the QMI \cite[Section 4.2]{LCA}).\smallskip

Unfortunately, the approach used in \cite{L&Sh-1,L&Sh-2} cannot be applied
to analysis of the regularized relative entropy
of $\pi$-entanglement  $E_{R,\pi}^{\infty}$ well defined by the expression (\ref{E-F-r-def})
for any set $\pi$ consisting of a single partition.  Analytical properties of the function $E^{\infty}_{R,\pi}$
can be explored by using the results from Section 3, since this function belongs to the class $II_n^{n-1}$ (in terms of Section 2.3) for any
$\pi$  consisting of a single partition (this is mentioned in Section 4.1.1).
\smallskip

The continuity properties of $E_{R,\pi}^{\infty}$ are described in the following\smallskip

\begin{property}\label{RE-LS}  \emph{Let $\,\widetilde{\S}^{n-1}_{II}(\H_{A_1...A_n})$ be the subset of $\,\S(\H_{A_1...A_n})$ defined in (\ref{S-II-n-1}). Let $\pi$ be the set consisting of a single partition of $\,\{1,2,..,n\}$.}\smallskip

A) \emph{The function $E^{\infty}_{R,\pi}$ is finite and lower semicontinuous on the set $\,\widetilde{\S}^{n-1}_{II}(\H_{A_1...A_n})$.  Moreover,
\begin{equation}\label{LS-r}
\liminf_{k\to+\infty} E^{\infty}_{R,\pi}(\rho_k)\geq E^{\infty}_{R,\pi}(\rho_0)
\end{equation}
for any sequence $\{\rho_k\}\subset\S(\H_{A_1...A_n})$ converging to a state $\rho_0\in\widetilde{\S}^{n-1}_{II}(\H_{A_1...A_n})$.}\smallskip

B) \emph{Let $\{\rho_k\}$ be an arbitrary sequence of states
in $\S(\H_{A_1...A_n})$ converging to a state $\rho_0\in\widetilde{\S}^{n-1}_{II}(\H_{A_1...A_n})$. The limit relation
\begin{equation}\label{rd-conv}
\lim_{k\to+\infty}E^{\infty}_{R,\pi}(\rho_k)= E^{\infty}_{R,\pi}(\rho_0)<+\infty
\end{equation}
holds provided one of the following conditions is valid:}\footnote{The finiteness of $\,E_{R,\pi}(\rho_0)$ and $\,I(A_1\!:...:\!A_n)_{\rho_0}$ follows from
the condition $\rho_0\in\widetilde{\S}^{n-1}_{II}(\H_{A_1...A_n})$.}

\begin{enumerate}
  \item [$\rm a)$] \emph{there exists a sequence $\{\omega_k\}\subset\S_{\rm sep}^{\pi}(\H_{A_1...A_m})$ converging to a state $\omega_0\in\S_{\rm sep}^{\pi}$  such that}
\begin{equation*}
\lim_{k\to+\infty}D(\rho_k\|\shs\omega_k)=D(\rho_0\|\shs\omega_0)<+\infty;
\end{equation*}
  \item [$\rm b)$] \emph{there exists}
  \begin{equation*}
\lim_{k\to+\infty}E_{R,\pi}(\rho_k)= E_{R,\pi}(\rho_0)<+\infty;
\end{equation*}
  \item  [$\rm c)$] \emph{there exists}
\begin{equation*}
\lim_{k\to+\infty} I(A_1\!:...:\!A_n)_{\rho_k}=I(A_1\!:...:\!A_n)_{\rho_0}<+\infty.
\end{equation*}
\end{enumerate}
\end{property}\smallskip

\begin{remark}\label{Th2-B} The conditions $\rm a)$,$\,\rm b)$ and $\rm c)$ in Proposition \ref{RE-LS}B  are related as follows: Theorem 1 in \cite{L&Sh-2} states that $\,\rm a)\Rightarrow b)$
in general and that $\,\rm a)\Leftrightarrow b)$ if the state $\rho_0$ is faithful; Proposition 2 in \cite{L&Sh-2} states that $\,\rm c)\Rightarrow b)$.

A criterion and different conditions for the validity of $\,\rm b)$ are described in \cite{L&Sh-2}.

Easy verifiable conditions for the validity of $\,\rm c)$ are presented in \cite[Section 4.2]{LCA}.
\end{remark}\smallskip

\emph{Proof.} A) It is mentioned at the end of Section 4.1.1  that the function $E^{\infty}_{R,\pi}$ belongs to the class $II_n^{n-1}$
in the settings $A'_1=A_{s(1)}$,...,$A'_n=A_{s(n)}$ for any permutation  $\{s(i)\}_{i=1}^n$  of $[n]$.
So, this claim can be derived from claim $\rm(i)$ of the $II_n^m$-part of Corollary \ref{main-c} in Section 3.2 by applying the arguments in Remark \ref{sym}. It suffices only to note that the function $E^{\infty}_{R,\pi}$ does not increase under action of local channels.

\smallskip

B) By Remark \ref{Th2-B} it suffices to prove that  condition $\rm b)$ implies (\ref{rd-conv}). Let $p\in\N$ be arbitrary. By Corollary 22 in \cite{L&Sh-1} the cone generated by the set $\S_{\rm sep}^{\pi}(\H_{A^p_1|...|A^p_n})$ is a weak* closed
subset of $\T_+(\H_{A^p_1|...|A^p_n})$. So, since the state $\omega'\otimes \omega''$ lies in $\S_{\rm sep}^{\pi}(\H_{A^p_1|...|A^p_n})$
for any states $\omega'\in\S_{\rm sep}^{\pi}(\H_{A^{t}_1|...|A^{t}_n})$ and $\omega''\in\S_{\rm sep}^{\pi}(\H_{A^{p-t}_1|...|A^{p-t}_n})$, $t=1,2,...,p-1$ (because $\pi$ consists of a single partition), the condition $\rm b)$  and the repeated use of Corollary 2G
in \cite{L&Sh-2} allow us to show that
\begin{equation*}
\lim_{k\to+\infty}E_{R,\pi}(\rho^{\otimes p}_k)= E_{R,\pi}(\rho^{\otimes p}_0)<+\infty\quad \forall p\in\mathbb{N}.
\end{equation*}
It follows that
\begin{equation}\label{ER-US+}
\limsup_{k\to+\infty}E_{R,\pi}^{\infty}(\rho_k)\leq  E_{R,\pi}^{\infty}(\rho_0)<+\infty
\end{equation}
because  $\,E_{R,\pi}^{\infty}(\rho_k)=\inf_p p^{-1}E_{R,\pi}(\rho^{\otimes p}_k)$ for all $k$. This relation and (\ref{LS-r}) imply (\ref{rd-conv}). $\Box$
\smallskip

\begin{remark}\label{ER-US} The proof of Proposition  \ref{RE-LS}B shows that the limit relation (\ref{ER-US+}) holds
for  an arbitrary sequence $\{\rho_k\}$ of states
in $\S(\H_{A_1...A_n})$ converging to \emph{any} state $\rho_0$ in $\S(\H_{A_1...A_n})$ provided that
one of the conditions $\rm a)$,$\,\rm b)$ and $\rm c)$ in Proposition \ref{RE-LS}B is valid.
\end{remark}\medskip

The convexity of the function $E_{R,\pi}^{\infty}$ (defined by the expression (\ref{E-F-r-def})
for any set $\pi$ consisting of a single partition) follows from the convexity and subadditivity of $E_{R,\pi}$ by Proposition 13 in \cite{UT-EM} (the proof remains valid
in the infinite-dimensional settings). Since the function $\rho\mapsto E_{R,\pi}(\rho)$ is lower semicontinuous on $\S(\H_{A_1...A_n})$ by Corollary 23 in \cite{L&Sh-1},
the function $\rho\mapsto k^{-1}E_{R,\pi}(\rho^{\otimes k})$ has the same property for any $k$. So, it follows from the definition
(\ref{E-F-r-def}) that $E_{R,\pi}^{\infty}$ is a measurable function on $\S(\H_{A_1...A_n})$ (w.r.t. the Borel $\sigma$-algebra in $\S(\H_{A_1...A_n})$).

Unfortunately, the convexity and measurability of $E_{R,\pi}^{\infty}$ does not imply that
\begin{equation}\label{ER-convex}
\,E_{R,\pi}^{\infty}(\bar{\rho}(\mu))\leq\int_{\S(\H_{A_1...A_n})}\,E_{R,\pi}^{\infty}(\rho)\mu(d\rho).
\end{equation}
for any Borel probability measure $\mu$ on $\S(\H_{A_1...A_n})$, where $\bar{\rho}(\mu)$ is the barycenter of $\mu$ defined in (\ref{bar}). This inequality could be deduced from a similar inequality for the function $\rho\mapsto k^{-1}E_{R,\pi}(\rho^{\otimes k})$ provided that  the convexity of the latter function is proved. But the arguments from the proof of Proposition 13 in \cite{UT-EM}
(implying the convexity of $E_{R,\pi}^{\infty}$)  do not imply the convexity of the function $\rho\mapsto k^{-1}E_{R,\pi}(\rho^{\otimes k})$ for a given finite $k$.
Nevertheless, the validity of (\ref{ER-convex}) under a certain condition  on $\bar{\rho}(\mu)$ can be proved by using the results from Section 3.2.
Since $E_{R,\pi}^{\infty}\in II_n^{n-1}$, the $II_n^m$-part of Corollary \ref{main-c} implies the following\smallskip

\begin{property}\label{RE-conv} \emph{Inequality (\ref{ER-convex}) holds for a Borel probability measure $\mu$ on\break $\S(\H_{A_1...A_n})$ provided that its barycenter $\bar{\rho}(\mu)$ belongs to the set $\,\widetilde{\S}^{n-1}_{II}(\H_{A_1...A_n})$ defined in (\ref{S-II-n-1}).}
\end{property}\smallskip

Note that the inequality (\ref{ER-convex}) with $E_{R,\pi}^{\infty}$ replaced by $E_{R,\pi}$ (where  $\pi$ is any set partitions) holds for \emph{any} Borel probability measure $\mu$ on $\S(\H_{A_1...A_n})$
because the nonnegative convex function $E_{R,\pi}$ is lower semicontinuous on $\S(\H_{A_1...A_n})$ by Corollary 23 in \cite{L&Sh-1} (the Jensen inequality (2) 
holds for any convex lower semicontinuous function $f$ on $\S(\H)$ and any Borel probability measure $\mu$ on $\S(\H)$, see, f.i., Proposition A-2 in \cite{EM}).

\smallskip

\begin{remark}\label{ER-PPT} The claims of Propositions \ref{RE-LS} and \ref{RE-conv} in the bipartite case $n=2$ are also valid
for the regularization $E_{R,PPT}^{\infty}$ of the relative entropy distance to the PPT states \cite{Rains-1,PPT}. This can be shown by repeating the arguments from the proofs of these claims, since the cone generated by the set of PPT states is weak* closed (by Lemma 24 in \cite{L&Sh-1}) and the function $E_{R,PPT}^{\infty}$ belongs to the class $II^1_2$  (a faithful $II_2^{1}$-type continuity bound for  $E_{R,PPT}^{\infty}$  is presented in Proposition 36 in \cite{L&Sh-1}).
\end{remark}

\subsubsection{Finite-dimensional approximation of $E_{R}$ and $E_{R}^{\infty}$ and its use}

Many results describing properties of the relative entropy of $\pi$-entanglement and its regularization in finite-dimensional multipartite quantum systems
remain valid in the infinite-dimensional case under some additional conditions. A simple way to establish the validity of these
results is to use the approximation technique based on the continuity conditions in \cite{L&Sh-2} and Proposition \ref{RE-LS} in Section 4.1.3.\smallskip

We illustrate this approach restricting attention to the relative entropy of entanglement $E_{R}$ and its regularization $E_{R}^{\infty}$
defined by formulae (\ref{ER-def}) and (\ref{E-F-r-def}) in which $\pi$ is a singleton set consisting of the finest partition $\left\{\{1\},\ldots, \{n\}\right\}$ (in this case $\S_{\rm sep}^{\pi}(\H_{A_1...A_n})$
coincides with the convex set $\S_{\rm sep}(\H_{A_1...A_n})$ of all fully separable states).\smallskip

\begin{property}\label{REE-FDA}  \emph{Let $\rho$ be a state from the set $\,\widetilde{\S}^{n-1}_{II}(\H_{A_1...A_n})$ defined in (\ref{S-II-n-1}). Let $\rho_k=Q_k\rho\shs Q_k [\Tr Q_k\rho\shs]^{-1}$, $Q_k=P_k^{1}\otimes...\otimes P_k^{n}$,
where $\{P_k^1\}\subset\B(\H_{A_1})$,..., $\{P_k^n\}\subset\B(\H_{A_n})$ are arbitrary sequences of projectors strongly converging to the unit operators
$I_{A_1}$,...,$I_{A_n}$ correspondingly.  Then
\begin{equation}\label{FDA-LR}
E^*_R(\rho_{A_1...A_m})=\lim_{k\to+\infty}E^*_R([\rho_k]_{A_1...A_m}),\quad E_R^*=E_R,E_R^{\infty},
\end{equation}
for any $\,m=2,3,...,n$, where $E_R$ and $E_R^{\infty}$ denote, respectively, the relative entropy of entanglement and its regularization of a state of the system $A_1...A_m$.}\smallskip

\emph{To guarantee the validity of (\ref{FDA-LR}) with $E_R^*=E_R$ it suffices to require that $\rho$ is any state in $\,\S(\H_{A_1...A_n})$
such that $E_R(\rho)<+\infty$.}
\end{property}\smallskip

\emph{Proof.}  Assume first that $\rho$ is any state in $\,\S(\H_{A_1...A_n})$ such that
$E_R(\rho)<+\infty$. By the monotonicity of the relative entropy of entanglement under local channels
this implies $E_R(\rho_{A_1...A_m})<+\infty$. Consider the sequence of
states
$$
\hat{\rho}_k=\hat{Q}_k\rho\shs \hat{Q}_k [\Tr \hat{Q}_k\rho\shs]^{-1},\quad \hat{Q}_k=P_k^{1}\otimes...\otimes P_k^{m}\otimes I_{A_{m+1}}\otimes...\otimes I_{A_{n}},
$$
tending to the state $\rho$.  By the monotonicity of the relative entropy of entanglement under selective LOCC we have
$$
\hat{c}_k E_R([\hat{\rho}_k]_{A_1...A_m})\leq E_R(\rho_{A_1...A_m})<+\infty,\quad \hat{c}_k=[\Tr \hat{Q}_k\rho\shs].
$$
Thus, by using the lower semicontinuity of $E_R$ (Corollary 20 in \cite{L&Sh-1}) we obtain
$$
\lim_{k\to+\infty}E_R([\hat{\rho}_k]_{A_1...A_m})=E_R(\rho_{A_1...A_m})<+\infty.
$$
Since $c_k[\rho_k]_{A_1...A_m}\leq \hat{c}_k[\hat{\rho}_k]_{A_1...A_m}$ for all $k$, where  $c_k=\Tr Q_k\rho$,
and  $c_k,\hat{c}_k\to1$ as $k\to+\infty$, this implies  (\ref{FDA-LR}) with $E_R^*=E_R$ by the dominated convergence condition for $E_R$ \cite[Proposition 2]{L&Sh-2}. \smallskip

Assume now that $\rho$ is any state in $\,\widetilde{\S}^{n-1}_{II}(\H_{A_1...A_n})$.
Then $E_R(\rho_{A_1...A_m})<+\infty$ by inequality (\ref{ER-UB}) because any state having the FA-property has finite entropy. So, it suffices
to prove (\ref{FDA-LR}) with $E_R^*=E_R^{\infty}$. Since the finiteness of $I(A_1\!:...:\!A_n)_{\rho}$ follows from  the upper bound (\ref{nMI-UB}), this can be done by applying  Proposition \ref{RE-LS}B in Section 4.1.3 with the help of  Lemma \ref{FDA-LR+}
below. $\Box$
\smallskip

\begin{lemma}\label{FDA-LR+} \emph{Let $\rho$ be a state in $\,\S(\H_{A_1...A_n})$ such that
$\,I(A_1\!:...:\!A_n)_{\rho}<+\infty\,$ and $\{\rho_k\}$ be the sequence defined
by the rule described in Proposition \ref{REE-FDA} via arbitrary sequences $\{P_k^1\}\subset\B(\H_{A_1})$,..., $\{P_k^n\}\subset\B(\H_{A_n})$ of projectors strongly converging to the unit operators
$I_{A_1}$,...,$I_{A_n}$ correspondingly.  Then}
\begin{equation}\label{mi-cont++}
\lim_{k\to+\infty} I(A_1\!:...:\!A_m)_{\rho_k}=I(A_1\!:...:\!A_m)_{\rho}<+\infty,\quad m=2,3,..
\end{equation}
\end{lemma}

\emph{Proof.} By using the monotonicity of the quantum mutual information (QMI) under local quantum operations it is easy to show that
$$
c_k I(A_1\!:...:\!A_n)_{\rho_k}\leq I(A_1\!:...:\!A_n)_{\rho},\quad c_k=\Tr Q_k\rho.
$$
Since $c_k\to1$ as $k\to+\infty$, it follows from this inequality and the lower semicontinuity of QMI that
\begin{equation}\label{mi-cont+}
\lim_{k\to+\infty} I(A_1\!:...:\!A_n)_{\rho_k}=I(A_1\!:...:\!A_n)_{\rho}<+\infty.
\end{equation}

By the second claim of Proposition 4.6 in \cite{LCA} the limit relation  (\ref{mi-cont+}) imply (\ref{mi-cont++}).
$\Box$\smallskip

Below we consider examples of using Proposition \ref{REE-FDA}.\smallskip

\begin{example}\label{one-ex}
Lemma 5 in \cite{PVP} and the additivity of the entropy imply that
\begin{equation}\label{LB-1}
E_R^{\infty}(\rho)\geq -S(A_i|A_j)_{\rho},\quad (i,j)=(1,2),(2,1),
\end{equation}
for any state $\rho$ of a bipartite finite-dimensional quantum system $A_1A_2$.\smallskip

Proposition \ref{REE-FDA} allows us to show that the inequalities in (\ref{LB-1}) remain valid
for any state $\rho$ of a bipartite infinite-dimensional quantum system provided that
the state $\rho_{A_i}$ has finite entropy and $S(\cdot|\cdot)$ is the extended conditional entropy defined in (\ref{ce-ext}).
Indeed, by the additivity of the entropy and the relative entropy it suffices to show that (\ref{LB-1}) holds with $E_R^{\infty}(\rho)$
replaced by $E_R(\rho)$. Due to the inequality (\ref{ER-UB}) this can be done by applying the last claim of Proposition \ref{REE-FDA}. By symmetry we
have only to show that
$$
S(A_1|A_2)_{\rho}=\lim_{k\to+\infty}S(A_1|A_2)_{\rho_k},\quad \rho_k=Q_k\rho\shs Q_k [\Tr Q_k\rho\shs]^{-1},\quad Q_k=P_k^{1}\otimes P_k^{2},
$$
where $P_k^s$ is the spectral projector of $\rho_{A_s}$ corresponding to its $k$ maximal eigenvalues, $s=1,2$.
This can be done by using Theorem \ref{main}A, since the function $\rho\mapsto S(A_1|A_2)_{\rho}$
belongs to the class $I_2^1$ (a faithful $I^1_2$-type continuity bound for this function is
obtained by Winter in \cite{W-CB}).

Note that direct proof of (\ref{LB-1}) in the infinite-dimensional case
requires technical efforts (especially, if $S(\rho)=S(\rho_{A_j})=+\infty$).
\end{example}\smallskip

\begin{example}\label{two-ex} It is shown in \cite{ERUB} that
\begin{equation}\label{LB-2}
E_R^*(\rho)\geq E_R^*(\rho_{A_iA_j})+S(\rho_{A_iA_j}),\;\; E_R^*=E_R,E_R^{\infty},\;\;(i,j)=(1,2),(2,3),(3,1),
\end{equation}
for any pure state $\rho$ in a tripartite finite-dimensional quantum system $A_1A_2A_3$.

Proposition \ref{REE-FDA} allows us to show that
\begin{itemize}
  \item all the inequalities in (\ref{LB-2}) with $E_R^*=E_R$ remain valid (with possible value $+\infty$ in one or both sides)
for any pure state $\rho$ of a tripartite infinite-dimensional system;
  \item all the inequalities in (\ref{LB-2}) with  with $E_R^*=E_R^{\infty}$ remain valid
for any pure state $\rho$ of a tripartite infinite-dimensional system provided that
any two of the  states $\rho_{A_1}$, $\rho_{A_2}$ and $\rho_{A_3}$ have  the FA-property.\footnote{This assumption and the purity of the state $\rho$ imply, by Theorem 1 in \cite{FAP} and inequality (\ref{ER-UB}), the finiteness
of all terms in (\ref{LB-2}).}
\end{itemize}

Indeed, note first that in the case
$E_R(\rho)=+\infty$ all the inequalities in (\ref{LB-2}) with $E_R^*=E_R$ hold trivially.
Thus, to derive both claims from Proposition \ref{REE-FDA} it suffices (by symmetry) to show that
$$
\lim_{k\to+\infty}S([\rho_k]_{A_1A_2})=S(\rho_{A_1A_2})\leq+\infty,\quad \rho_k=Q_k\rho\shs Q_k [\Tr Q_k\rho\shs]^{-1},\quad Q_k=P_k^{1}\otimes P_k^{2}\otimes P_k^{3},
$$
for any pure state $\rho$ of $A_1A_2A_3$, where $P_k^s$ is the spectral projector of $\rho_{A_s}$ corresponding to its $k$ maximal eigenvalues, $s=1,2,3$.
This can be done by noting that $S(\rho_{A_1A_2})=S(\rho_{A_3})$ and $S([\rho_k]_{A_1A_2})=S([\rho_k]_{A_3})$ as the states $\rho$ and $\rho_k$ are pure. Since $c_k[\rho_k]_{A_3}\leq \rho_{A_3}$ for each $k$ for some positive number $c_k$
tending to $1$ as $k\to+\infty$, the required limit relation can be proved easily by using concavity and lower semicontinuity of the entropy.
\end{example}\smallskip\pagebreak

\begin{example}\label{three-ex}
If $ABC$ is a finite-dimensional tripartite quantum system then
\begin{equation}\label{I-E_R}
  I(A\!:\!B|C)_{\rho}\geq E^{\infty}_R(\rho_{A|BC})-E^{\infty}_R(\rho_{A|C})\quad \forall\rho\in \S(\H_{ABC}),
\end{equation}
where $I(A\!:\!B|C)_{\rho}=S(\rho_{AC})+S(\rho_{BC})-S(\rho_{ABC})-S(\rho_{C})\,$ is the quantum conditional
mutual information and  $E^{\infty}_R(\rho_{X|YZ})$ denotes the regularization of the bipartite relative entropy of entanglement
of a state $\rho_{XYZ}$ w.r.t. the partition $X|YZ$ \cite[Lemma 1]{Brandao}.  The proof of this result is based on
the state redistribution protocol proposed in \cite{D&J,D&J+}. So, if we want to upgrade this proof to the case of infinite-dimensional tripartite quantum system
$ABC$ it is necessary to upgrade  the results from \cite{D&J,D&J+} to this case. It obviously requires
serious technical efforts (to take the possibility of infinite values of the marginal entropies of a state into account, the necessity to use of extended definition of \break $I(A\!:\!B|C)_{\rho}$, etc.).

Proposition \ref{REE-FDA} allows us to directly derive an infinite-dimensional generalization of Lemma 1 in \cite{Brandao} from its original version. Namely, it allows us to prove that \emph{inequality  (\ref{I-E_R}) is valid for a state
$\rho$ of infinite-dimensional tripartite quantum system $ABC$ provided that  the FA-property  holds either for the state $\rho_A$ (case $(a)$)
or for both states $\rho_B$ and $\rho_C$ (case $(b)$) and}
\begin{equation}\label{I-fun}
I(A\!:\!B|C)_{\rho}=
\left\{\begin{array}{l}
        I(A\!:\!BC)_{\rho}-I(A\!:\!C)_{\rho}\;\; \textit{in case }\; (a)\\
        I(B\!:\!AC)_{\rho}-I(B\!:\!C)_{\rho}\;\; \textit{in case }\; (b).
        \end{array}\right.
\end{equation}
Since $\,\rho_X\in\S_{\rm \textsf{FA}\!}(\H_X)\,\Rightarrow S(\rho_X)<+\infty\,\Rightarrow I(X\!:\!Y)_{\rho}<+\infty\,$ by the Theorem in \cite{FAP} and due to (\ref{nMI-UB}),
all the expressions in (\ref{I-fun}) are well defined. If all the marginal entropies of a state $\rho$ are finite then these expressions
coincide with the above finite-dimensional formula for $I(A\!:\!B|C)_{\rho}$. They also agree with the extension of $I(A\!:\!B|C)_{\rho}$
to the set of all states of an infinite-dimensional tripartite quantum system $ABC$  \cite[Theorem 2]{CMI}.

To prove the above generalization of Lemma 1 in \cite{Brandao} it suffices to take arbitrary sequences $\{P_k^X\}\subset\B(\H_{X})$ of finite-rank projectors strongly converging to the unit operators $I_{X}$, $X=A,B,C$. Theorem 2 in \cite{CMI} and  Proposition \ref{REE-FDA} imply, respectively, that
\begin{equation*}
\lim_{k\to+\infty}I(A\!:\!B|C)_{\rho_k}=I(A\!:\!B|C)_{\rho}<+\infty,
\end{equation*}
and
\begin{equation*}
\lim_{k\to+\infty}E_R^{\infty}([\rho_k]_{A|X})=E_R^{\infty}(\rho_{A|X})<+\infty,\quad X=B, BC
\end{equation*}
in both cases $(a)$ and $(b)$.  In case $(b)$ this is true, since in this case $\rho_{BC}\in\S_{\rm \textsf{FA}\!}(\H_{BC})$ by Proposition 1 in \cite{FAP}.
By Lemma 1 in \cite{Brandao} the inequality (\ref{I-E_R}) holds for the state
$\rho_k\doteq Q_k\rho\shs Q_k [\Tr Q_k\rho\shs]^{-1}$, $Q_k=P_k^A\otimes P_k^B\otimes P_k^C$, for any $k$. So,
the above limit relations imply that (\ref{I-E_R}) holds for the state $\rho$.
\end{example}

\subsubsection{On the energy-constrained versions of $E_{R,\pi}$}

Dealing with infinite-dimensional quantum systems we have to take into account the existence of quantum states with infinite energy which can not be
produced in a physical experiment. This motivates an idea to define the relative entropy of $\pi$-entanglement of a state $\rho$
with finite energy by formula (\ref{ER-def}) in which the infimum is taken over all $\pi$-separable states with the energy not exceeding some (sufficiently large) bound $E$ \cite{DNWL}.
This gives the following energy-constrained version of the relative entropy of $\pi$-entanglement
\begin{equation}\label{ree-H-E-def}
  E^{H_{\!A^n}}_{R,\pi}(\rho\shs|E)=\inf_{\sigma\in\S_{\mathrm{sep}}^{\pi}(\H_{A_1...A_n}),\Tr  H_{\!A^n}\sigma\leq E}D(\rho\shs\|\shs\sigma),\quad
\end{equation}
where $H_{\!A^n}$ is the Hamiltonian of  $n$-partite  quantum system $A^n\doteq A_1...A_n$ \cite{DNWL}.\footnote{In \cite{DNWL} the case when $E_{R,\pi}=E_R$ (i.e. when $\pi$ consists of the single finest partition of $\{1,2,...,n\}$) is considered.}

An obvious drawback of definition (\ref{ree-H-E-def})  is its dependance of the bound $E$. Note also that
$E^{H_{\!A^n}}_{R,\pi}(\rho\shs|E)\neq0$ for any $\pi$-separable state $\rho$ such that $\Tr H_{\!A^n}\rho>E$ and that
the monotonicity of $E^{H_{\!A^n}}_{R,\pi}(\rho\shs|E)$ under local operations can be shown only under the appropriate restrictions on
the energy amplification factors of such operations.

A less restrictive way to take the energy constraints into account is to take the infimum in
formula (\ref{ER-def})  over all $\pi$-separable states with finite energy, i.e. to consider the following quantity
\begin{equation}\label{ree-H-def}
  E^{H_{\!A^n}}_{R,\pi}(\rho)=\inf_{\sigma\in\S^{\pi}_{\mathrm{sep}}(\H_{A_1...A_n}),\Tr H_{\!A^n}\sigma<+\infty}D(\rho\shs\|\shs\sigma)=\lim_{E\rightarrow+\infty}E^{H_{\!A^n}}_{R,\pi}(\rho\shs|E),
\end{equation}
where $H_{\!A^n}$ is the Hamiltonian of  $A^n\doteq A_1...A_n$. The limit in (\ref{ree-H-def}) can be replaced by the infimum over all $E>0$, since the function $E\mapsto E^{H_{\!A^n}}_{R,\pi}(\rho\shs|E)$ is non-increasing.  If the Hamiltonian $H_{\!A^n}$  has the form
\begin{equation}\label{Hn}
H_{\!A^n}=H_{A_1}\otimes I_{A_2}\otimes...\otimes I_{A_n}+\cdots+I_{A_1}\otimes... \otimes I_{A_{n-1}}\otimes H_{A_n},
\end{equation}
where $H_{A_s}$ is a  positive operator on $\H_{A_s}$, $s=\overline{1,n}$, then the set of $\pi$-separable states with finite energy is dense in $\S^{\pi}_{\mathrm{sep}}(\H_{A_1...A_n})$. So, in this case it is reasonable
to assume the coincidence of $E^{H_{\!A^n}}_{R,\pi}(\rho)$ and $E_{R,\pi}(\rho)$ for any state $\rho$ with finite energy.\footnote{This coincidence is not obvious, since the infima of a lower semicontinuous function
over a closed set and over a dense subset if this set may be different.} In the following proposition  we establish the validity of this assumption under the particular condition
on the Hamiltonians $H_{A_1}$,...,$H_{A_n}$.\smallskip

\begin{property}\label{EC-REE} \emph{Let $H_{\!A^n}$ be the Hamiltonian of a composite system $A_1...A_n$ expressed by formula (\ref{Hn}) via the Hamiltonians $H_{A_1}$,...,$H_{A_n}$ of the subsystems $A_1$,...,$A_n$. If at least $\,n-1\,$ of them satisfy condition (\ref{H-cond+}) then
\begin{equation*}
  E^{H_{\!A^n}}_{R,\pi}(\rho)=E_{R,\pi}(\rho)
\end{equation*}
for any state $\rho$ in $\S(\H_{A_1...A_n})$ such that $\Tr H_{\!A^n}\rho=\sum_{s=1}^n\Tr H_{A_s}\rho_{A_s}<+\infty$. Moreover,
for any such  state $\rho$ the infimum in the definition (\ref{ER-def}) of $E_{R,\pi}(\rho)$ can be taken only over all finitely-decomposable
$\pi$-separable states $\sigma$ (i.e. the states having the form (\ref{f-d-s})) such that  $\Tr H_{\!A^n}\sigma=\sum_{s=1}^n\Tr H_{A_s}\sigma_{A_s}<+\infty$.}\smallskip
\end{property}\smallskip

\emph{Proof.} By the arguments in Remark \ref{sym} we may assume that the Hamiltonians $H_{A_1}$,...,$H_{A_{n-1}}$ satisfy condition (\ref{H-cond+}).
We may also assume that all the Hamiltonians $H_{A_1}$,...,$H_{A_{n}}$ are densely defined operators.

Let $\S_0$ be the convex subset of
$\S(\H_{A_1...A_n})$ consisting of all states $\rho$ with finite value of $\Tr H_{\!A^n}\rho=\sum_{s=1}^n\Tr H_{A_s}\rho$.
Consider the function
\begin{equation*}
  \widetilde{E}^{H_{\!A^n}}_{R,\pi}(\rho)=\inf_{\sigma\in\S^{\mathrm{\pi,f}}_{\mathrm{sep}}(\H_{A_1...A_n})\cap\S_0}D(\rho\shs\|\shs\sigma),
\end{equation*}
on the set $\S_0$, where $\S^{\mathrm{\pi,f}}_{\mathrm{sep}}(\H_{A_1...A_n})$ is the set of finitely-decomposable $\pi$-separable states in $\S(\H_{A_1...A_n})$. Since
the set $\S^{\mathrm{\pi,f}}_{\mathrm{sep}}(\H_{A_1...A_n})\cap\S_0$ is convex, the joint convexity of the relative entropy and  Proposition 5.24 in \cite{O&P}
imply that the function $\widetilde{E}^{H_{\!A^n}}_{R,\pi}$ is convex and that the inequality (\ref{RE-LAA})  with $E_{R,\pi}$ replaced by $\widetilde{E}^{H_{\!A^n}}_{R,\pi}$ holds for any states $\rho$ and $\sigma$ in $\S_0$. Since
$$
 \widetilde{E}^{H_{\!A^n}}_{R,\pi}(\rho)\leq D(\rho\shs\|\shs\rho_{A_1}\otimes...\otimes\rho_{A_n})=I(A_1:...:A_n)_{\rho}\leq 2\sum_{s=1}^{n-1}S(\rho_{A_s})\quad \forall \rho\in\S_0,
$$
where the second inequality follows from (\ref{nMI-UB}), the function $\widetilde{E}^{H_{\!A^n}}_{R,\pi}$ belongs to the class $L_n^{n-1}(2,1|\shs\S_0)$ in terms of Remark 9 in \cite{CBM}. It is clear that $\,E_{R,\pi}(\rho)\leq E^{H_{\!A^n}}_{R,\pi}(\rho)\leq\widetilde{E}^{H_{\!A^n}}_{R,\pi}(\rho)\,$ for any state $\rho$ in $\S_0$.\smallskip

For each natural $s$ in $[1,n-1]$ let $P_r^s$ be the spectral projector of the operator $H_{A_s}$ corresponding to its $r$ minimal eigenvalues (taking the multiplicity into account).
Let $P_r^n=\sum_{i=1}^r|i\rangle\langle i|$, where $\{|i\rangle\}$ is a basis in $\H_{A_n}$ such that $\langle  i|H_{A_n}|i\rangle<+\infty$ for all $i$.

By the proof of Theorem 1 in \cite{AFM} the set $\S_0$ has the invariance property stated in Remark 9 in \cite{CBM}. By this remark Theorem 5 in \cite{CBM} is generalized
to all the functions from the class $L_n^{n-1}(2,1|\shs\S_0)$ containing the function $\widetilde{E}^{H_{\!A^n}}_{R,\pi}$. Note also that  the function $E_{R,\pi}$ belongs to the class  $L_n^{n-1}(1,1)$ \cite[the proof of Proposition 39]{L&Sh-1}. Thus, Theorem 5 in \cite{CBM}, its generalization mentioned before and the arguments from the proof of Theorem \ref{main} in Section 3 show that
\begin{equation*}
\lim_{r\to+\infty} E_{R,\pi}(\rho_r)=E_{R,\pi}(\rho)\quad \textrm{and} \quad \lim_{r\to+\infty} \widetilde{E}^{H_{\!A^n}}_{R,\pi}(\rho_r)=\widetilde{E}^{H_{\!A^n}}_{R,\pi}(\rho),\quad \forall\rho\in\S_0,
\end{equation*}
where $\rho_r=c_r^{-1}Q_r\rho\shs Q_r$,  $\,Q_r=P_r^1\otimes...\otimes P_r^{n}$, $c_r=\Tr Q_r\rho$. Hence, to prove that $\widetilde{E}^{H_{\!A^n}}_{R,\pi}(\rho)=E^{H_{\!A^n}}_{R,\pi}(\rho)=E_{R,\pi}(\rho)$ it suffices to show that
$\widetilde{E}^{H_{\!A^n}}_{R,\pi}(\rho_r)=E_{R,\pi}(\rho_r)$ for any $r$. This can be done by using Lemma \ref{loc-l} in Section 4.1.2 and by noting that $\Tr H_{\!A^n}\sigma=\sum_{s=1}^n\Tr H_{A_s}\sigma_{A_s}<+\infty$ for any $\pi$-separable state
$\sigma$ in $\S(\H_{A_1...A_n})$ supported by the subspace $P_r^1\otimes...\otimes P_r^{n}(\H_{A_1...A_n})$. $\square$

\subsection{Rains bound and its regularization}

\subsubsection{Definitions and basic properties}

The Rains bound (also called Rains Divergence) is an important characteristic of a bipartite quantum state
that gives an upper bound on the distillable entanglement of this state \cite{Rains-1,Rains-2,Rains-3}. By using the Lindblad's
extension of the quantum relative (defined in (\ref{qre-L})) the Rains bound  can be defined as
\begin{equation*}
R(\rho)=\inf_{\sigma\in \T_R}(D(\rho\|\sigma)+1-\Tr\sigma),
\end{equation*}
where
\begin{equation}\label{T-R-def}
\T_R=\left\{\sigma\in \T(\H_{AB})\,|\,\sigma\geq0,\|\sigma^\Gamma\|_1\leq1\right\},
\end{equation}
$\Gamma$ is the partial transposition map in $\T(\H_{AB})$ \cite{Wilde-new}. The Rains bound possesses many properties
of entanglement measures, in particular, it is a convex function equal to zero on the set of separable states and non-increasing under LOCC operations.
However, the Rains bound may be equal to zero at some entangled states (f.i., it is equal to zero at all the entangled states with positive partial transpose).
The main advantage of the Rains bound as a quasi-entanglement measure consists in the fact that it gives a tighter upper bound on the distillable entanglement
than the relative entropy of entanglement. It is essential also that it has a semi-definite program formulation \cite{Rains-2,Wilde-new}.

Since the set $\T_{R}$ (defined  in (\ref{T-R-def})) contains all separable states
in $\S(\H_{AB})$, the Rains bound $R(\rho)$ does not exceed the relative entropy of entanglement $E_R$:
\begin{equation*}
R(\rho)\leq E_R(\rho)\quad \forall \rho\in\S(\H_{AB}).
\end{equation*}

\subsubsection{Continuity and convexity properties of $R$ and $R^{\infty}$}

Despite the fact that the Rains bound definition is very similar to the definition of the relative entropy of entanglement,
the technique developed in \cite{L&Sh-1} for analysis of the latter function (and other "relative entropies of resource") in the infinite-dimensional case
can not be applied to
the Rains bound (by the reason described at the end of Section 4.3 in \cite{L&Sh-1}). So, the good analytical properties (achievability and lower semicontinuity) of the function  $E_R$ and of the relative entropy of PPT-entanglement in infinite-dimensional bipartite systems established in \cite{L&Sh-1} are not yet proved for the Rains bound. Accordingly, the convergence criterion and all its corollaries  obtained in \cite{L&Sh-2} can not also be applied to the Rains bound. So, one can say that for all its advantages, the Rains bound is \emph{more difficult} for  analysis in the infinite-dimensional case than other entanglement measures of the relative entropy distance type.

The approximation technique developed in Section 3 gives a way to partially overcome the problems with analysis of the
Rains bound mentioned before. It also allows us to establish analytical properties for the regularized Rains bound defined as
\begin{equation*}
R^{\infty}(\rho)=\lim_{k\rightarrow+\infty}k^{-1}R(\rho^{\otimes k})=\inf_{k\in\mathbb{N}}k^{-1}R(\rho^{\otimes k}),
\end{equation*}
where $\rho^{\otimes k}$ is treated as a state of the bipartite quantum system $(A^{k})(B^{k})$. The above limit exists due to Fekete’s lemma (cf.\cite{Fekete}), because the
Rains bound is subadditive on tensor products. The nonadditivity of the Rains bound implies that $R^{\infty}\neq R$ \cite{Rains-n}.

The results in Section 3 can be applied to  the functions $R$ and $R^{\infty}$ because both these functions belong to the class $II_2^{1}$:
a common faithful $II_2^{1}$-type continuity bound for $R$ and $R^{\infty}$ is presented in Proposition 36 in \cite{L&Sh-1}.

\smallskip
\begin{property}\label{RB-LS}  \emph{Let
\begin{equation}\label{S-II-1}
\widetilde{\S}_{II}^1(\H_{AB})=\left\{\shs\rho\in\S(\H_{AB})\,|\,\textit{either } \rho_{A}\!\!\in\S_{\rm\! \textsf{FA}\!}(\H_{A}) \textit{ or } \rho_{B}\!\!\in\S_{\rm\! \textsf{FA}\!}(\H_{B})\shs\right\}
\end{equation}
be the subset of $\,\S(\H_{AB})$ consisting of  states $\rho$
such that the FA-property holds for at least one of
the marginal states of $\rho$.}\smallskip

A) \emph{The functions $R$ and $R^{\infty}$ are finite and lower semicontinuous on the set $\,\widetilde{\S}_{II}^1(\H_{AB})$.  Moreover,
\begin{equation*}
\liminf_{k\to+\infty} R^*(\rho_k)\geq R^*(\rho_0),\quad R^*=R,R^{\infty},
\end{equation*}
for any sequence $\{\rho_k\}\subset\S(\H_{AB})$ converging to a state $\rho_0\in\widetilde{\S}_{II}^1(\H_{AB})$.}\smallskip

B) \emph{Let $\{\rho_k\}$ be an arbitrary sequence of states
in $\S(\H_{AB})$ converging to a state $\rho_0\in\widetilde{\S}_{II}^1(\H_{AB})$. The limit relations
\begin{equation}\label{RB-conv}
\lim_{k\to+\infty}R^*(\rho_k)= R^*(\rho_0)<+\infty,\quad R^*=R,R^{\infty},
\end{equation}
hold provided that one of the following conditions is valid:}\footnote{The finiteness of $E_R(\rho_0)$ and $\,I(A\!:\!B)_{\rho_0}$ follows from
the condition $\rho_0\in\widetilde{\S}_{II}^1(\H_{AB})$.}

\begin{enumerate}
  \item [$\rm a)$] \emph{there exists a sequence $\{\omega_k\}\subset\T_{R}$ converging to an operator $\omega_0\in\T_{R}$  such that}
\begin{equation}\label{RB-re-c}
\lim_{k\to+\infty}D(\rho_k\|\shs\omega_k)=D(\rho_0\|\shs\omega_0)<+\infty;
\end{equation}
 \item [$\rm b)$] \emph{there exists}
  \begin{equation}\label{re-c+}
\lim_{k\to+\infty}E_{R}(\rho_k)= E_{R}(\rho_0)<+\infty;
\end{equation}
\item  [$\rm c)$] \emph{there exists}
\begin{equation}\label{RB-mi-cont}
\lim_{k\to+\infty} I(A\!:\!B)_{\rho_k}=I(A\!:\!B)_{\rho_0}<+\infty.
\end{equation}
\end{enumerate}
\end{property}\smallskip

 \begin{remark}\label{Th2-B+} A criterion and different sufficient conditions for the validity of $\,\rm b)$ are described in \cite{L&Sh-2}.
Easy verifiable sufficient conditions for the validity of $\,\rm c)$ are presented in \cite[Section 4.2]{LCA}.
\end{remark}\smallskip

\emph{Proof.} A) It is mentioned before the proposition that the functions $R$ and $R^{\infty}$ belong to the class $II_2^{1}$.
Hence, this claim follows directly from  claim $\rm(i)$ of the\break $II_n^m$-part of Corollary \ref{main-c} in Section 3 applied to each of the functions $R$ and $R^{\infty}$, since
these functions are symmetrical w.r.t. the subsystems $A$ and $B$ and do not increase under action of local channels.
\smallskip

B) We prove first that (\ref{RB-conv}) holds provided that condition (\ref{RB-re-c}) is valid. By part A it suffices to show
that (\ref{RB-re-c}) implies that
\begin{equation}\label{RB-US}
\limsup_{k\to+\infty}R^*(\rho_k)\leq R^*(\rho_0),\quad R^*=R,R^{\infty}.
\end{equation}

The validity of (\ref{RB-US}) the case $R^*=R$ can be shown modifying the arguments from the proof of Theorem 1 in \cite{L&Sh-2}.

For given $\lambda\in(0,1)$ and $k\geq0$ consider the quantity
$$
Y_{\lambda,k}=\inf_{\sigma\in\T_{R}}D(\rho_k\shs\|\shs (1-\lambda)\sigma+\lambda\omega_k)+1-(1-\lambda)\Tr\sigma-\lambda\Tr\omega_k.
$$
The convexity of $\T_{R}$ implies that $Y_{\lambda,k}(\rho_k)\geq R(\rho_k)$, because $(1-\lambda)\sigma+\lambda\omega_k\in \T_{R}$ for all $k$. On the other hand, by the joint convexity of the relative entropy we have
$$
\begin{array}{c}
\displaystyle Y_{\lambda,k}\leq(1-\lambda)\inf_{\sigma\in\T_{R}}(D(\rho_k\shs\|\shs\sigma)+1-\Tr\sigma)+
\lambda (D(\rho_k\shs\|\,\omega_k)+1-\Tr\omega_k)\\\\=(1-\lambda)R(\rho_k)+\lambda(D(\rho_k\shs\|\,\omega_k)+1-\Tr\omega_k).
\end{array}
$$
Since we may assume that $D(\rho_k\|\shs\omega_k)$ is finite for all $k$, we have
$R(\rho_k)=\inf_{\lambda\in (0,1)}Y_{\lambda,k}$ for all $k\geq0$.
Thus, to prove (\ref{RB-US}) with $R^*=R$ it suffices\footnote{We use the well-known fact that the infimum of a family of upper semicontinuous functions is an upper semicontinuous function. A more detailed justification for this step can be found in the proof of Theorem 1 in \cite{L&Sh-2}.} to show that
\begin{equation*}
\limsup_{k\to+\infty}Y_{\lambda,k}\leq Y_{\lambda,0}\quad \forall\lambda\in(0,1).
\end{equation*}
Since $\Tr\omega_k$ tends to $\Tr\omega_0$ as $k\to+\infty$, this can be done  by proving that
\begin{equation*}
\lim_{k\to+\infty}D(\rho_k\shs\|\shs(1-\lambda)\sigma+\lambda\omega_k)=D(\rho_0\shs\|\shs(1-\lambda)\sigma+\lambda\omega_0)<+\infty
\end{equation*}
for any operator $\sigma$ in $\T_{R}$ and any given $\lambda\in(0,1)$.  This limit relation follows from (\ref{RB-re-c}) by Proposition~2 in \cite{DTL},
because $\lambda\omega_k\leq(1-\lambda)\sigma+\lambda\omega_k$ for all $k$.\smallskip

Now it is not hard to show that (\ref{RB-re-c}) implies the validity of (\ref{RB-US}) with $R^*=R^{\infty}$.

Let $p\in\N$ be arbitrary. It follows from (\ref{RB-re-c}) that
\begin{equation*}
\lim_{k\to+\infty}D(\rho^{\otimes p}_k\|\shs\omega^{\otimes p}_k)=D(\rho^{\otimes p}_0\|\shs\omega^{\otimes p}_0)<+\infty.
\end{equation*}
By the above part of the proof this limit relation implies that
\begin{equation*}
\limsup_{k\to+\infty}R(\rho^{\otimes p}_k)\leq R(\rho^{\otimes p}_0)<+\infty.
\end{equation*}
Hence, (\ref{RB-US}) with $R^*=R^{\infty}$ holds, because $R^{\infty}(\rho_k)=\inf_{p\in\mathbb{N}}p^{-1}R(\rho_k^{\otimes p})$ for all $k\geq0$.\smallskip

Assume that condition (\ref{re-c+}) holds. Then for an arbitrary subsequence $\{\rho_{k_t}\}$ of $\{\rho_{k}\}$ Theorem 1B in \cite{L&Sh-2}
guarantees the existence of a sequence
of $\{\omega_t\}$ of separable states in $\S(\H_{AB})$ converging to a separable state $\omega_0\in\S(\H_{AB})$ such that
\begin{equation*}
\lim_{t\to+\infty}D(\rho_{k_t}\|\shs\omega_t)=D(\rho_0\|\shs\omega_0)<+\infty.
\end{equation*}
By the above part of the proof this implies that
\begin{equation*}
\lim_{t\to+\infty}R^{*}(\rho_{k_t})=R^{*}(\rho_0),\quad R^*=R,R^{\infty},
\end{equation*}
because $\{\omega_t\}_{t\geq0}\subset\T_{R}$. Since the subsequence $\{\rho_{k_t}\}$ is arbitrary, we obtain (\ref{RB-conv}).
\smallskip

If  condition (\ref{RB-mi-cont}) holds then condition (\ref{re-c+}) is valid by Proposition 2 in \cite{L&Sh-2}.
So, the above part of the proof implies the validity of (\ref{RB-conv}). $\Box$
\smallskip

\begin{remark}\label{RB-US-r} The proof of Proposition \ref{RB-LS}B shows that the limit relations in (\ref{RB-US}) hold
for  an arbitrary sequence $\{\rho_k\}$ of states
in $\S(\H_{AB})$ converging to \emph{any} state $\rho_0$ in $\S(\H_{AB})$ provided that
one of the conditions $\rm a)$,$\,\rm b)$ and $\rm c)$  is valid.
\end{remark}\smallskip

\begin{remark}\label{RB-US-r+} The properties of the functions $R$ and $R^{\infty}$ established of Proposition  \ref{RB-LS}
coincide with the properties of the function  $E_R^{\infty}$ given by Proposition \ref{RE-LS} in Section 4.1.3 in the bipartite case $n=2$.
\end{remark}\smallskip

The convexity of the function $R$ follows from its definition and the joint convexity of the relative entropy,
the convexity of the function $R^{\infty}$ follows from the convexity and subadditivity of $R$ by Proposition 13 in \cite{UT-EM}.
In contrast to $E_{R,\pi}$ and $E_{R,\pi}^{\infty}$, the measurability of $R$ and $R^{\infty}$ is an open question (since $R$
is defined via the infimum over a non-countable set of states). So, to prove the inequalities
\begin{equation}\label{R-convex}
R^{*}(\bar{\rho}(\mu))\leq\int_{\S(\H_{AB})}\,R^{*}(\varrho)\mu(d\varrho),\quad   R^{*}=R,R^{\infty},
\end{equation}
for a Borel probability measure $\mu$ on $\S(\H_{A_1...A_n})$  (where $\bar{\rho}(\mu)$ is the barycenter of $\mu$ defined in (\ref{bar})) we have first to verify the existence of the
integrals in their right hand sides. \smallskip

Since $R,R^{\infty}\in II_2^{1}$, the $II_m^n$-part of Corollary \ref{main-c} in Section 3 implies the following\smallskip

\begin{property}\label{R-conv} \emph{The functions $R$ and $R^{\infty}$ are  $\mu$-integrable w.r.t.  a Borel probability measure $\mu$ on $\S(\H_{AB})$ provided that its barycenter $\bar{\rho}(\mu)$ (defined in (\ref{bar})) belongs to the set $\,\widetilde{\S}^{1}_{II}(\H_{AB})$ defined in (\ref{S-II-1}). In this case both inequalities in  (\ref{R-convex}) hold.}
\end{property}\smallskip

\subsubsection{Other results about the functions $R$ and $R^{\infty}$}

\emph{Finite-dimensional approximation of $R$ and $R^{\infty}$.} Proposition \ref{RB-LS} in the previous subsection and Lemma \ref{FDA-LR+} in Section 4.1.4
allows us to prove the following \smallskip

\begin{property}\label{RB-FDA}  \emph{Let $\rho$ be a state from the set $\,\widetilde{\S}^{1}_{II}(\H_{AB})$ defined in (\ref{S-II-1}). Let $\rho_k=Q_k\rho\shs Q_k [\Tr Q_k\rho\shs]^{-1}$, $Q_k=P_k^{A}\otimes P_k^{B}$,
where $\{P_k^X\}\subset\B(\H_{X})$ is an arbitrary sequences of projectors strongly converging to the unit operator
$I_{X}$, $X=A,B$.  Then}
\begin{equation*}
R^*(\rho)=\lim_{k\to+\infty}R^*(\rho_k),\quad R^*=R,R^{\infty}.
\end{equation*}
\end{property}

\emph{Energy-constrained definition of $R$.}  Assume that $H_{AB}$ is the Hamiltonian of an infinite-dimensional bipartite system $AB$ and
\begin{equation*}
\T^{H_{\!AB}}_R=\left\{\sigma\in \T(\H_{AB})\,|\,\sigma\geq0,\|\sigma^\Gamma\|_1\leq1, \Tr H_{AB}\sigma<+\infty \right\}
\end{equation*}
is the subset of $\T_R$ consisting of operators with finite "mean energy". Then we
may defined the energy-constrained version of the Rains bounds as follows
\begin{equation*}
  R^{H_{\!AB}}(\rho)=\inf_{\sigma\in\T^{H_{\!AB}}_R}(D(\rho\shs\|\shs\sigma)+1-\Tr\sigma)
\end{equation*}

Using the arguments from the proof of Proposition \ref{EC-REE} in Section 4.1.5 one can prove the following\smallskip

\begin{property}\label{EC-RB} \emph{Let $H_{A}$ and $H_{B}$ be positive operators on the spaces $\H_A$ and $\H_B$ treated as Hamiltonians  of the subsystems $A$ and $B$. If at least one of the operators $H_{A}$ and $H_{B}$ satisfies condition (\ref{H-cond+}) and $H_{AB}=H_{A}\otimes I_{B}+I_{A}\otimes H_{B}$
then
\begin{equation*}
  R^{H_{\!AB}}(\rho)=R(\rho)
\end{equation*}
for any state $\rho$ in $\S(\H_{AB})$ such that $\,\Tr H_{AB}\rho=\Tr H_{A}\rho_{A}+\Tr H_{B}\rho_{B}<+\infty$.}
\end{property}\smallskip

\subsection{Conditional entanglement of mutual information}

\subsubsection{Definitions and basic properties}

The \emph{conditional entanglement of
mutual information} is an additive entanglement measure in a multipartite quantum system $A_1...A_n$ with an operational meaning in terms of the process of partial
state merging \cite{CE-II}. For  a state $\rho$ of a finite-dimensional system $A_1...A_n$ it is defined as
\begin{equation*}
  E_{I}(\rho)=\frac{1}{2}\displaystyle\inf_{\hat{\rho}\in \mathfrak{M}_1(\rho)}\left[I(A_1A'_1\!:...:\!A_nA'_n)_{\hat{\rho}}-I(A'_1\!:...:\!A'_n)_{\hat{\rho}}\right],
\end{equation*}
where $\mathfrak{M}_1(\rho)$ is the set of all extensions $\hat{\rho}\in\mathfrak{S}(\mathcal{H}_{A_1..A_nA'_1..A'_n})$ of the state
$\rho$ \cite{CE-II,CE-I}. To generalize this definition to an arbitrary state $\rho$ of an infinite-dimensional multipartite system $A_1...A_n$ note that
the quantity $I(A_1A'_1\!:...:\!A_nA'_n)_{\varrho}-I(A'_1\!:...:\!A'_n)_{\varrho}$ well defined for any state $\varrho$ with finite $I(A'_1\!:...:\!A'_n)_{\varrho}$
coincides at any such state with the nonnegative lower semicontinuous function
\begin{equation}\label{delta-def}
\Delta(\varrho)=I(A_1\!:\!A'_2...A'_n|A'_1)_{\varrho}+\sum_{k=2}^n
I(A_k\!:\!A_1...A_{k-1}A'_1...A'_{k-1}A'_{k+1}...A'_n|A'_k)_{\varrho}
\end{equation}
well defined for any state $\varrho$ state of the infinite-dimensional system $A_1...A_nA'_1...A'_n$, where
all the summands are the extended tripartite QCMI introduced in \cite[Theorem 2]{CMI}. Expression (\ref{delta-def}) is obtained in \cite{NQD}
in the finite-dimensional settings and generalized to the infinite-dimensional case in \cite[Proposition 8]{CMI}.

It is mentioned in \cite[Section IV-C]{CBM} that the function $f=\Delta$  belongs to the class $L_{2n}^n(2,n+1)$ (in the notation used in  \cite{CBM}). It follows that the function
\begin{equation}\label{E-I-def}
  E_{I}(\rho)=\frac{1}{2}\inf_{\hat{\rho}\in \mathfrak{M}_1(\rho)}\Delta(\hat{\rho}),
\end{equation}
on the set of all states of the infinite-dimensional system $A_1...A_n$ belongs to the class $N^{n}_{n,1}(1,(n+1)/2)\subset II_n^{n}$. Faithful $II_n^n$-type continuity
bounds for this function are obtained in \cite[Proposition 22]{CBM}. If fact, a stronger claim is valid. \smallskip

\begin{lemma}\label{E-I-p} \emph{The function $E_{I}$ defined in (\ref{E-I-def}) has a faithful $I_n^{n}$-type continuity bound, i.e. it belongs to the class
$I_n^{n}$.}
\end{lemma}\smallskip

\emph{Proof.}  Let $H_{A_1}$,....,$H_{A_n}$
be  arbitrary positive operators on $\H_{A_1}$,....,$\H_{A_n}$
satisfying condition (\ref{H-cond}). By using the continuity bound for
the QCMI presented in \cite[Lemma 25]{QC} it is easy to show the existence
of a faithful $I_n^{n}$-type continuity bound for the function $\Delta$ defined in (\ref{delta-def}), i.e.
such a function  $\,\widehat{B}_\Delta(\varepsilon,E|H_{A_1},...,H_{A_n})\,$ that
$$
\lim_{\varepsilon\to0^+}\widehat{B}_\Delta(\varepsilon,E|H_{A_1},...,H_{A_n})=0\quad \forall E>0
$$
and $\,|\Delta(\rho)-\Delta(\sigma)|\leq\widehat{B}_\Delta(\varepsilon,E|H_{A_1},...,H_{A_n})\,$ for any states $\rho$ and $\sigma$ in $\S(\H_{A_1...A_n})$ satisfying the conditions
\begin{equation}\label{b-cond}
 \sum_{i=1}^n\Tr H_{A_i}\rho_{A_i},\sum_{i=1}^n\Tr H_{A_i}\sigma_{A_i}\leq nE\quad\textrm{and}\quad\frac{1}{2}\|\rho-\sigma\|_1\leq\varepsilon.
\end{equation}

By using the isometrical
equivalence of all purifications of a given state one can show (see Section IV in \cite{C&W}) that
\begin{equation}\label{Lambda}
E_I(\rho)=\frac{1}{2}\inf_{\Lambda}\shs\Delta(\id_{A_1...A_n}\otimes \Lambda(\bar{\rho})),
\end{equation}
where $\bar{\rho}$ is any pure state in $\,\S(\H_{A_1..A_nR})$ such that $\Tr_R\shs\bar{\rho}=\rho$ and the infimum is over
all channels $\Lambda:\T(\H_R)\rightarrow\T(\H_{A'_1..A'_n})$.

Let $\rho$ and $\sigma$ be states in $\,\S(\H_{A_1..A_n})$ satisfying the conditions in (\ref{b-cond}).
Then there exist pure states $\bar{\rho}$ and $\bar{\sigma}$ in $\,\S(\H_{A_1..A_nR})$ such that
$\,\frac{1}{2}\|\shs\bar{\rho}-\bar{\sigma}\|_1\leq\sqrt{2\varepsilon}$, $\Tr_R\shs\bar{\rho}=\rho$ and $\Tr_R\shs\bar{\sigma}=\sigma$ \cite{H-SCI,Wilde}.
Since for any quantum channel $\Lambda:\T(\H_R)\rightarrow\T(\H_{A'_1..A'_n})$ we have $\frac{1}{2}\|\id_{A_1...A_n}\otimes \Lambda(\bar{\rho})-\id_{A_1...A_n}\otimes \Lambda(\bar{\sigma)}\|_1\leq\sqrt{2\varepsilon}$ by monotonicity of the
trace norm under action of a channel, by applying the $I_n^{n}$-type continuity bound $\widehat{B}_\Delta(\varepsilon,E|H_{A_1},...,H_{A_n})$ mentioned before
we obtain
$$
|\Delta(\id_{A_1...A_n}\otimes \Lambda(\bar{\rho}))-\Delta(\id_{A_1...A_n}\otimes \Lambda(\bar{\sigma}))|\leq\widehat{B}_\Delta(\sqrt{2\varepsilon},E|H_{A_1},...,H_{A_n})
$$
for any channel $\Lambda:\T(\H_R)\rightarrow\T(\H_{A'_1..A'_n})$. So, it follows from (\ref{Lambda}) that
$$
|E_I(\rho)-E_I(\sigma)|\leq\frac{1}{2}\widehat{B}_\Delta(\sqrt{2\varepsilon},E|H_{A_1},...,H_{A_n}).
$$
The r.h.s. of this inequality tends to zero as $\varepsilon$ tends to zero. So, it is a  faithful $I_n^{n}$-type continuity bound
for the function $E_I$. $\Box$

\subsubsection{Continuity and convexity properties}

The results of Section 3 allows us to prove  the following\smallskip

\begin{property}\label{E-I-LS}  \emph{Let $A_1,..,A_n$ be arbitrary quantum systems and
\begin{equation}\label{S-I-n}
\S_I^n(\H_{A_1...A_n})=\left\{\shs\rho\in\S(\H_{A_1...A_n})\,|\,S(\rho_{A_{s}})<+\infty \textrm{ for all }  \shs s=\overline{1,n}\shs\right\}
\end{equation}
be the subset of $\,\S(\H_{A_1...A_n})$ consisting of states with finite marginal entropies.}\smallskip

A) \emph{The function $E_I$ is finite and lower semicontinuous on the set $\S_I^n(\H_{A_1...A_n})$.  Moreover,
\begin{equation*}
\liminf_{k\to+\infty}E_I(\rho_k)\geq E_I(\rho_0)
\end{equation*}
for any sequence $\{\rho_k\}\subset\S(\H_{A_1...A_n})$ converging to a state $\rho_0\in\S_I^n(\H_{A_1...A_n})$.}\smallskip

B) \emph{If $\,\{\rho_k\}$ is a sequence of states
in $\S(\H_{A_1...A_n})$ converging to a state $\rho_0\in\S_I^n(\H_{A_1...A_n})$
such that
\begin{equation*}
\lim_{k\to+\infty} I(A_1\!:...:\!A_n)_{\rho_k}=I(A_1\!:...:\!A_n)_{\rho_0}
\end{equation*}
then}
\begin{equation*}
\lim_{k\to+\infty}E_I(\rho_k)= E_I(\rho_0)<+\infty.
\end{equation*}
\end{property}\smallskip

\emph{Proof.} A) By Lemma \ref{E-I-p} the function $E_I$ belongs to the class $I_n^{n}$. So, this claim follows directly from claim $\rm(i)$ of the $I_n^m$-part of Corollary \ref{main-c} in Section 3.2, since the function $E_I$ does not increase under action of local channels.
\smallskip

B) The convex function $E_I$ satisfies  the local monotonicity condition (\ref{m-cond}) and does not exceed the QMI $I(A_1\!:...:\!A_n)$ (because it does not
not exceed the c-squashed entanglement $E_{\rm c\textup{-}sq}$ \cite{CE-I}). Thus, to prove this claim it suffices, by Corollary 2 in Section 3.2, to prove that
\begin{equation}\label{E-I-US}
\limsup_{k\to+\infty}E_I(\rho_k)\leq E_I(\rho_0)
\end{equation}
assuming that
\begin{equation}\label{S-I-C}
\lim_{k\to+\infty}S([\rho_k]_{A_i})=S([\rho_0]_{A_i})<+\infty\quad\textrm{ for all }\;i=\overline{1,n}.
\end{equation}

For a given $\epsilon>0$ let $\hat{\rho}_0$ be a state in $\S(\H_{A_1..A_nA'_1...A'_n})$ such that
$\frac{1}{2}\Delta(\hat{\rho}_0)\leq E_I(\rho_0)+\epsilon$ (the finiteness of $E_I(\rho_0)$ follows from claim A).

Let $\bar{\rho}_0$ be a pure state in $\S(\H_{A_1..A_nA'_1...A'_nR})$, where $R$ is an infinite-dimensional quantum system, such that $\Tr_R\bar{\rho}_0=\hat{\rho}_0$.
The well known results of the purification theory (cf. \cite{H-SCI,Wilde})  imply the existence of a
sequence  $\{\bar{\rho}_k\}$ of pure states in $\S(\H_{A_1..A_nA'_1...A'_nR})$ converging to the state $\bar{\rho}_0$ such that $\Tr_{A'_1...A'_nR}\bar{\rho}_k=\rho_k$
for all $k$. Since the sequence of states $\hat{\rho}_k\doteq\Tr_{R}\bar{\rho}_k$ in $\S(\H_{A_1..A_nA'_1...A'_n})$
converges to the state $\hat{\rho}_0$, the assumed validity of  (\ref{S-I-C}) for all $i=\overline{1,n}$
implies that
$$
\lim_{k\to+\infty}\Delta(\hat{\rho}_k)=\Delta(\hat{\rho}_0)
$$
by Proposition 8 in \cite{CMI}. So, it follows from the definition of $E_I$ that
$$
\limsup_{k\to+\infty}E_I(\rho_k)\leq\frac{1}{2}\lim_{k\to+\infty}\Delta(\hat{\rho}_k)=\frac{1}{2}\Delta(\hat{\rho}_0)\leq E_I(\rho_0)+\epsilon
$$
by the construction of the state $\hat{\rho}_0$. This implies (\ref{E-I-US}) because $\epsilon$ is arbitrary. $\Box$
\medskip

The convexity of the function $E_I$ in the infinite-dimensional composite system
$A_1...A_n$ is proved by a simple upgrading the finite-dimensional arguments from \cite{CE-II}. Namely, one should establish
the following identity
\begin{equation}\label{E-I-eq}
  \Delta\left(\sum_{i\in I} p_i \varrho_i\otimes|\psi^1_i\rangle\langle\psi^1_i|\otimes...\otimes|\psi^n_i\rangle\langle\psi^n_i|\right)=\sum_{i\in I} p_i\Delta(\varrho_i)
\end{equation}
valid for any (finite or countable) set $\{\varrho_i\}_{i\in I}$ of states in $\S(\H_{A_1A'_1...A_nA_n'})$ and any probability distribution $\{p_i\}_{i\in I}$, where $\Delta$ in the r.h.s.
is the function defined in (\ref{delta-def}),  while $\Delta$ in the l.h.s is the function defined in (\ref{delta-def}) with $A'_1$,...,$A_n'$ replaced, respectively,  by $A'_1A''_1$,...,$A_n'A''_1$, $\{\psi^1_i\}_{i\in I}$,...$\{\psi^n_i\}_{i\in I}$ are orthonormal bases in the auxiliary spaces $\H_{A''_1}$,...,$\H_{A''_n}$ (their dimension coincides with the cardinality of $I$). If all the states $\rho_i$ have finite marginal entropies and the
Shannon entropy of the probability distribution $\{p_i\}_{i\in I}$ is finite then (\ref{E-I-eq}) follows from the
coincidence of the r.h.s. of (\ref{delta-def}) with $\,I(A_1A'_1\!:...:\!A_nA'_n)_{\varrho}-I(A'_1\!:...:\!A'_n)_{\varrho}$. The validity of (\ref{E-I-eq}) in the general case is proved via approximation using Proposition 8B in \cite{CMI}.

Using (\ref{E-I-eq}) the convexity of $E_I$ is proved by simple arguments  based on the definition of $E_I$ (because the argument of $\Delta$ in the l.h.s.
of (\ref{E-I-eq}) is an extension of $\sum_{i\in I} p_i \rho_i$ provided that $\varrho_i$ is an extension of $\rho_i$). Moreover,
since the identity (\ref{E-I-eq}) holds also for a countable set of indexes $I$, by this way one can prove the $\sigma$-\emph{convexity} of $E_I$, i.e. the inequality
$$
E_I\left(\sum_{i=1}^{+\infty} p_i \rho_i\right)\leq\sum_{i=1}^{+\infty} p_i E_I(\rho_i)
$$
for any countable set $\{\rho_i\}_{i=1}^{+\infty}$ of states in $\S(\H_{A_1...A_n})$ and any probability distribution $\{p_i\}_{i=1}^{+\infty}$.

It is not clear how to prove the "continuous" Jensen inequality for the function $E_I$ in general. Moreover,
it is not clear how to prove that this function is measurable w.r.t. the Borel $\sigma$-algebra in $\S(\H_{A_1...A_n})$.
However, claim $\rm(v)$ of the $I_n^m$-part of Corollary \ref{main-c} in Section 3.2 and the above Lemma \ref{E-I-p} allow us to prove the following\smallskip

\begin{property}\label{E-I-conv} \emph{Let $A_1,..,A_n$ be arbitrary quantum systems. The function $E_I$ defined in (\ref{E-I-def}) is  $\mu$-integrable w.r.t.  a Borel probability measure $\mu$ on $\S(\H_{A_1...A_n})$ provided that its barycenter $\bar{\rho}(\mu)$ (defined in (\ref{bar})) belongs to the set $\S^n_{I}(\H_{A_1...A_n})$ defined in (\ref{S-I-n}). In this case}
\begin{equation*}
E_I(\bar{\rho}(\mu))\leq\int_{\S(\H_{A_1...A_n})}E_I(\varrho)\mu(d\varrho).
\end{equation*}
\end{property}

\subsubsection{Other results about the function $E_I$}

\emph{Simplified definition.} The following proposition implies, in particular,  that for any state $\rho$ with finite marginal entropies $S(\rho_{A_1}),...,S(\rho_{A_n})$  we may assume in the definition
(\ref{E-I-def}) of $E_I(\rho)$ that $A'_1...A'_n$ are quantum systems with arbitrary finite dimensions.\smallskip

\begin{property}\label{E_I-s-d} \emph{Let $A_1,..,A_n$ be arbitrary quantum systems. If $\rho$ is a state in the set $\S^n_{I}(\H_{A_1...A_n})$ defined in (\ref{S-I-n}) then the infimum in the definition (\ref{E-I-def})  can be taken over any dense subset of
$\mathfrak{M}_1(\rho)$, in particular, over the set}
$$
\mathfrak{M}^{\rm f}_1(\rho)=\{\hat{\rho}\in\S(\H_{A_1...A_nA'_1...A'_n})\,|\,\hat{\rho}_{A_1...A_n}=\rho,\;\rank\hat{\rho}_{A'_1},...,\rank\hat{\rho}_{A'_n}<+\infty\}.
$$
\end{property}

\emph{Proof.} The claim of the proposition follows from Proposition 8D in \cite{CMI} which implies that the function
$\varrho\mapsto\Delta(\varrho)$ is  continuous
on the set $\mathfrak{M}_1(\rho)$ provided that $S(\rho_{A_1}),...,S(\rho_{A_n})$ are finite.

The possibility to take the infimum in the definition (\ref{E-I-def}) over the set $\mathfrak{M}^{\rm f}_1(\rho)$ can be also
deduced from claim $\rm (iv)$ of Corollary \ref{main-c} in Section 3.2 by considering the function
\begin{equation*}
  \widetilde{E}_{I}(\rho)=\frac{1}{2}\displaystyle\inf_{\hat{\rho}\in \mathfrak{M}^{\rm f}_1(\rho)}\Delta(\hat{\rho}).
\end{equation*}
The obvious modification of the proof of Lemma \ref{E-I-p} allows us to show that $\widetilde{E}_{I}$ belongs to the class
$I_n^{n}$. Thus, to prove that $\,E_{I}(\rho)=\widetilde{E}_{I}(\rho)$
for any state $\rho$ in $\,\S^n_I(\H_{A_1...A_n})$ it suffices, by claim $\rm (iv)$ if the $I_n^m$-part of Corollary \ref{main-c}, to note that
$\,E_{I}(\rho)=\widetilde{E}_{I}(\rho)$ for any  state $\rho\in\S(\H_{A_1...A_n})$ such that $\rank \rho_{A_i}<+\infty$ for all $i=1,2,...,n$ (because
we may treat $\rho$ as a state  of finite-dimensional system $A_1...A_n$ in this case). $\Box$\smallskip

\emph{About the null set of $E_{I}$.} In the finite-dimensional settings the function $E_{I}$ is equal
to zero at a state $\rho$ of $A_1...A_n$ if and only if $\rho$ is a (fully)
separable state. The "if" part of this claim is proved in \cite{CE-II} in the case $n=2$. Generalization of this proof
to the case $n>2$ is obvious.  The "only if" part can be deduced  from the similar property of the squashed
entanglement $E_{\rm sq}$ established in \cite{Brandao}, since it is known that $E_{\rm sq}\leq E_I$ \cite{CE-I}.

The main problem of the infinite-dimensional case is to prove that $E_I(\rho)=0$ for any (fully) separable state $\rho$ in $\S(\H_{A_1...A_n})$, since  the  arguments from \cite{CE-II} mentioned above imply that $E_I(\rho)=0$ only for countably-decomposable  separable states $\rho$ of a system $A_1...A_n$, i.e. the states that can be represented
as $\sum_{i}p_i\,\sigma^1_i\otimes...\otimes\sigma^n_i$, where $\{\sigma^k_i\}\subset\S(\H_{A_k})$, $k=\overline{1,n}$, and $\{p_i\}$ is a probability distribution. It is not clear how to prove that $E_I$ is equal to zero
on the set of countably-non-decomposable (fully) separable states in $\S(\H_{A_1...A_n})$. Using the approximation technique developed in Section 3 this problem can be solved for all
separable states in $\S(\H_{A_1...A_n})$ with finite marginal entropies.  \smallskip

\begin{property}\label{E-I-N}  \emph{Let $A_1,..,A_n$ be infinite-dimensional quantum systems. If $\rho$ is an arbitrary state in $\,\S(\H_{A_1...A_n})$ then
\begin{equation}\label{E-I-B}
  E_I(\rho)=0\quad \Rightarrow \quad \rho\,\textit{ is a (fully) separable state}.
\end{equation}
If $\,\rho$ is a state in the set $\S^n_{I}(\H_{A_1...A_n})$ defined in (\ref{S-I-n}) then $\,"\Leftrightarrow"\,$ holds in (\ref{E-I-B}).}
\end{property} \smallskip

\emph{Proof.} The first and the second  claims of Proposition \ref{E-I-N} are deduced, respectively,  from claims $(\rm iii)$ and $(\rm ii)$ of Corollary \ref{main-c} in Section 3.2 (applied to the function $E=E_I$)
by using the above Lemma \ref{E-I-p}, the monotonicity of $E_I$ under local channels and the properties of $E_I$ in the finite-dimensional settings mentioned before. $\Box$
\smallskip

\emph{Finite-dimensional approximation of $E_{I}$.} Using Proposition \ref{E-I-LS}B in Section 4.3.2 one can  establish the following analogue
of Proposition \ref{REE-FDA} in Section 4.1.4.\smallskip

\begin{property}\label{I-E-FDA}  \emph{Let $\rho$ be a state from the set $\,\S^n_I(\H_{A_1...A_n})$ defined in (\ref{S-I-n}). Let $\rho_k=Q_k\rho\shs Q_k [\Tr Q_k\rho\shs]^{-1}$, $Q_k=P_k^{1}\otimes...\otimes P_k^{n}$,
where $\{P_k^1\}\subset\B(\H_{A_1})$,..., $\{P_k^n\}\subset\B(\H_{A_n})$ are arbitrary sequences of projectors strongly converging to the unit operators
$I_{A_1}$,...,$I_{A_n}$ correspondingly.  Then
\begin{equation*}
E_I(\rho_{A_1...A_m})=\lim_{k\to+\infty}E_I([\rho_k]_{A_1...A_m})
\end{equation*}
for any $\,m=2,3,...,n$, where $E_I$  denotes the conditional entanglement of mutual information  of a state of the system $A_1...A_m$.}\smallskip
\end{property}\smallskip

\emph{Proof.} Since $I(A_1\!:...:\!A_n)_{\rho}<+\infty$  by the upper bound (\ref{nMI-UB}), the claim of the proposition  follows from
 Lemma \ref{FDA-LR+} in Section 4.1.4 and  Proposition \ref{E-I-LS}B in Section 4.3.2. $\Box$

\section{Concluding remarks}

In this article, we describe a method for studying characteristics of multipartite infinite-dimensional quantum systems
based on a finite-dimensional approximation of these characteristics. This method can be applied to any characteristic of a multipartite
quantum system provided that a uniform continuity bound with the energy-type constraint of one of two types (described in Section 2.3) is constructed
for this characteristic (or, at least, the existence of such a continuity bound is proved).

We consider applications of the new approximating technique for analysis of several  important characteristics
of composite quantum systems for which the methods proposed before can not be applied. In fact, one can use this technique
for studying any characteristics from the classes $I^{m}_n$ and $II^{m}_n$ introduced in Section 2.3 (an incomplete list of such characteristics is presented in the Appendix).

One of the advantages of the proposed technique is its applicability to analysis of the regularized versions of correlation and entanglement measures
in composite quantum systems (which are difficult to analyze using standard methods due to the complexity of their definition).  So, the results
on the properties of  the regularized relative entropy of $\pi$-entanglement and  the regularized Rains bound
obtained in the article are the most nontrivial applications of this technique.

The nonstandard use of the uniform continuity bounds with the energy-type constraints considered in the article shows the necessity to develop
new methods for deriving such  continuity bounds. They can be used for solving different questions
concerning a given characteristic of infinite-dimensional quantum systems not directly related to its uniform continuity.

\section*{Appendix: Basic characteristics from the classes  $I^{m}_n$ and $II^{m}_n$}

In the presentation of the main results of this article, the classes $I^{m}_n$ and $II^{m}_n$  introduced in Section 2.3 were significantly used.
Below is an (incomplete) list of characteristics of a quantum system $A_1...A_n$, indicating the concrete classes to which they belong with the justification.\footnote{There is a reason to believe that many of the characteristics marked as belonging to the class $II^{m}_n$  actually belong to the class $I^{m}_n$. But to proof this conjecture a rigorous proof is needed. For example,
it is pointed below that the $n$-partite squashed entanglement and c-squashed entanglement belong to the class $II^{n-1}_n$, while the bipartite squashed entanglement
is marked as a characteristic from $I^{1}_2$, because a $I^{1}_2$-type faithful uniform continuity bound for the latter is constructed in \cite{SE}.
It is natural to assume that updating the arguments in \cite{SE} it is possible to prove the existence of $I^{n-1}_n$-type faithful uniform continuity bounds for the $n$-partite squashed entanglement and c-squashed entanglement.}

\bigskip

\centerline{\textbf{Class $I_n^m$}}

\begin{itemize}
  \item the von Neumann entropy $S(\rho)$ of a state $\rho$ of $A_1$ ($I_1^1$, Section 2.3);
  \item the (extended) quantum conditional entropy $S(A_1|A_2)$ defined in (\ref{ce-ext}) ($I_2^1$, Section 2.3);
  \item the quantum  mutual information $I(A_1\!:...:\!A_n)$  ($I^{n-1}_{n}$, Section 2.3);
  \item the (extended) quantum conditional mutual information $I(A_1\!:...:\!A_n|A_{n+1})$ ($I^{n-1}_{n+1}$, Section 2.3);
  \item the one-way classical correlation $C_{A_2}$ in $A_1A_2$ with measured system $A_2$ ($I^{1}_{2}$, Section 2.3));
  \item the regularized one-way classical correlation $C^{\infty}_{A_2}$ in $A_1A_2$ with measured system $A_2$ ($I^{1}_{2}$, Section 2.3);
  \item the quantum discord in $A_1A_2$ with measured system $A_2$  ($I^{1}_{2}$, Section 2.3);
  \item the entanglement of formation $E_F$ in $A_1A_2$ ($I^{1}_{2}$, Section 2.3);
  \item the regularized entanglement of formation $E^{\infty}_F$ in $A_1A_2$ ($I^{1}_{2}$, Section 2.3);
  \item the squashed entanglement $E_{\rm sq}$ in $A_1A_2$ ($I^{1}_{2}$, Section 2.3);
  \item the topological entanglement entropy in $A_1A_2A_3$ ($I_3^1$, the representation in \cite[Remark 9]{CMI} and the properties of the QCMI mentioned in \cite[Section 4.2]{QC}).
\end{itemize}

\bigskip\pagebreak

\centerline{\textbf{Class $II_n^m$}}
\medskip

\begin{itemize}
  \item all the characteristics belonging to the class $I_n^m$ (described before);
  \item the one-way classical correlation $C_{A_1}$ in $A_1A_2$ with measured system $A_1$ ($II^{1}_{2}$, Section 2.3);
  \item the regularized one-way classical correlation $C^{\infty}_{A_1}$ in $A_1A_2$ with measured system $A_1$ ($II^{1}_{2}$, Section 2.3);
  \item the quantum discord in $A_1A_2$ with measured system $A_1$ ($II^{1}_{2}$, Section 2.3);
  \item the constrained Holevo capacity  $\chi_{A_1}$ of the  partial trace over $A_2$ in $A_1A_2$  ($II^{1}_{2}$, \cite[Proposition 21]{QC});
  \item the relative entropy  of $\pi$-entanglement $E_{R,\pi}$ in $A_1...A_n$ for any set of partitions $\pi$ ($II^{n-1}_{n}$, Section 2.3);
  \item the regularized relative entropy  of $\pi$-entanglement $E^{\infty}_{R,\pi}$ in $A_1...A_n$ for any singleton set of partitions $\pi$ ($II^{n-1}_{n}$, Section 2.3);
  \item the relative entropy of NPT-entanglement $E_{R,PPT}^{\infty}$ in $A_1A_2$ ($II^{1}_{2}$,  Remark \ref{ER-PPT});
  \item the regularized relative entropy of NPT-entanglement $E_{R,PPT}^{\infty}$ in $A_1A_2$ ($II^{1}_{2}$,  Remark \ref{ER-PPT});
  \item the Rains bound in $A_1A_2$ ($II^{1}_{2}$, Section 4.22);
  \item the regularized Rains bound in $A_1A_2$ ($II^{1}_{2}$, Section 4.22);
  \item the squashed entanglement $E_{\rm sq}$ in $A_1...A_n$ ($II^{n-1}_{n}$, \cite[Proposition 24]{QC});
  \item the c-squashed entanglement $E_{\rm c\textup{-}sq}$ in $A_1...A_n$ ($II^{n-1}_{n}$, \cite[Proposition 24]{QC});
  \item the entropy reduction of a local measurement in $A_1A_2$ with measured system $A_1$ ($II^{1}_{2}$, \cite[Corollary 3]{CMC});
  \item the function $\rho\mapsto I(B_1\!:...:\!B_n)_{\Phi_1\otimes...\otimes\Phi_n(\rho)}$ on $\S(\H_{A_1...A_n})$, where $\Phi_1:A_1\rightarrow B_1$,..., $\Phi_n:A_n\rightarrow B_n$ are arbitrary quantum channels ($II^{n-1}_{n}$, \cite[Proposition 24]{CBM}).\footnote{The observation in \cite[Section V]{CBM}
      shows that many other functions of this type belong to the class $II^{m}_{n}$.}
\end{itemize}

\bigskip

I am grateful to A.S.Holevo, G.G.Amosov and E.R.Loubenets for useful discussion. I am grateful to M.Plenio and to the participants
of his seminar at Ulm University (where the initial version of this work was reported) for useful comments.
I am also grateful to L.Lami for the collaboration that motivated the final stage of this  research. Special thanks to S.Becker for construction of a counterexample showing that the condition (\ref{FA-SC+})
is not a criterion of the FA-property.

\end{document}